\documentclass[acmtog,authorversion]{acmart}
\acmSubmissionID{230}

\usepackage{booktabs} % For formal tables
\usepackage{amsmath}
\usepackage{amssymb}
\usepackage{amsfonts}
\usepackage{url}
\usepackage{subfigure}
\usepackage{here}
\usepackage{algorithmic}
\renewcommand{\algorithmicrequire}{\textbf{Input:}}
\renewcommand{\algorithmicensure}{\textbf{Output:}}

\newcommand{\argmin}{\mathop{\rm arg~min}\limits}

\usepackage{color}
\newcommand{\chkA}[1]{#1}
\newcommand{\chkB}[1]{#1}
\newcommand{\chkC}[1]{#1}
\newcommand{\chkD}[1]{#1}
\newcommand{\chkE}[1]{#1}
\newcommand{\chkK}[1]{#1}
\newcommand{\chkG}[1]{#1}
\newcommand{\chkH}[1]{#1}
\newcommand{\chkL}[1]{#1}
\newcommand{\chkM}[1]{#1}
\newcommand{\chkN}[1]{#1}
\newcommand{\chkO}[1]{#1}
\newcommand{\chkP}[1]{#1}
\newcommand{\chkQ}[1]{#1}
\newcommand{\chkR}[1]{#1}
\newcommand{\chkS}[1]{#1}
\newcommand{\chkT}[1]{#1}
\newcommand{\chkU}[1]{#1}
\newcommand{\chkV}[1]{#1}
\newcommand{\chkURL}[1]{\textcolor{magenta}{#1}}

\setcitestyle{square}

\usepackage[ruled]{algorithm2e} % For algorithms

\SetAlFnt{\small}
\SetAlCapFnt{\small}
\SetAlCapNameFnt{\small}
\SetAlCapHSkip{0pt}
\IncMargin{-\parindent}

\acmJournal{TOG}
\acmYear{2019}\acmVolume{38}\acmNumber{6}\acmArticle{175}\acmMonth{11} \acmDOI{10.1145/3355089.3356523}

\begin{document}
% Title portion
\title{\chkQ{Animating Landscape: Self-Supervised Learning of \chkT{Decoupled} Motion and Appearance for Single-Image Video Synthesis}} 

\author{Yuki Endo}
\affiliation{%
  \institution{University of Tsukuba \& Toyohashi University of Technology}
}
\email{endo@cs.tsukuba.ac.jp}

\author{Yoshihiro Kanamori}
\affiliation{%
  \institution{University of Tsukuba}
}
\email{kanamori@cs.tsukuba.ac.jp}

\author{Shigeru Kuriyama}
\affiliation{%
  \institution{Toyohashi University of Technology}
}
\email{sk@tut.jp}

\renewcommand\shortauthors{Y. Endo, Y. Kanamori, and S. Kuriyama}

\begin{teaserfigure}
   \includegraphics[width=1.\linewidth, clip]{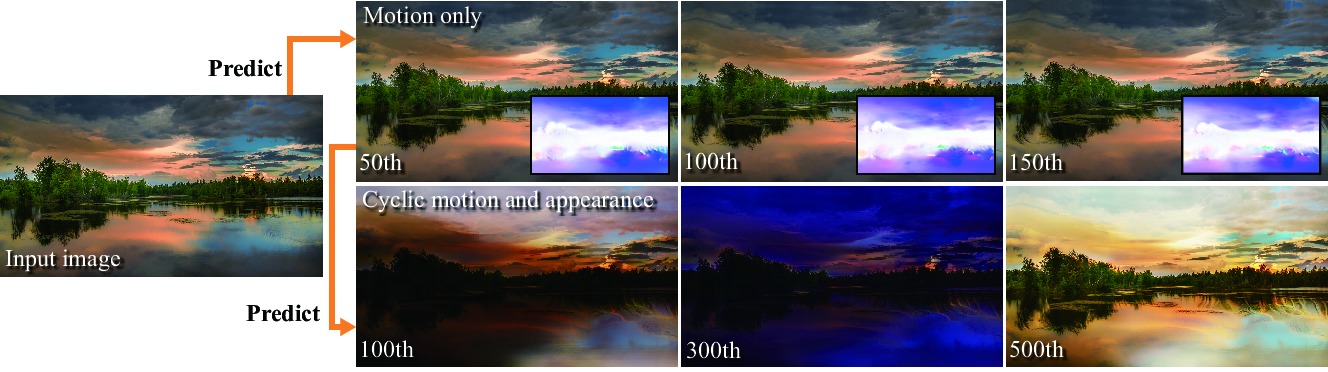}
   \caption{
   Given a single scenery image, our method predicts \chkE{the} motion (e.g., moving clouds) and appearance (e.g., time-varying colors) \chkR{separately} \chkE{to generate a cyclic animation via \chkR{self-supervised} learning of} time-lapse videos using our convolutional neural networks that infer backward flow fields (insets) and color transfer functions for \chkE{converting} the input image. \chkQ{The flow fields are visualized using the colormap shown in Figures~\ref{fig:res1} and~\ref{fig:res1_2}. }The output frame size is $\chkD{1,024}\times576$. Please see the \chkD{supplemental video for} \chkE{the resultant \chkR{animations}}.
   \chkV{Input photo: Pixabay/Pexel.com.} 
   }\label{fig:teaser}
\end{teaserfigure}

\begin{abstract}
\chkH{Automatic} generation \chkH{of a high-quality video} from a single image remains a challenging \chkH{task despite} the recent advances in deep generative models.
%This paper proposes a method that can create \chkH{a high-\chkR{resolution}, long-term \chkR{animation}} from a single \chkH{scenery} image using convolutional neural networks (CNNs).
This paper proposes a method that can create \chkH{a high-\chkR{resolution}, long-term \chkR{animation}} using convolutional neural networks (CNNs) \chkT{from a single landscape image where we mainly focus on skies and waters}.
\chkL{\chkM{Our key \chkH{observation is} that} \chkH{the} {\em motion} (e.g., moving clouds) and {\em appearance} (e.g., time-varying colors in the sky) \chkH{in natural scenes have different time scales. 
%\chkM{We \chkH{thus} learn} them asynchronously and \chkQ{predict} them with separate control while handling future uncertainty in both predictions by introducing latent codes}}.
\chkM{We \chkH{thus} learn} them \chkR{separately} and \chkQ{predict} them with \chkT{decoupled} control while handling future uncertainty in both predictions by introducing latent codes}}.
\chkH{Unlike previous methods that infer output frames directly, our CNNs predict spatially-smooth intermediate data, i.e., for motion,} flow \chkH{fields for warping, and for appearance,} color transfer maps\chkH{, via \chkR{self-supervised} learning, i.e., without \chkR{explicitly-provided} ground truth.}
% ↑ self-supervised の説明で「人手で与えた正解データ」という説明だと、本論文で出てくる「人工正解データ」が該当しないことになるので不適です。
%\chkH{These intermediate outputs are applied not to each previous output frame, but to the input image only once for each output frame.
%This design is crucial to avoid error accumulation in long-term predictions, which is the \chkL{essential} problem in previous recurrent approaches.}
\chkH{These intermediate \chkR{data} are applied not to each previous output frame, but to the input image only once for each output frame.
% 改めて読むと、この時点ではまだ提案手法 (のモーション予測) が recurrent approach だということを言っていないのに「前のフレームは入力しない」と言っていて、初見では意味不明だなと思いました。
This design is crucial to \chkR{alleviate} error accumulation in long-term predictions, which is the \chkL{essential} problem in previous recurrent approaches.}
\chkH{The output frames can be looped like cinemagraph, and also be controlled directly by specifying latent codes or indirectly via \chkL{visual} annotations.
We demonstrate the effectiveness of our method through comparisons with the state-of-the-arts on video prediction as well as appearance manipulation. Resultant videos, codes, and datasets will be available at~\href{http://www.cgg.cs.tsukuba.ac.jp/~endo/projects/AnimatingLandscape}{\chkURL{http://www.cgg.cs.tsukuba.ac.jp/\~{}endo/projects/AnimatingLandscape}}.
}
\end{abstract}




\maketitle

\section{Introduction}
\label{sec:Introduction}

\chkH{From a scenery image, humans can imagine how the clouds move and the sky color changes as time goes by. Reproducing such transitions in scenery images is a common subject of not only artistic contents called {\em cinemagraph}\chkN{~\cite{DBLP:journals/tog/BaiAAR12,DBLP:journals/tog/LiaoJH13,DBLP:conf/iccv/OhJJWKK17}} but also various techniques for image manipulation \chkN{(e.g., scene completion~\cite{DBLP:journals/tog/HaysE07}, time-lapse mining~\cite{DBLP:journals/tog/Seitz15}, attribute editing~\cite{DBLP:journals/tog/ShihPDF13,DBLP:journals/tog/LaffontRTQH14}, and sky replacement~\cite{DBLP:journals/tog/TsaiSLSY16}).} 
However, creating a natural animation from a scenery image remains a challenging task in the fields of computer graphics and computer vision.}

\chkH{Previous methods in this topic} can be grouped into two categories. 
The first \chkA{category} is \chkA{the} example-based approach \chkA{that} can create \chkA{a} realistic animation by \chkA{transferring exemplars}, 
e.g., fluid motion~\cite{DBLP:journals/cgf/OkabeAIS09,DBLP:journals/cgf/PrashnaniNVS17} \chkA{or time-varying scene appearance}~\cite{DBLP:journals/tog/ShihPDF13}.
\chkA{This approach}, \chkK{however}, heavily \chkA{relies} on \chkA{reference} videos \chkA{that match the target scene}. 
The other \chkA{category} is \chkA{the} learning-based approach, \chkA{which is typified by the recent remarkable} techniques using {\em Deep Neural Networks} (DNNs).

DNN-based techniques have achieved great success in image generation tasks, particularly thanks to {\em Generative Adversarial Networks} (GANs)~\cite{PGGAN,wang2018pix2pixHD}  \chkE{and other generative} models, e.g., {\em Variational \chkE{Auto-Encoders}} (VAEs), \chkE{which} were also used to generate a video~\cite{xiong2018learning,Prediction-ECCV-2018} \chkK{from a single image.} 
\chkA{Unfortunately, the resolution and quality of the \chkD{resulting} videos are far lower than those generated in image generation tasks.}
\chkA{One reason for the poor results is that the spatiotemporal domain of videos is too large for \chkE{generative models} to learn, compared to the domain of images}.
\chkA{Another reason is the uncertainty in future frame predictions;} for example, \chkA{imagine clouds in the sky in a single still image.
The clouds might move left, right, forward\chkD{,} or backward in the next frames according to the} environmental factors \chkK{such as} wind.
\chkA{Due to} \chkK{such} uncertainty, learning \chkA{a} unique output from a single input (i.e., one-to-one mapping) is intractable and unstable. 
\chkA{The} recent work \chkA{using} \chkE{VAEs} to handle \chkA{the} uncertainty is still insufficient \chkD{for generating} realistic and diverse results~\cite{Prediction-ECCV-2018}.

In this paper, we propose a \chkA{learning-based approach} that can create a high-\chkR{resolution} \chkQ{video from a single outdoor} image \chkA{using} DNNs. 
This is accomplished by \chkR{self-supervised} learning with \chkA{a} training \chkA{dataset} of time-lapse videos.
%\chkT{We mainly focus on generating landscape animations especially for skies and waters, and this} is accomplished by \chkR{self-supervised} learning with \chkA{a} training \chkA{dataset} of time-lapse videos.
\chkA{Our key idea is to learn the} {\em motion} (e.g., \chkA{moving} clouds in the sky and \chkA{ripples} on \chkD{a} lake) and \chkA{the} {\em appearance} (e.g., \chkA{time-varying colors in daytime, sunset,} and night) separately, \chkK{by considering} their spatiotemporal differences.
\chkK{For example, clouds move rapidly on the scale of seconds, whereas sky color changes slowly on the scale of tens of minutes, as shown in the riverside scene of Figure~\ref{fig:teaser}.
Moreover, the moving clouds exhibit detailed patterns, whereas the sky color varies overall smoothly.}

\chkA{With this observation in mind, we learn/predict the motion and appearance \chkR{separately} using two types of DNN models (Figure~\ref{fig:Overview}) as follows.}
%\chkE{For motion, because large-timestep prediction is difficult, our motion predictor learns the differences between two successive frames as a backward flow field.
\chkE{For motion, because \chkR{one-shot} prediction \chkR{of complicated motion} is difficult, our motion predictor learns the differences between two successive frames as a backward flow field.
Long-term prediction} is achieved by inputting the predicted frames recurrently.
\chkE{Motion-added} images \chkK{are then} generated \chkE{at high resolution} by reconstructing pixels from the input image after tracing back the flow fields.
For appearance, our predictor learns \chkE{the differences between the input frame and arbitrary frames in each training video} \chkA{as spatially-smooth color transfer functions}.
In the prediction phase, \chkE{color transfer functions are predicted at sparse frames and are applied to the motion-predicted frames via temporal interpolation.}
\chkT{We assume that the motion and color variations in landscape time-lapse videos are spatiotemporally-smooth, and enforce such regularization in our training, which works well particularly with the motions of clouds in the sky and waves on water surfaces as well as the color variations of dusk/sunset in the sky.}
\chkH{The output animation can be looped, inspired by cinemagraph.}

\chkA{To combat the uncertainty of future prediction, we also extract latent codes both for motion and appearance, which depict potential future variations and enable \chkD{the learning of} one-to-many mappings.
The user can manipulate the latent codes to control the motion and appearance smoothly in the latent space.}
\chkA{Note that the backward flow fields, color transfer functions, and latent codes are learned in \chkR{a  self-supervised} manner because their ground-truth data are not available in general.
Unlike previous techniques using 3D convolutions~\cite{DBLP:conf/nips/VondrickPT16,xiong2018learning,Prediction-ECCV-2018} for predictions with fixed numbers of frames, our networks \chkK{adopt} 2D convolutional layers. \chkK{This approach allows} fast learning and prediction and \chkE{abolishes} the limit \chkD{on} the number of predicted frames by recurrent feeding.}

\chkK{Our main contributions \chkE{are} summarized as follows:
\begin{itemize}
\item A framework for automatic synthesis \chkQ{of animation from a single outdoor image} with fully convolutional neural network (CNN) models for motion and appearance prediction\chkE{,}
\item Higher-resolution and longer-term movie generation by training with only hundreds to thousands of time-lapse videos in \chkR{a self-supervised} manner\chkE{, and}
\item \chkT{Decoupled} control mechanism for the variations of motion and appearance that change at different time intervals \chkE{based on latent codes}.
\end{itemize}
We demonstrate these advantages by comparing \chkP{with} various methods of \chkE{video prediction} as well as \chkE{attribute transfer}.}
\chkR{Our user study reveals that our results are subjectively evaluated as competitive or superior to those of previous methods or commercial software (see Appendix~\ref{sec:UserStudy}).}
\chkE{We also show applications for controlling the motion and appearance of output frames.}

\section{Related Work}
\label{sec:RelatedWork}
\chkK{Here we briefly review the related work of our technical components; optical flow \chkP{prediction}, color transfer, style transfer, video prediction, and so forth.}

\subsection{\chkH{Optical Flow} Prediction}
\label{sec:flow}
\chkH{Optical flow} prediction from a single image has been \chkA{studied \chkH{with} various \chkH{approaches}}.
\chkA{Supervised approaches using CNNs have also been proposed~\cite{DBLP:conf/iccv/WalkerGH15,DBLP:journals/corr/abs-1712-04109}.}
%\chkA{Supervised approaches for estimating optical flow using CNNs have also been proposed~\cite{DBLP:conf/iccv/WalkerGH15,DBLP:journals/corr/abs-1712-04109}.}
The point is how to prepare \chkE{ground-truth flow fields} for supervised learning.
The above-mentioned methods exploited existing techniques (e.g., {\em FlowNet}~\cite{DBLP:conf/iccv/DosovitskiyFIHH15}, {\em DeepFlow}~\cite{DBLP:conf/iccv/WeinzaepfelRHS13,DBLP:journals/ijcv/RevaudWHS16}, and {\em SpyNet}~\cite{DBLP:conf/cvpr/RanjanB17}) for generating ground-truth flow fields \chkE{synthetically}.
However, we confirmed that \chkE{previous methods} \chkK{relying \chkE{on} such synthetic data} \chkE{yield} poor prediction\chkD{s} for time-lapsed videos (see Section~\ref{sec:comparison1}) for which no genuine ground-truth is available.

\chkK{Recently, \chkR{self-supervised} approaches have been proposed for estimating flow fields between two input images~\cite{DBLP:conf/aaai/RenYNLYZ17, Wang_2018_CVPR}.}
We also adopt \chkR{a self-supervised} approach where a flow field is computed between two consecutive frames.
The main difference against the existing approaches is that we input only a single image in the inference phase and handle the prediction uncertainty by introducing latent codes.

\begin{figure*}
   \includegraphics[width=1.\linewidth, clip]{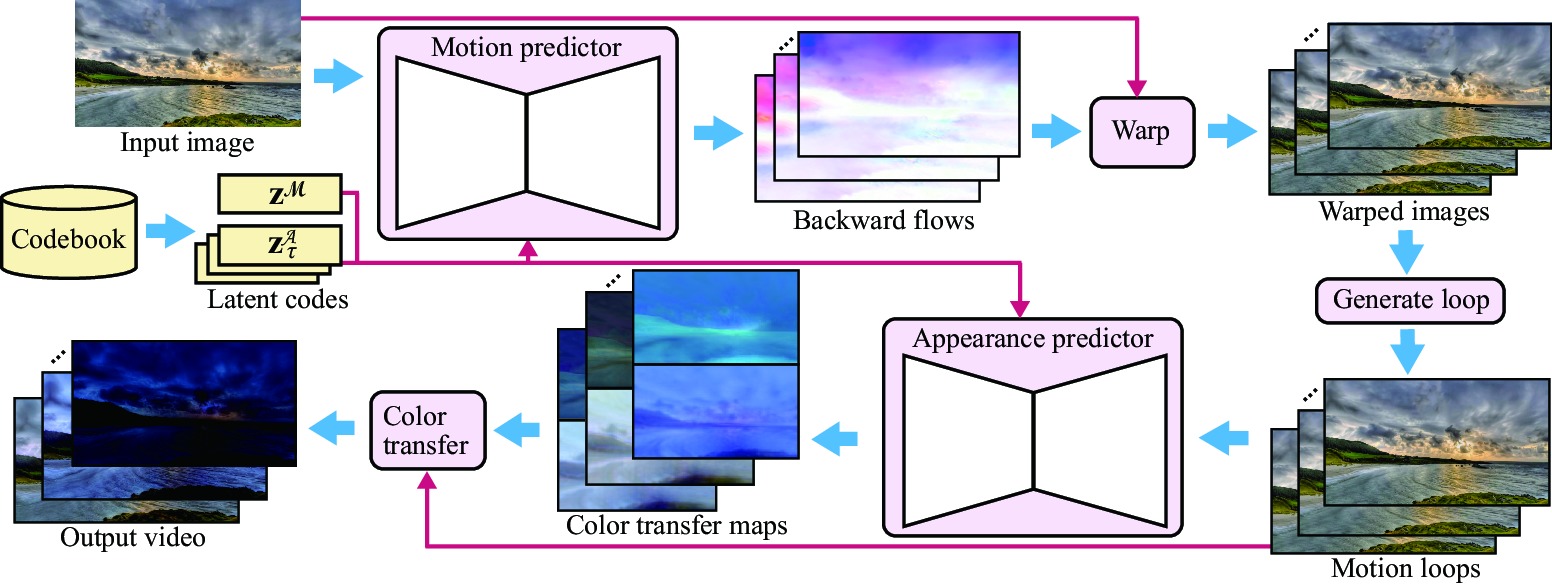}
   \caption{
   Overview of \chkR{our inference pipeline}. Given \chkE{the input image and latent codes that control future variations}, the motion predictor generates future backward flows\chkE{. The flows} are used to warp the input image \chkE{to synthesize motion-added images, which are then converted to a cyclic motion loop. The} appearance predictor generates color transfer maps, which are finally used \chkE{for color transfer} to obtain \chkE{the} output video. 
   \chkV{Input photo: Per Erik Sviland (Vvuxdqn-0vo)/Youtube.com.} 
   }\label{fig:Overview}
\end{figure*}

\subsection{Appearance Manipulation}
\chkA{Color transfer~\cite{DBLP:journals/cga/ReinhardAGS01} is a fundamental technique} for changing color appearance.
This technique \chkA{\chkD{makes} the overall color of a target image \chkD{conform} to that of a reference image} while retaining the scene structure of \chkA{the target, by matching the statistics (i.e., the mean and standard deviation) of the two images.
The original method~\cite{DBLP:journals/cga/ReinhardAGS01} is enhanced by respecting local color distributions} using soft clustering~\cite{DBLP:conf/cvpr/TaiJT05} or semantic region correspondence~\cite{DBLP:journals/cgf/WuDKMPZ13}. 
\chkG{There is also a color tone transfer method specialized in sky replacement~\cite{DBLP:journals/tog/TsaiSLSY16}. }

Style transfer can convey richer information\chkD{,} \chkA{including textures using DNNs}\chkD{,} than color transfer.
The \chkA{original} work~\cite{DBLP:conf/cvpr/GatysEB16} \chkE{in this literature} optimizes an output image via backpropagation of the perceptual loss for retaining the source content and style loss for transferring the target style.
\chkA{Faster \chkE{transfer} is accomplished by pre-training autoencoders} for specific \chkE{styles}~\cite{DBLP:conf/eccv/JohnsonAF16} and \chkA{using whitening and coloring transforms (WCTs)} for arbitrary \chkE{styles}~\cite{DBLP:conf/nips/LiFYWLY17}. 
Semantic region correspondence \chkA{can also be} \chkA{integrated}~\cite{DBLP:conf/cvpr/LuanPSB17}. 
%\chkA{However, the strong expression power of style transfer works negatively for our purpose; it yields unnatural results for various scenes due to overfitting, as shown in our experiments in Section~\ref{sec:Experiment}.}
\chkA{However, the strong expression power of style transfer works negatively for our purpose; it yields unnatural results for various scenes due to overfitting \chkR{(see Section~\ref{sec:Experiment})}.}
\chkK{We instead delegate the texture transfer to our motion predictor and change the color appearance using the transfer functions that can avoid overfitting.}

\chkA{A recent arXiv paper~\cite{DBLP:journals/corr/abs-1808-07413}} presents a method to manipulate attributes of natural scenes (e.g., night, sunset, and winter) via \chkA{style transfer~\cite{DBLP:conf/cvpr/LuanPSB17} and image synthesis using \chkD{a} conditional GAN.
From a semantic layout of the input image and a target attribute vector, the method first synthesizes an intermediate style image, which is then used for style transfer with the input image.
Animations can be generated by gradually changing the attribute vector, but enforcing temporal coherence is difficult with this two-step synthesis.
In contrast, our method offers smooth appearance \chkO{transitions} via latent-space} interpolation, as we demonstrate in Section~\ref{sec:Experiment}.

\subsection{Video Generation from a Still Image}
\chkA{An} early attempt to animate a natural scene in a single image \chkD{was} \chkA{a procedural approach called} stochastic motion texture~\cite{DBLP:journals/tog/ChuangGZCSS05}. 
\chkD{This approach \chkE{generates}} simple quasi-periodic \chkA{motions of individual components\chkD{,} such as} swaying trees, rippling water, and bobbing boats\chkD{,} \chkA{with parameter tuning for each \chkD{component}}.

\chkA{Example-based approaches} can reproduce realistic \chkA{motion} \chkC{or appearance} without complex parameters by directly transferring \chkA{reference videos~\cite{DBLP:journals/cgf/OkabeAIS09,DBLP:journals/cgf/OkabeAO11,DBLP:journals/vc/OkabeDA18,DBLP:journals/cgf/PrashnaniNVS17,DBLP:journals/tog/ShihPDF13}.
However, their results become unnatural without an appropriate reference video similar to the input image.}
\chkK{This issue can be alleviated \chkE{at the cost of} larger database and \chkE{larger} computational \chkE{resource}.}
\chkK{Also, existing techniques often impose tedious manual processes for specifying\chkO{,} e.g., alpha mattes\chkP{,} flow fields and regions for fluid.}
%\chkK{Also, existing techniques often impose tedious manual processes for specifying\chkO{,} e.g., alpha mattes and flow fields and regions for fluid.}
Our method can generate \chkA{high-\chkR{resolution}} \chkQ{videos} using only hundreds of \chkA{megabytes of pre-trained data within} a few minutes \chkA{on} a single GPU.
\chkE{Our method can run automatically yet can also be controlled using latent codes.}

\chkK{Example-based appearance transfer~\cite{DBLP:journals/tog/ShihPDF13} can reproduce the time-varying color variations in a static image with \chkP{a} reference video. However, simple frame-by-frame transfer suffers from flickering artifacts for dynamic objects in the scene.}
\chkO{Key-frame} interpolation alleviates such flickering, which is not directly applicable if the outputs are videos containing dynamic objects\chkE{,} unlike ours.
\chkE{The method by Laffont et al.~\shortcite{DBLP:journals/tog/LaffontRTQH14} achieves appearance transfer using a manually-annotated database whereas our training datasets do not require manual annotations.}

\chkA{The past few years have witnessed the \chkD{dramatic} advance\chkD{s in} learning-based approaches\chkH{, particularly using DNN.}
\chkH{For example, DNN architectures used for video prediction include} not only 2DCNN \cite{DBLP:conf/nips/Xue0BF16,ICLR06Mathieu,ICLR07Lotter,ICLR08Babaeizadeh,DBLP:conf/cvpr/HaoHB18} but also} convolutional Recurrent Neural Network\chkD{s} (cRNN\chkD{s})~\cite{DBLP:journals/corr/RanzatoSBMCC14}, Long Short-Term Memory (LSTM) \cite{DBLP:conf/icml/SrivastavaMS15,DBLP:conf/eccv/ZhouB16,DBLP:conf/nips/DentonB17,DBLP:conf/eccv/ByeonWSK18}, and 3DCNNs~\cite{DBLP:conf/nips/VondrickPT16,xiong2018learning,Prediction-ECCV-2018}.
However, even with \chkP{the} state-of-the-art \chkD{techniques}~\cite{xiong2018learning,Prediction-ECCV-2018}, the frame length and \chkA{resolution} of generated videos are \chkA{quite} limited (\chkE{i.e.,} up to 16 or 32 frames at $128 \times 128$) due to \chkA{the} training complexity \chkA{and architecture design}. 
\chkA{In a sharp contrast,} our method can generate \chkA{much higher-resolution} videos with \chkA{an unlimited number of} frames \chkA{by} leveraging intermediate flow fields and color transfer \chkA{functions, as we \chkD{discuss} in Section~\ref{sec:Experiment}}.
Note that \chkA{a recent work by Li et al.~\shortcite{Prediction-ECCV-2018}} also predicts flow fields like our method.
The key differences \chkD{of our method} are that i) \chkD{their method} requires ground-truth flow fields\chkD{,} whereas ours does not (i.e., \chkD{learning is \chkR{self-supervised});} ii) \chkD{their method} uses \chkE{3DCNN}\chkD{,} whereas ours uses \chkE{2DCNN}\chkD{, which reduces} \chkA{the} training complexity\chkD{;} \chkA{and iii) \chkD{their method} cannot provide direct control over appearance transition\chkD{,} whereas ours can 
%\chkD{because the training of motion and appearance is decoupled and asynchronous}}.
\chkD{because we employ \chkT{decoupled} training of motion and appearance}}.

\section{Method Overview}
\label{sec:Overview}
\chkE{Figure~\ref{fig:Overview} shows the whole pipeline of our \chkQ{video synthesis}, where our method first generates motion-added frames from the single input image, \chkP{optionally} makes them looped by linear blending, and then applies color transfer to each frame.
As we explained in Section~\ref{sec:Introduction}, our motion predictor infers backward flows recurrently, whereas our appearance predictor infers a color transfer function for each frame.
This design is crucial for \chkR{handling} the well-known problem in recurrent inference where error accumulates in the cycled output frames~\cite{DBLP:conf/nips/ShiCWYWW15}; 
%in our motion prediction, error accumulates in the backward flows, which we assume are spatially-smooth and thus less sensitive to error, and each predicted frame is reconstructed by tracing back to the input image. 
in our motion prediction, error accumulates in the backward flows, which we assume are spatially-smooth and thus less sensitive to error. 
\chkR{Each predicted frame is reconstructed by tracing back to the input image to avoid error accumulation in RGB values due to repetitive color sampling.}
In our appearance prediction, \chkR{on the other hand,} we avoid recurrent feeding and infer time-varying color transfer maps from the input image directly.
Blur artifacts and error accumulation in output RGB values can be avoided because the per-pixel RGB value in the input image is sampled only once for each output frame in both predictions.}

\chkE{We handle the future uncertainty in both predictions using latent codes extracted in the training phases.
By assuming that the overall motion throughout an animation sequence is similar, we control the motion in a single animation only with a single latent code. 
On the other hand, because our appearance predictor is trained with frame pairs between an input image and arbitrary frames in each training video, we require a latent code to control the appearance of each frame.
Consequently, for appearance control of an animation sequence, we require a sequence of latent codes, which has the same length as the output frame length.
The latent codes can be specified automatically or manually, from latent codes stored during training (hereafter we refer to \chkH{them as a {\em codebook})}.}

\section{Models}
\label{sec:Model}
\chkA{Hereafter\chkD{,} we describe our network models and distinguish the \chkP{notations} \chkD{between} motion and appearance with \chkD{the} superscripts $\mathcal{M}$ and $\mathcal{A}$, respectively.
%\chkA{Hereafter\chkD{,} we describe our network models and distinguish\chkD{, in} the notation \chkD{between} motion and appearance with \chkD{the} superscripts $\mathcal{M}$ and $\mathcal{A}$, respectively.
Our motion predictor $P^\mathcal{M}$ and appearance predictor $P^\mathcal{A}$} are encoder-decoders \chkA{with the same architecture} of \chkD{a fully CNN}. \chkA{The inputs of the predictors are i)} a linearly-normalized RGB image $\chkA{\textbf{I}} \in [-1,1]^{w \times h \times 3}$ (where $w$ and $h$ are image width and height) \chkA{and ii) a latent code $\textbf{z}$ to account for uncertainty of future prediction\chkD{.}
\chkD{Code} $\textbf{z}^\mathcal{M}$ controls the motion in a whole sequence\chkD{,} whereas $\textbf{z}^\mathcal{A}$ \chkD{controls} the appearance of only a single frame. 
The outputs of the predictors are multi-channel intermediate maps that are then used to convert the input image \textbf{I} into} an output RGB frame $\hat{\textbf{O}} \in [-1,1]^{w \times h \times 3}$, where we use \chkD{a circumflex} (~$\hat{ }$~) to indicate \chkA{an} inferred output. 

\chkA{In the following subsections, for motion and appearance, we first explain the inference phase to illustrate the use cases of the predictors and then describe how to train \chkD{the network\chkP{s}}.} 

\begin{figure}
   \includegraphics[width=1.\linewidth, clip]{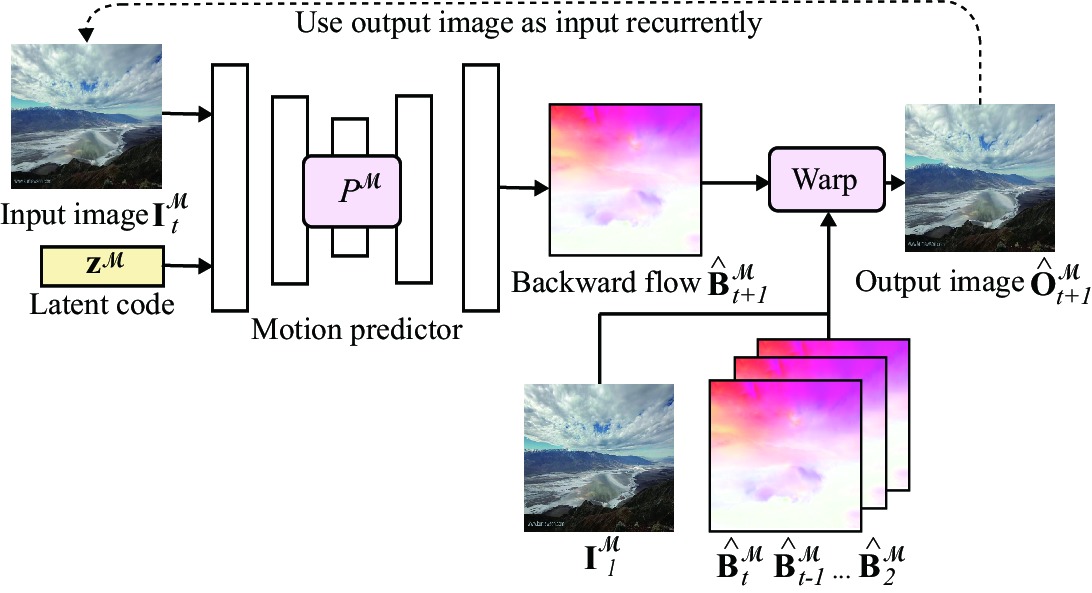}
   \caption{
   \chkE{Recurrent inference using} the motion predictor \chkE{$P^\mathcal{M}$}. \chkE{A} backward flow $\hat{\textbf{B}}_{t+1}^\mathcal{M}$ at \chkE{time} $t+1$ \chkE{is predicted} from an image $\textbf{I}_t^\mathcal{M}$ at \chkE{time} $t$ and a latent code $\textbf{z}^\mathcal{M}$. An output frame $\hat{\textbf{O}}_{t+1}^\mathcal{M}$ is obtained by warping $\textbf{I}_t^\mathcal{M}$ using  $\hat{\textbf{B}}_{t+1}^\mathcal{M}$. $\hat{\textbf{O}}_{t+1}^\mathcal{M}$ is \chkE{then} used as \chkE{the next input} $\textbf{I}_{t+1}^\mathcal{M}$. This procedure is \chkD{repeated} to obtain multiple frames. 
   \chkV{Input photo: echoesLA (zleuiAR2syI)/Youtube.com.}
   }\label{fig:WPnet_test}
\end{figure}

\subsection{Motion \chkD{Predictor}}
\label{sec:motion_predictor}
\paragraph*{Inference.} %
Given an input image $\mathbf{I}_{t=1}^\mathcal{M}$ ($= \mathbf{I}$, where \chkE{$t$} indicates the \chkE{time, i.e., the frame number}) \chkE{and a latent code $\textbf{z}^\mathcal{M}$}, the motion predictor $P^\mathcal{M}$ infers \chkE{a} backward flow \chkE{field} $\chkE{\hat{\textbf{B}}_{t+1}^\mathcal{M}} \in [-1,1]^{w \times h \times 2}$ using $tanh$ \chkB{for} normalization. 
\chkE{Here the \chkB{pixel positions} of $\textbf{I}^\mathcal{M}_t$ \chkB{are normalized} as $[-1, 1]^2$.} 
\chkC{The pixel value at position $\mathbf{p}$ \chkE{in} the output frame \chkE{$\hat{\textbf{O}}_{t+1}^\mathcal{M}$} is then \chkE{reconstructed} by sampling \chkE{that in} the \chkE{current frame $\mathbf{I}_{t}^\mathcal{M}$} at $\mathbf{p} + \chkE{\hat{\textbf{B}}_{t+1}^\mathcal{M}(\mathbf{p})}$ \chkE{via bilinear} interpolation, where $\chkE{\hat{\mathbf{B}}_{t+1}^\mathcal{M}}(\mathbf{p})$ is the flow vector at $\mathbf{p}$\chkE{. We} call this \chkE{reconstruction} operation as {\em warping} \chkE{in this paper}.}
\chkE{We} recurrently use the predicted frames $\hat{\textbf{O}}_{t+1}^\mathcal{M}$ as the next motion predictor input $\textbf{I}_{t+1}^\mathcal{M} $\chkE{(see Figure~\ref{fig:WPnet_test})}. 
%\chkE{However, na\"ive recurrent feeding causes blurry output frames, as explained in Section~\ref{sec:Overview}. 
\chkE{However, \chkR{if we warp the current frame to synthesize the next frame na\"ively, the output frames will become gradually blurry}, as explained in Section~\ref{sec:Overview}. 
Therefore, we instead warp flow fields $\hat{\mathbf{B}}_{t=2}^\mathcal{M}$, $\hat{\mathbf{B}}_{t=3}^\mathcal{M}$, $\dots$,  $\hat{\mathbf{B}}_{t+1}^\mathcal{M}$ sequentially to accumulate flow vectors so that we can reconstruct each output frame $\hat{\textbf{O}}_{t+1}^\mathcal{M}$ from the input image $\mathbf{I}_{t=1}^\mathcal{M}$ directly.}

\begin{figure}
   \includegraphics[width=1.\linewidth, clip]{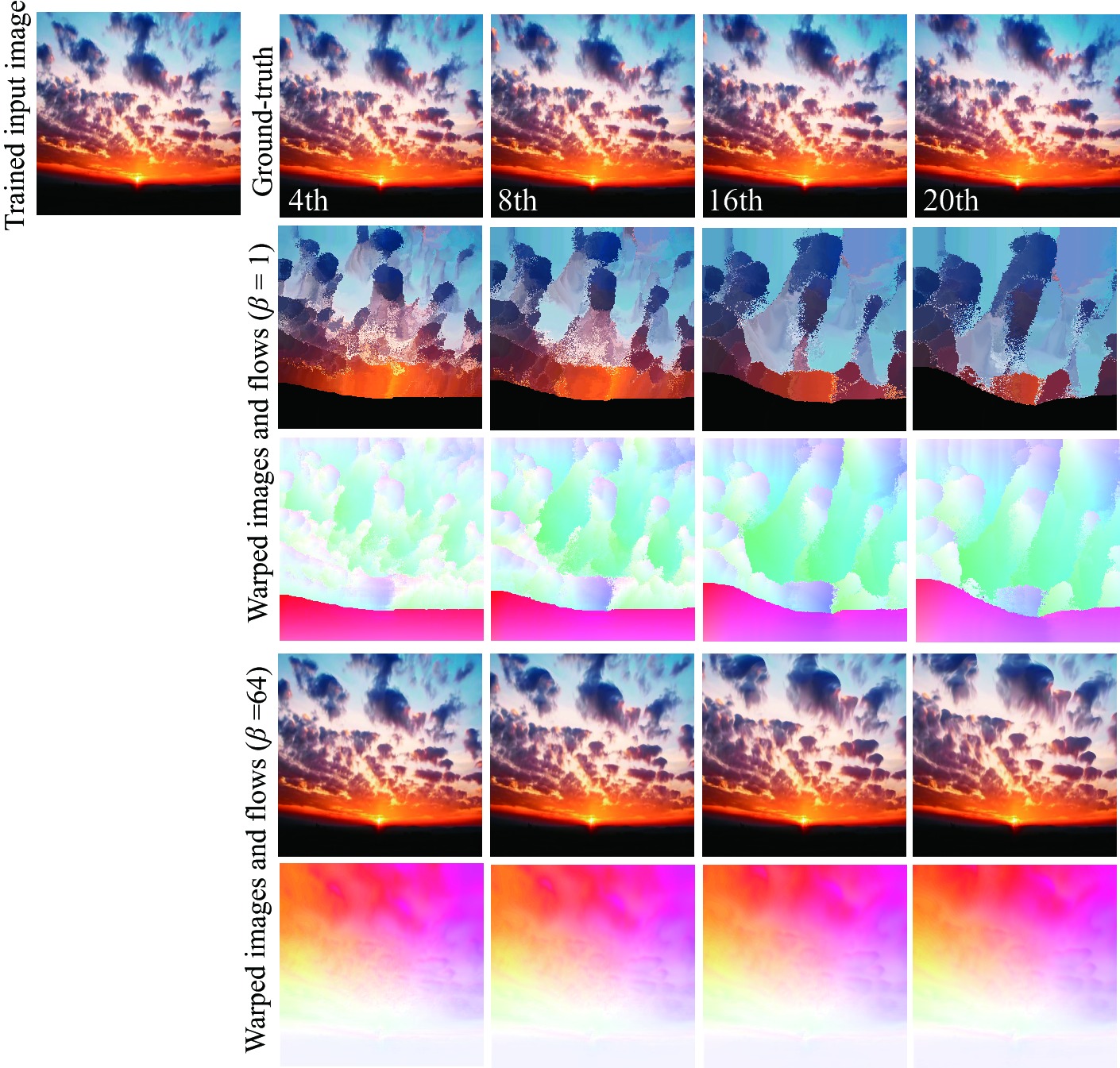}
   \caption{
   \chkE{Motion restriction with constant $\beta$. The inferred flows become inconsistent without restriction (i.e., $\beta=1$) but yield warped images close to ground-truth frames with restriction (i.e., $\beta=64$).} 
   \chkV{Input photos: Melania Anghel (rM7aPu9WV2Q)/Youtube.com.}
   }\label{fig:restrict_flow_range}
\end{figure}

\chkE{Predicting flow fields in our \chkR{self-supervised} setting} is challenging because \chkE{it is essentially to find correspondences between two consecutive frames with large degrees of freedom, which is easily trapped into local optima yielding inconsistent flow fields.}
\chkE{We thus} restrict the range of the output flow fields \chkE{both in prediction and training phases} \chkD{by assuming} that the objects do not move \chkE{significantly} in a single timestep. 
Specifically, we divide inferred flow fields by \chkE{a constant $\beta>1$} to restrict the range of their magnitudes to $[-1/\beta,1/\beta]^2$.
%Figure~\ref{fig:restrict_flow_range} demonstrates the effectiveness for using $\beta$.
Figure~\ref{fig:restrict_flow_range} demonstrates the effectiveness \chkP{of} $\beta$\chkP{, with the results obtained after training only using the single image shown at the top left}.
Without this restriction (\chkH{i.e.,} $\beta=1$), the estimated flow fields are inconsistent and \chkD{the} reconstructed images are corrupted. 
With this restriction (\chkL{e.g.,} $\beta=64$), \chkB{the reconstructed frames \chkD{match} to the} ground-truth \chkP{more closely, thanks} \chkD{to \chkP{the} consistent flow field estimation}. 
%With this restriction (\chkL{e.g.,} $\beta=64$), \chkB{the reconstructed frames \chkD{more closely match} to the} ground-truth \chkD{due to consistent flow field estimation}. 

\paragraph*{Training.} %
\chkK{A straightforward \chkE{way} for training the motion predictor is to minimize the difference between inferred and ground-truth flow fields, \chkE{as done in} \cite{DBLP:conf/iccv/WalkerGH15,DBLP:journals/corr/abs-1712-04109,Prediction-ECCV-2018}.
Our motion predictor, in contrast, learns future flow fields in \chkR{a self-supervised} manner only from time-lapse videos that have no ground-truth.}

\begin{figure}
   \includegraphics[width=1.\linewidth, clip]{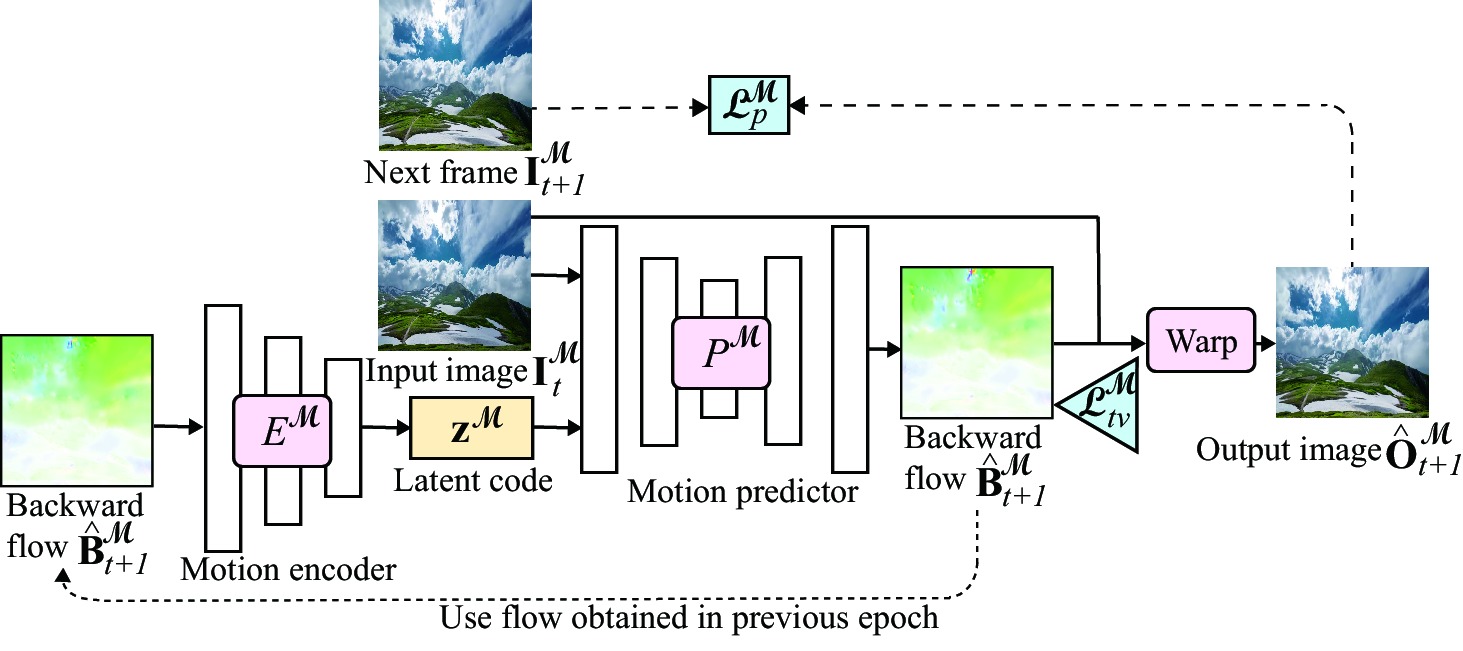}
   \caption{
  \chkE{Training of} the motion predictor \chkE{$P^\mathcal{M}$. The training is done using} consecutive frames $\textbf{I}^\mathcal{M}_{t}$ and \chkG{$\textbf{I}^\mathcal{M}_{t+1}$} \chkE{such that the loss $\mathcal{L}^\mathcal{M}_p$ between $\textbf{I}^\mathcal{M}_{t+1}$ and the output $\hat{\textbf{O}}^\mathcal{M}_{t+1}$ is minimized}. $\hat{\textbf{O}}^\mathcal{M}_{t+1}$ \chkE{is} obtained by warping $\textbf{I}^\mathcal{M}_{t}$ using \chkE{the backward flow} $\hat{\textbf{B}}^\mathcal{M}_{t+1}$\chkE{, which} is \chkE{regularized with the} loss $\mathcal{L}^\mathcal{M}_{tv}$. $\textbf{z}^\mathcal{M}$ is obtained by encoding \chkE{the previously} inferred flow using the motion encoder $E^\mathcal{M}$ \chkE{in our \chkR{self-supervised} setting where ground-truth flows do not exist}. 
  \chkV{Input photos: Akio Terasawa (gRnKhf9Kw1Q)/Youtube.com.}
   }\label{fig:WPnet}
\end{figure}

Figure~\ref{fig:WPnet} \chkP{outlines the training of the} motion predictor. 
%Figure~\ref{fig:WPnet} \chkD{provides an} overview of the motion predictor \chkD{training}. 
We first define \chkP{a} L2 loss for \chkP{the} network output $\hat{\textbf{O}}^\mathcal{M}_{t+1}$ obtained from \chkP{the} input \chkC{image} $\textbf{I}^\mathcal{M}_{t}$ and the next frame $\textbf{I}^\mathcal{M}_{t+1}$: 
\begin{equation}
  \mathcal{L}^\mathcal{M}_p = \|\textbf{I}^\mathcal{M}_{t+1}-\hat{\textbf{O}}^\mathcal{M}_{t+1}\|_2^2 \; , 
\end{equation}
\chkC{where $\|\cdot\|_2$ means L2 norm. }
Also, \chkB{a weighted} total variation loss is applied to the output flow field \chkS{for edge-preserving smoothing}\chkD{:}
\begin{equation}
\chkQ{
    \mathcal{L}^\mathcal{M}_{tv} = \displaystyle \sum_{\textbf{p}, \textbf{q} \in N(\textbf{p})}w(\textbf{I}^\mathcal{M}_{t+1}(\textbf{p}), \textbf{I}^\mathcal{M}_{t+1}(\textbf{q}))\|\hat{\mathbf{B}}_{t+1}^\mathcal{M}(\textbf{p})-\hat{\mathbf{B}}_{t+1}^\mathcal{M}(\textbf{q})\|_1, \label{eq:L_M_tv}
    }
\end{equation}
\begin{equation}
\chkQ{
    w(\textbf{x}, \textbf{y}) = \exp \left(-\frac{\|\textbf{x}-\textbf{y}\|_1}{\sigma}\right), 
    }
\end{equation}
%where \chkQ{$\textbf{q} \in N(\textbf{p})$ is a neighboring pixel of \textbf{p},} and $\sigma$ is a constant to determine influence of this term.
where \chkR{$N(\textbf{p})$ indicates the right and above neighbors of \textbf{p},} and $\sigma$ is a constant to determine influence of this term.
%where \chkB{$(i, j)$ are pixel} coordinates, $c$ denotes the channel index. Also, \chkC{$w(\mathbf{x})$ takes a gradient of RGB color as $\mathbf{x}$, and $\sigma$ is a constant to determine influence of this term. }
The output flow field $\hat{\mathbf{B}}^\mathcal{M}_{t+1}$ is smoothed \chkE{using the weighting function $w$ \chkP{such} that $\hat{\mathbf{B}}^\mathcal{M}_{t+1}$ respects} the color variations of the next frame $\mathbf{I}^\mathcal{M}_{t+1}$. 
Using weights $\lambda^\mathcal{M}_{p}$ and $\lambda^\mathcal{M}_{tv}$, our \chkA{total} training loss function is defined by
\begin{equation}
  \label{eq:loss_mt}
  \mathcal{L}^\mathcal{M} = \lambda^\mathcal{M}_{p}~\mathcal{L}^\mathcal{M}_p + \lambda^\mathcal{M}_{tv}~\mathcal{L}^\mathcal{M}_{tv} \; .
\end{equation}

To \chkP{handle} future uncertainty \chkA{and extract latent codes $\mathbf{z}^\mathcal{M}$}, we \chkC{simultaneously} train the motion encoder $E^\mathcal{M}$.
%To consider future uncertainty \chkA{and extract latent codes $\mathbf{z}^\mathcal{M}$ that control \chkD{the} output flow fields}, we \chkC{simultaneously} train the motion encoder $E^\mathcal{M}$.
\chkK{Problems similar to this one-to-many mapping were tackled in BicycleGAN~\cite{DBLP:conf/nips/ZhuZPDEWS17}, where latent codes are learned \chkE{from} ground-truth images.
In our case, the latent codes $\mathbf{z}^\mathcal{M}$ should be learned \chkE{from} the flow fields $\mathbf{B}^\mathcal{M}_{t+1}$, whose ground-truth are not available.}

\chkE{To overcome this chicken-and-egg problem, we initialize the input flow field of our motion encoder $E^\mathcal{M}$ as zero tensor in the first epoch, and gradually update it with $\textbf{B}^\mathcal{M}_{t+1}$ during the training phase.
Another problem is that\chkP{, because} a pair of consecutive frames for training is selected randomly from each training video for each epoch (see Section~\ref{sec:ImplementationDetailsInBody}), a na\"ive approach would initialize the input of $E^\mathcal{M}$ for each pair, which yields slow convergence.
We thus re-use the input of $E^\mathcal{M}$ for each training video, assuming that frames throughout the video exhibit a similar motion. 
We refer to this re-used input as a {\em common motion field} for the training video and condition it on a single latent code $\mathbf{z}^\mathcal{M}$. \chkS{A common motion field of each training video is stored in each epoch and used in the next epoch to extract the latent code $\mathbf{z}^\mathcal{M}$ of the corresponding video.
In this way, we finally store the code $\mathbf{z}^\mathcal{M}$}} in a codebook \chkE{for the use in the inference phase.
A \chkO{pseudo-code} of this training procedure is shown in Appendix~\ref{sec:TrainingDetails}.}

\subsection{Appearance Predictor}
\label{sec:AppearancePredictor}

\begin{figure}
   \includegraphics[width=1.\linewidth, clip]{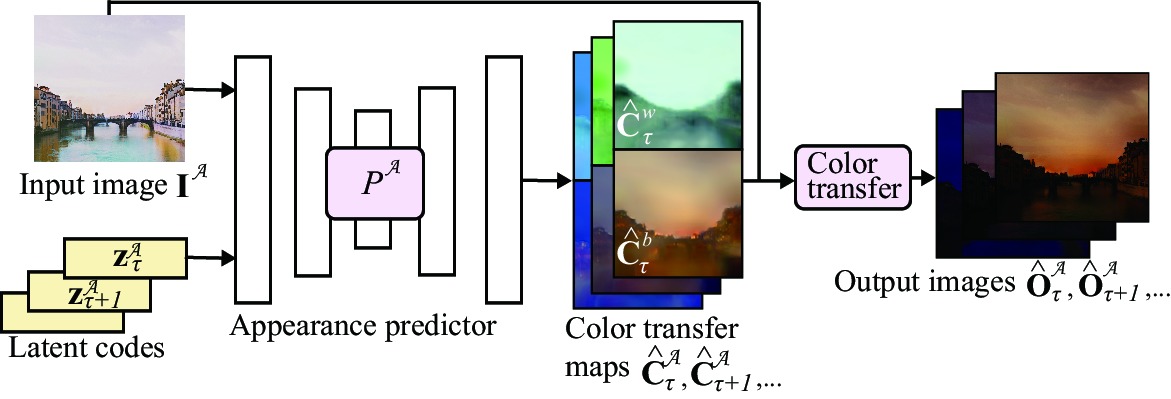}
   \caption{
   \chkE{Inference using} the appearance predictor \chkE{$P^\mathcal{A}$}. \chkE{Color} transfer maps $\hat{\textbf{C}}_{\tau}^\mathcal{A} = \{\hat{\textbf{C}}_{\tau}^{w}, \hat{\textbf{C}}_{\tau}^{b}\}$ at \chkE{time} $\tau$ \chkE{are computed} from an input image $\textbf{I}^\mathcal{A}$ \chkE{and} latent code $\textbf{z}^\mathcal{A}_{\tau}$. An output frame $\hat{\textbf{O}}_{\tau}^\mathcal{A}$ is obtained by applying \chkE{color transfer} to $\textbf{I}^\mathcal{A}$. Multiple frames are obtained using latent code sequence $\{\textbf{z}^\mathcal{A}_{\tau}\}$\chkE{, unlike the} recurrent \chkE{feeding} \chkD{\chkR{used} in} the motion predictor \chkE{$P^\mathcal{M}$}. 
   \chkV{Input photo: Domenico Loia/Unsplash.com. }
   }\label{fig:CTnet_test}
\end{figure}

\paragraph*{Inference.} %
Given \chkE{$\mathbf{I}^\mathcal{A}$ (equals a motion-added frame $\textbf{I}^\mathcal{A}_{t}$ at time $t$), our} appearance predictor $P^\mathcal{A}$ infers \chkE{a} color transfer \chkE{map} $\hat{\textbf{C}}_{\tau}^\mathcal{A} = \{\hat{\mathbf{C}}_{\tau}^{w}, \hat{\mathbf{C}}_{\tau}^{b}\}$ \chkE{(where $\hat{\mathbf{C}}_{\tau}^{w}, \hat{\mathbf{C}}_{\tau}^{b} \in [-1, 1]^{w \times h \times 3}$)} for \chkE{an arbitrary frame} $\tau$ \chkE{(Figure~\ref{fig:CTnet_test})}.
Each color transfer map $\hat{\textbf{C}}_{\tau}^\mathcal{A}$ is controlled by \chkK{the latent code $z^\mathcal{A}_{\tau}$ at frame $\tau$.}
\chkK{The output frame \chkE{$\hat{\textbf{O}}_{\tau}^\mathcal{A}$} is \chkE{then} computed by \chkE{applying the map $\hat{\textbf{C}}_{\tau}^\mathcal{A}$} to \chkE{the} input image \chkE{$\mathbf{I}^\mathcal{A}$} as follows:}
\begin{align}
  \hat{\textbf{O}}_{\tau}^\mathcal{A} & = \textsc{ColorTransfer}\left(\hat{\textbf{C}}_{\tau}^\mathcal{A}, \textbf{I}^\mathcal{A} \right) \label{eq:ColorTransfer} \\
  & = \tanh \left(\hat{\textbf{C}}_{\tau}^{w} \circ \textbf{I}^\mathcal{A} + \hat{\textbf{C}}_{\tau}^{b}\right), 
\end{align}
where $\circ$ denotes Hadamard product and $tanh$ is used to restrict the pixel values of  \chkE{$\hat{\textbf{O}}_{\tau}^\mathcal{A}$ within $[-1, 1]$}. 

\chkE{In the final video generation (Section~\ref{sec:CinemagraphGeneration}), we first interpolate the latent code sequence $\{\mathbf{z}_{\tau}^\mathcal{A}\}$ linearly, and then apply color transfer to each frame.}

\begin{figure}
   \includegraphics[width=1.\linewidth, clip]{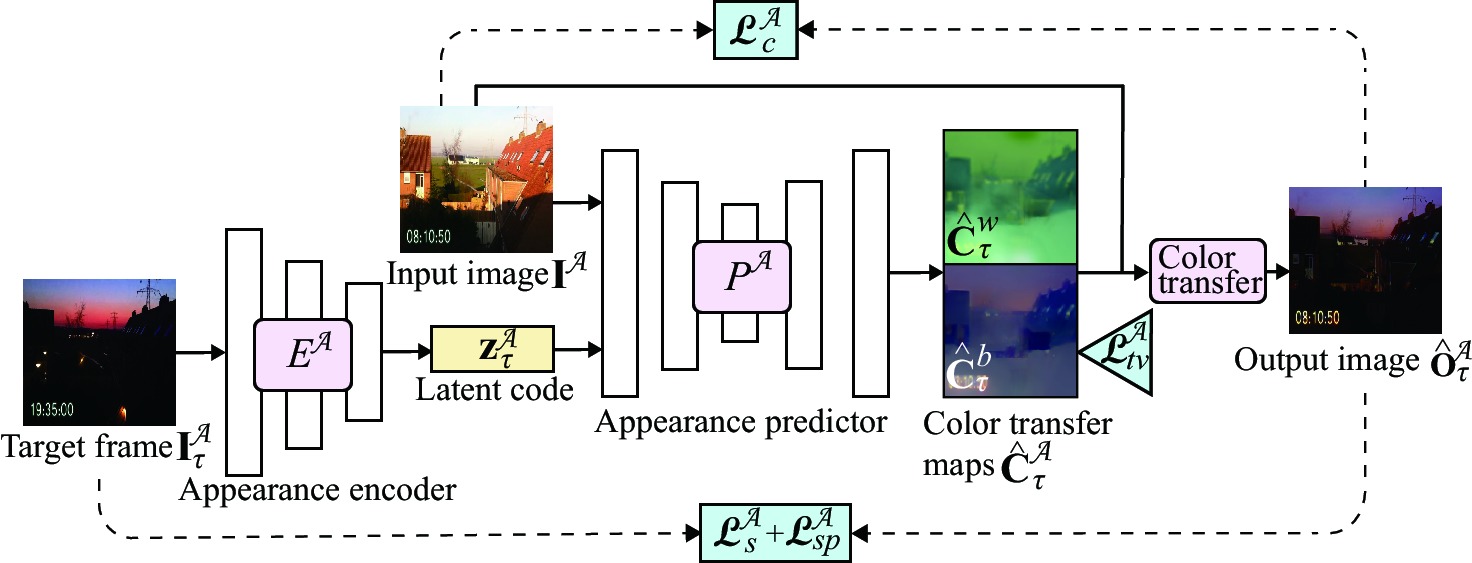}
   \caption{
   \chkE{Training of} the appearance predictor \chkE{$P^\mathcal{A}$}. \chkE{The training uses each pair} of a source image $\textbf{I}^\mathcal{A}$ and a target image $\textbf{I}^\mathcal{A}_{\tau}$ \chkE{such that the} losses $\mathcal{L}^\mathcal{A}_{s}$ and $\mathcal{L}^\mathcal{A}_{sp}$ between $\textbf{I}^\mathcal{A}_{\tau}$ and $\hat{\textbf{O}}^\mathcal{A}_{\tau}$ \chkE{are minimized.} $\hat{\textbf{O}}^\mathcal{A}_{\tau}$ \chkE{is} obtained \chkE{via} color transfer based on $\hat{\textbf{C}}^\mathcal{A}_{\tau}$ to $\textbf{I}^\mathcal{A}$. 
   \chkE{The} losses $\mathcal{L}^\mathcal{A}_c$ and $\mathcal{L}^\mathcal{A}_{tv}$ \chkE{impose that the content of $\textbf{I}^\mathcal{A}$ be preserved in $\hat{\textbf{O}}^\mathcal{A}_{\tau}$ and $\hat{\textbf{C}}^\mathcal{A}_{\tau}$ be regularized, respectively.} 
   $\textbf{z}^\mathcal{A}_{\tau}$ is obtained by encoding the target image $\textbf{O}^\mathcal{A}_{\tau}$ using the appearance encoder $E^\mathcal{M}$. 
   \chkV{Input photos: Anonymous (a8CTqQAxBzI)/Youtube.com. }
   }\label{fig:CTnet}
\end{figure}

\paragraph*{Training.} %
\chkG{Figure~\ref{fig:CTnet} \chkP{outlines the training} of the appearance predictor. }
%\chkG{Figure~\ref{fig:CTnet} provides an overview of the appearance predictor training. }
We first define loss functions between \chkC{two} frames with different \chkD{appearances} sampled from the training dataset. 
To learn style conversion for \chkP{the} entire image, we use \chkP{a} style loss between \chkP{the} inferred output frame $\hat{\textbf{O}}^\mathcal{A}_{\tau}$ and \chkC{\chkP{the} ground-truth target frame} $\textbf{I}^\mathcal{A}_{\tau}$: 
\begin{equation}
	\mathcal{L}^\mathcal{A}_s = \displaystyle \sum_l \|G(F_l(\textbf{I}^\mathcal{A}_{\tau}))-G(F_l(\hat{\textbf{O}}^\mathcal{A}_{\tau}))\|_2^2 \; , 
\end{equation}
where the function $F_l$ outputs feature maps obtained from the $l$-th layer of the pre-trained VGG16~\cite{Simonyan14c}. The function $G$ outputs the \chkA{Gram} matrix of the features maps. 
%For \chkD{each layer} $l$, 
\chkS{Inspired by the existing style transfer algorithm~\cite{DBLP:conf/eccv/JohnsonAF16}}, we use \texttt{relu\_2\_2}, \texttt{relu\_3\_3}, and \texttt{relu\_4\_3} \chkS{as the layers $l$}. 
\chkA{Note that the style loss is insensitive to spatial color distributions due to the Gram matrix, which makes\chkD{, for example,} a \chkD{partially red sky during sunset} difficult to handle.}
Therefore, \chkD{an} additional weak constraint is imposed on the output frame $\hat{\textbf{O}}^\mathcal{A}_{\tau}$ to roughly \chkP{conform to the} spatial color distributions: 
%Therefore, \chkD{an} additional weak constraint is imposed on the output frame $\hat{\textbf{O}}^\mathcal{A}_{\tau}$ to roughly match spatial color distributions: 
\begin{equation}
  \mathcal{L}^\mathcal{A}_{sp} = \|SP(\textbf{I}^\mathcal{A}_{\tau})-SP(\hat{\textbf{O}}^\mathcal{A}_{\tau})\|_2^2 \; , 
\end{equation}
where $SP$ indicates the spatial pyramid pooling function~\cite{DBLP:journals/pami/HeZR015}\chkP{, which} \chkB{outputs} fixed-size feature maps by dividing an image into multi-level grids. We set the pyramid height as one and divide the image into a $32 \times 32$ grid, where average pooling is applied to each cell. 
\chkP{Whereas} the above losses are defined \chkP{against} the \chkA{ground-truth target} frame $\textbf{I}^\mathcal{A}_{\tau}$, \chkP{a} content loss is defined \chkP{against} the input \chkE{$\textbf{I}^\mathcal{A}$} to keep the input scene structure:
\begin{equation}
  \mathcal{L}^\mathcal{A}_c =  \displaystyle \sum_l\|F_l(\chkE{\mathbf{I}^\mathcal{A}})-F_l(\hat{\textbf{O}}^\mathcal{A}_{\tau})\|_2^2 \; . 
\end{equation}
%For \chkD{each layer} 
\chkP{As the layer} 
$l$ in this loss function, we use \texttt{relu\_1\_2} only to \chkD{retain} high-frequency components of the input scene. 
Finally, the inferred color transfer map $\hat{\textbf{C}}_{\tau}^\mathcal{A}$ \chkP{is regularized} to improve the generalization ability of the model: %Finally, regularization is applied to the inferred color transfer map $\hat{\textbf{C}}_{\tau}^\mathcal{A}$ to improve the generalization ability of the model: 
\begin{equation}
\chkQ{
  \mathcal{L}^\mathcal{A}_{tv} = \displaystyle \sum_{\textbf{p}, \textbf{q} \in N(\textbf{p})}w(\textbf{I}^\mathcal{A}(\textbf{p}), \textbf{I}^\mathcal{A}(\textbf{q}))\|\hat{\mathbf{C}}^\mathcal{A}_{\tau}(\textbf{p})-\hat{\mathbf{C}}^\mathcal{A}_{\tau}(\textbf{q})\|_1. 
}
\end{equation}
Note that, unlike \chkC{Equation~(\ref{eq:L_M_tv}), these color transfer maps are smoothed \chkE{\chkP{such} that $\hat{\textbf{C}}_{\tau}^\mathcal{A}$ respects} the scene structure of the input image $\textbf{I}^\mathcal{A}_{t}$.}
Total loss $\mathcal{L}^{\mathcal{A}}$ is then given by the summation of the above losses with weights $\lambda^\mathcal{A}_{s}$, $\lambda^\mathcal{A}_{sp}$, $\lambda^\mathcal{A}_{c}$, and $\lambda^\mathcal{A}_{tv}$: 
\begin{equation}
\label{eq:loss_ap}
  \mathcal{L}^{\mathcal{A}} = \lambda^\mathcal{A}_{s}~\mathcal{L}^\mathcal{A}_s + \lambda^\mathcal{A}_{sp}~\mathcal{L}^\mathcal{A}_{sp} + \lambda^\mathcal{A}_{c}~\mathcal{L}^\mathcal{A}_c + \lambda^\mathcal{A}_{tv}~\mathcal{L}^\mathcal{A}_{tv} \; .
\end{equation}

\chkE{We also train the} appearance encoder $E^\mathcal{A}$ \chkP{to extract} latent codes $\textbf{z}^\mathcal{A}_{\tau}$ \chkE{simultaneously using $\mathcal{L}^{\mathcal{A}}$.
The} input of the appearance encoder $E^\mathcal{A}$ \chkE{is the target frame $\textbf{I}^\mathcal{A}_{\tau}$ so that the inferred output $\hat{\textbf{O}}_{\tau}^\mathcal{A}$ is conditioned on $\textbf{I}^\mathcal{A}_{\tau}$.
After the training, a sequence of latent codes $\{\mathbf{z}^\mathcal{A}_{\tau}\}$ for each training video is extracted using $E^\mathcal{A}$ and} stored in a codebook,
\chkK{\chkE{similarly} to \chkE{the} motion predictor}.
%\chkK{\chkE{similarly} to \chkE{the} motion predictor (see Appendix~\ref{sec:TrainingDetails})}.

\subsection{Implementation}
\label{sec:ImplementationDetailsInBody}
\chkP{The network architectures of our predictors are summarized in Appendix~A.1.} 
\chkP{The} motion and appearance predictors $P^{\mathcal{M}}$ and $P^{\mathcal{A}}$ are fully CNNs, each of which consists of three downsampling layers, five residual blocks, and three upsampling layers. The networks contain skip connections also used in U-Net~\cite{RFB15a}. 
Our motion and appearance encoders $E^{\mathcal{M}}$ and $E^{\mathcal{A}}$ adopt \chkA{the} same network structure \chkD{as that in} \texttt{resnet\_128}~\cite{DBLP:conf/nips/ZhuZPDEWS17}\chkD{, which consists} of six layers for convolution, pooling, and linear transformation. 

%To avoid training biases for video clips containing more frames, 
To avoid training biases \chkR{by longer} video clips, 
we train \chkD{each} pair of frames sampled randomly for each video clip in each epoch. 
%While the motion predictor $P^{\mathcal{M}}$ trains a pair of consecutive frames, the appearance predictor $P^{\mathcal{A}}$ trains \chkD{any pair of frames}. 
\chkR{Whereas} the motion predictor $P^{\mathcal{M}}$ \chkR{learns from} a pair of consecutive frames, the appearance predictor $P^{\mathcal{A}}$ \chkR{learns from} \chkD{any pair of frames}. 
\chkP{The pseudo-codes of the} training procedures are described in \chkP{Appendix} A.2.
%The network architectures and training procedures are described \chkD{more completely} in \chkA{Appendices~A.1 and A.2}. 

The training image size $w \times h$ was set to $256 \times 256$ for the predictors and $128 \times 128$ for the encoders \chkA{for both motion and appearance}. The number of \chkD{dimensions} of the latent codes $\textbf{z}$ was set to 8. We used the Adam optimizer~\cite{DBLP:journals/corr/KingmaB14} with a learning rate of $1.0 \times 10^{-4}$, two coefficients of $\{0.5, 0.999\}$, and \chkD{a batch size} of 8 for backpropagation. Regarding the weights of the loss functions, we \chkS{empirically chose} $\lambda^\mathcal{M}_p = 1$, $\lambda^\mathcal{M}_{tv} = 1$, $\lambda^\mathcal{A}_s = 1$, $\lambda^\mathcal{A}_{sp} = 1 \times 10^{-2}$, $\lambda^\mathcal{A}_{c} = 1 \times 10^{-5}$, $\lambda^\mathcal{A}_{tv} = 0.1$, and $\sigma=0.1$. 

\section{Single-image Video Generation}
\label{sec:CinemagraphGeneration}
\chkH{Now we explain how to generate a video from a single image by integrating the two predictors. 
Inspired by cinemagraph, the output animation can be looped as an option. 
Here we explain the looped version.}

\chkE{Algorithm~1 summarizes} the procedure of \chkK{our} \chkH{video} generation. 
\chkK{The motion prediction first generates \chkE{a sequence} of frames $\mathcal{V^M}$, which is then converted to a \chkE{looped one} $\mathcal{V}^\mathcal{M}_{loop}$.
\chkE{A sequence} of output frames $\mathcal{V}$ are finally generated from $\mathcal{V}^\mathcal{M}_{loop}$ through the appearance prediction.}
\chkH{Note that $\mathcal{V}^\mathcal{M}$ is used instead of $\mathcal{V}^\mathcal{M}_{loop}$ if the looping process is not required.}
To make \chkE{a} \chkR{motion} \chkE{loop} $\mathcal{V}^\mathcal{M}_{loop}$ from the non-periodic \chkE{sequence} $\mathcal{V^M}$, various methods \chkE{can be used}~\cite{DBLP:conf/siggraph/SchodlSSE00,DBLP:journals/tog/LiaoFH15}. Among \chkD{the} several methods \chkD{that we tested}, simple \chkD{cross-fading} ~\cite{DBLP:conf/siggraph/SchodlSSE00} 
worked relatively well for making plausible \chkD{animations} without \chkE{significant discontinuities}. 
\chkE{Whereas} the \chkR{resolutions} of \chkK{images} $\mathbf{I}$, \chkK{the output frames \chkE{in} $\mathcal{V^M}$, $\mathcal{V}^\mathcal{M}_{loop}$, and the final video} $\mathcal{V}$ are not limited, \chkE{the inputs to the predictors and encoders are} resized to \chkB{fixed \chkR{resolutions} for training}. 
The inferred flow fields and color transfer maps are resized to the original size and then applied to the original input image. \chkA{We do not magnify output frames directly to avoid \chkD{blurring}. }
\chkE{To handle sampling outside of previous flow fields during the reconstruction of output frames, reflection padding is applied to the input image and previous flow fields. }

\begin{figure}
\begin{center}
\begin{tabular}{l}\hline
\textbf{Algorithm 1.} \chkE{Single-image \chkH{Video Generation}} \\\hline
\textbf{Input}: Input image $\textbf{I}$, latent codes $z^\mathcal{M}$, $\{z^\mathcal{A}_{\tau}\}$\\
\textbf{Output}: Output video $\mathcal{V} = \{\textbf{I}_{1}, \textbf{I}_{2}, ...\}$\\ 
//Motion \chkA{prediction}\\
~1: $\mathcal{V^M} \leftarrow \{\textbf{I}\} $\\
~2: $\textbf{I}^\mathcal{M}_{t=1} \leftarrow \textbf{I}$\\
~3: \textbf{for each} frame \textbf{do}\\
~4: \hspace{1em} $\hat{\textbf{B}}^\mathcal{M}_{t+1} \leftarrow$ {\sc Resize}$(P^{\mathcal{M}}$({\sc Resize}$(\textbf{I}^\mathcal{M}_{t}), \textbf{z}^\mathcal{M}))$\\
~5: \hspace{1em} $\hat{\textbf{O}}^\mathcal{M}_{t+1} \leftarrow$  {\sc Warp}($ \{\hat{\textbf{B}}^\mathcal{M}_{t+1},\hat{\textbf{B}}^\mathcal{M}_{t}, \ldots,\hat{\textbf{B}}^\mathcal{M}_{t=2}\}, \textbf{I}^\mathcal{M}_{t=1})$\\
~6: \hspace{1em} $\mathcal{V^M} \leftarrow \mathcal{V^M} \bigcup \hat{\textbf{O}}^\mathcal{M}_{t+1}$\\
~7: \hspace{1em}  $\textbf{I}^\mathcal{M}_{t+1} \leftarrow \hat{\textbf{O}}^\mathcal{M}_{t+1}$\\
~8: \textbf{endfor}\\
~9: $\mathcal{V}^\mathcal{M}_{loop} \leftarrow$ {\sc GenerateLoop}($\mathcal{V^M}$)\\
//Appearance \chkA{prediction} \\
10: $\mathcal{V} \leftarrow \phi $\\
11: $\{\textbf{z}^\mathcal{A}_{t}\} \leftarrow$ {\sc InterpolateLatent\chkC{Codes}}$(\{\textbf{z}^\mathcal{A}_{\tau}\})$\\
12: \textbf{for all} $\textbf{z}^\mathcal{A}_{t}$ in $\{\textbf{z}^\mathcal{A}_{t}\}$ \textbf{do}\\
13: \hspace{1em} $\textbf{I}^\mathcal{A} \leftarrow$ {\sc GetNextFrameCyclically}$(\mathcal{V}^\mathcal{M}_{loop})$\\
14: \hspace{1em} $\hat{\textbf{C}}^\mathcal{A}_{t} \leftarrow P^{\mathcal{A}}(${\sc Resize}$(\textbf{I}^\mathcal{A}), \textbf{z}^\mathcal{A}_{t})$\\
15: \hspace{1em} $\hat{\textbf{O}}^\mathcal{A}_{t} \leftarrow$ {\sc ColorTransfer}({\sc Resize}$(\hat{\textbf{C}}^\mathcal{A}_{t}$),$ \textbf{I}^\mathcal{A})$\\
16: \hspace{1em}$\mathcal{V} \leftarrow \mathcal{V} \bigcup \hat{\textbf{O}}^\mathcal{A}_{t}$ \\
17: \textbf{endfor}\\
\hline
\end{tabular}
\end{center}
\end{figure}

\chkE{We can control the future variations of output frames with latent codes $z^\mathcal{M}$ and $\{z^\mathcal{A}_{\tau}\}$, and also adjust the speeds of motion and appearance.
The latent codes can be selected randomly (in this case, automatically) or manually from the codebook. 
We also show some applications to control latent codes indirectly in Section~\ref{sec:OutputControllability}.
The motion speed can be adjusted} by simply multiplying \chkE{flow fields by an arbitrary} scalar value.
Meanwhile, \chkE{the} appearance speed is determined \chkH{in two ways; by \chkP{adjusting} the number of latent codes in a sequence obtained from the codebook, or by repeating } the motion loop $\mathcal{V}^\mathcal{M}_{loop}$ an integer number of times during one cycle of appearance variation.
\chkH{We \chkP{adopt} the latter for all the looped videos.}
\chkE{The latent code sequence for appearance at \chkO{key-frames} are linearly interpolated to generate latent codes for the whole frames.}
\chkE{We also interpolate the final and initial} latent codes to generate a cycle.

\section{Experiments}
\label{sec:Experiment}
We implemented our system \chkK{with PyTorch library running} on a PC with NVIDIA GeForce GTX 1080 Ti GPUs. We stopped training after 5,000 epochs, and the computation time was about \chkD{one} week on a single GPU. \chkD{Motion} and appearance inferences \chkA{to generate a $640 \times 320$ frame} took 0.054 seconds and 0.058 seconds, respectively. Overall computation time to generate a cinemagraph \chkK{of} 1,010 frames was 98 seconds, \chkD{which included} trained model parameter \chkD{loadings} to a GPU \chkD{of} 9 seconds, motion inference \chkD{of} 11 seconds, \chkR{motion} loop generation \chkD{of} 6 seconds, \chkD{and} appearance inference \chkD{of} 59 seconds. \chkD{The} other processes \chkD{consumed} the remaining time. 

\begin{figure*}[t]
   \includegraphics[width=1.\linewidth, clip]{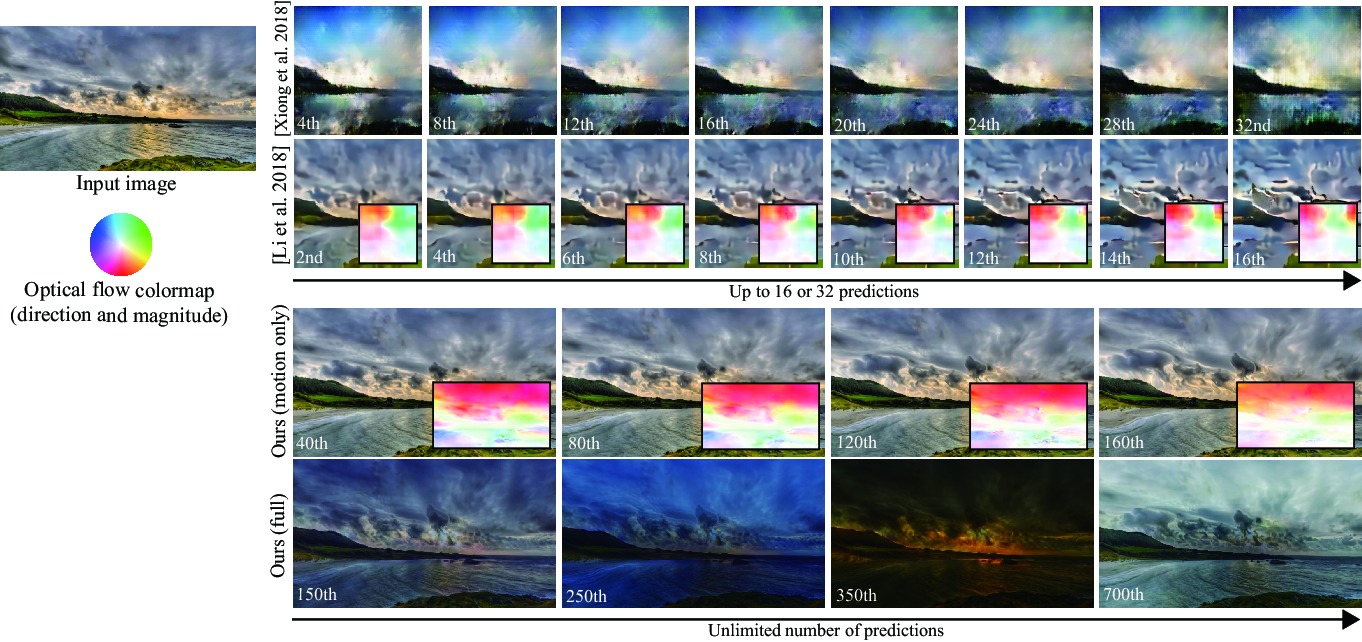}
   \caption{
   Qualitative comparison with the state-of-the-art video generation \chkQ{by Xiong et al.~\shortcite{xiong2018learning} and} Li et al.~\shortcite{Prediction-ECCV-2018}. \chkE{Their and our} output \chkE{resolutions} are $128\times128$ and $640\times360$, respectively. 
   \chkV{Input photo: Per Erik Sviland (Vvuxdqn-0vo)/Youtube.com.}
   }\label{fig:res1}
\end{figure*}

The results in our paper are demonstrated in the supplemental video.
The directions and magnitudes of optical flow vectors are visualized using the pseudo colors shown in Figure~\ref{fig:res1}.
%We will release the} resultant videos, codes, and datasets on our project page. 

\subsection{Dataset Generation}
\label{sec:Datasets}
For training the motion predictor and encoder, we used \chkA{the time-lapse video dataset published by Xiong et al.~\shortcite{xiong2018learning}}. 
The dataset \chkB{\chkD{was divided into} 1,825 video clips for training and 224 clips for \chkD{testing at the} resolution of $640\times360$}. 
\chkA{To avoid learning motions \chkD{that were too subtle}, we first sampled} every other frame from \chkP{each} \chkK{training} video clip and then automatically omitted pairs of frames in which the average of differences of pixel values of consecutive frames \chkD{was} less than 0.02. The resultant video clips contain \chkA{227} frames on average. 

\chkS{Because the videos used for motion modeling are too short to observe appearance transitions}, we collected 125 one-day video clips from YouTube and \chkA{the dataset published by Shih et al.~\shortcite{DBLP:journals/tog/ShihPDF13}} \chkS{for appearance modeling}. 
Because appearance changes more slowly than motion, we omitted more redundant frames from the dataset. 
\chkE{Specifically}, we first sampled \chkA{frames about every 10 minutes in real-world time} for each video clip\chkE{, and then omitted} consecutive frames containing \chkD{smaller} appearance variations.
\chkA{To do this, we computed} the sum of the RGB differences of the average of the pixel values for the consecutive frames, and adjacent frames were automatically omitted \chkP{if} the corresponding sum was less \chkE{than} 0.3. 
With this sampling process, the number of frames for each training clip is reduced to 15 on average.
Note that the input images shown in \chkD{this} paper \chkD{were} not included in the training data unless otherwise noted. 
%Note that the input images shown in \chkD{this} paper \chkD{were} \chkT{downloaded from the Internet (e.g., Unsplash.com and Pexels.com) and} not included in the training data unless otherwise noted. 
\begin{figure*}[t]
   \includegraphics[width=1.\linewidth, clip]{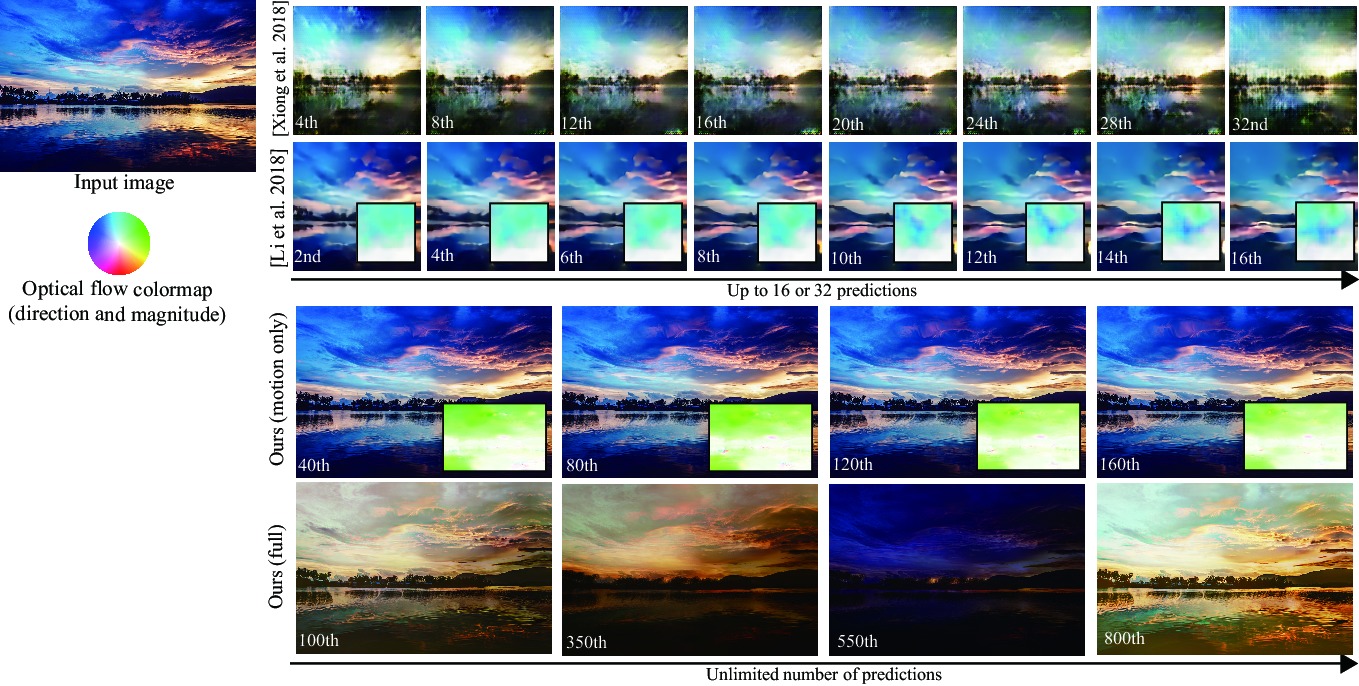}
   \caption{
   Another qualitative comparison with the state-of-the-art video generation \chkQ{by Xiong et al.~\shortcite{xiong2018learning} and} Li et al.~\shortcite{Prediction-ECCV-2018}. \chkE{Their and our} output \chkE{resolutions} are $128\times128$ and $1024\times683$, respectively. 
   \chkV{Input photo: Fancycrave.com/Pexels.com.} 
   }\label{fig:res1_2}
\end{figure*}

%\begin{figure*}[t]
%   \includegraphics[width=1.\linewidth, clip]{images/Comparison_Xiong.jpg}
%   \caption{
%   Qualitative comparisons \chkD{of} \chkE{the} GAN-based video generation \chkE{by Xiong et al.}~\shortcite{xiong2018learning}. \chkE{Their and our} output \chkE{resolutions} are $128\times128$ and $640\times360$, respectively.
%   \chkE{Because their} \chkD{input images} \chkD{were not provided in \chkE{their} paper,} we used \chkE{images similar to} the first \chkE{frames in their results as our inputs}. 
%   }\label{fig:Comparison_Xiong}
%\end{figure*}

\subsection{Comparisons with Video Prediction Models}
\label{sec:comparison1}

To clarify the advantages of our method, we compared it with \chkP{the} state-of-the-art video prediction models for a single input image. 
The \chkD{comparison} models are 3DCNN encoder-decoders \cite{Prediction-ECCV-2018} that predict flow fields \chkA{for a fixed number frames} from an input image and generate future frames based on the predicted flows. 
To train \chkD{the comparison} models, we used the same \chkP{training} data~\cite{xiong2018learning} \chkP{as ours}, 
%To train \chkD{the comparison} models, we used the same data~\cite{xiong2018learning} \chkD{used to train as our model}, 
and ground-truth flow fields were created using SpyNet~\cite{DBLP:conf/cvpr/RanjanB17} based on \chkD{the} authors' codes~\cite{Prediction-ECCV-2018}.
We used \chkP{their default} parameters and image size. 
%We used the same parameters and image size as \chkD{those in the comparison} codes. 
The number of \chkA{epochs} was also \chkA{the} same as ours (5,000), 
%and more epochs did not improve the results. 
and \chkP{improvement was not observed with} more epochs. 
\chkQ{We also compared our method with \chkD{other} recent GAN-based models~\cite{xiong2018learning}. }

Figures~\ref{fig:res1} \chkQ{and~\ref{fig:res1_2} show} qualitative comparisons. The right images are generated frames and flow fields using each method from the upper-left image. As shown \chkD{by} the insets in \chkQ{the second and third} rows, 
our method generates \chkA{more plausible} flow fields than the previous method according to \chkA{the input scene structure}\chkP{; f}or example, \chkP{whereas the} \chkD{clouds and the} water surface move differently, \chkD{the} lands \chkQ{remain} static overall. 
\chkQ{In the first row, the GAN-based method severely suffers from artifacts even in low-resolution images.} In the second row, \chkA{the \chkP{generated} frames by \chkP{Li et al.}} are unnaturally abstracted despite \chkA{\chkD{the model's} two-phase design that first predicts flow fields and then generates future frame pixels}. 
%In the first row, \chkA{the frames generated by \chkD{Li et al.'s method}} are unnaturally abstracted despite \chkA{\chkD{the model's} two-phase design that first predicts flow fields and then generates future frame pixels}. 
\chkP{Our results are} \chkA{clearer} \chkD{and} higher-resolution as demonstrated in the third row. 
%The proposed method can generate \chkD{both} \chkA{clearer} \chkD{and} higher-resolution images as demonstrated in the third row. 
Moreover, \chkA{our method can theoretically generate \chkB{an} unlimited \chkB{number of} frames.} 
Finally, as shown in the \chkA{third} row, our method can generate \chkD{a} \chkR{looped} animation \chkD{that} also contains appearance variations\chkD{,} thanks to \chkT{decoupled} learning, whereas \chkK{the comparative method cannot handle it \chkE{sufficiently}}.

%We also compared our method with \chkD{other} recent GAN-based models~\cite{xiong2018learning}. 
%\chkK{Because accurately reproducing their results \chkE{is difficult}}, 
%we used the results shown in their paper and project page for qualitative comparisons. 
%As shown in Figure~\ref{fig:Comparison_Xiong}, our method can generate more plausible and diverse results \chkP{at} higher resolution.

\begin{figure}
   \includegraphics[width=1.\linewidth, clip]{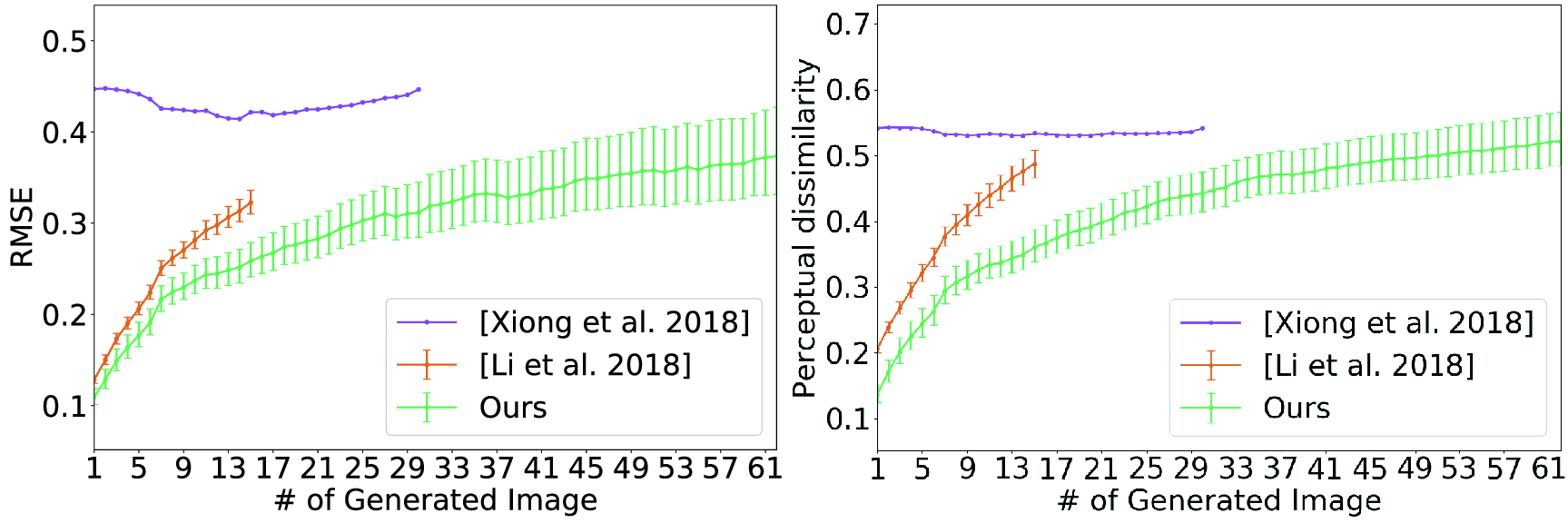}
   \caption{
   Quantitative comparisons \chkE{with} ground-truth for 224 test video clips. 
   RMSE \chkP{(left)} and perceptual dissimilarity~\cite{Zhang_2018_CVPR} \chkP{(right)} are computed for each predicted frame. The solid lines and error bars denote average, minimum, and maximum values of the metrics\chkE{, respectively}. 
   }\label{fig:graph}
\end{figure}

\chkK{In addition, we conducted quantitative \chkD{evaluations} using 224 test video clips \chkP{with the methods by \chkQ{Xiong et al.~\shortcite{xiong2018learning} and} Li et al.~\shortcite{Prediction-ECCV-2018}, regarding} the accuracy \chkP{compared to} ground-truth successive frames.}
We compared differences between generated sequences and ground-truth \chkK{ones} frame-by-frame. 
As evaluation metrics, we used RMSE and perceptual dissimilarity~\cite{Zhang_2018_CVPR} based on Alex-Net~\cite{DBLP:conf/nips/KrizhevskySH12}. 
Because \chkP{our results depend on latent codes}, we \chkD{compared} average, minimum, and maximum values of evaluation metrics for \chkG{five latent codes \chkE{sampled} from the codebook}. 
%\chkA{Because \chkK{the} \chkD{videos generated} by our method vary depending on the motion type} parameter $t^\mathcal{M}$, we \chkD{compared} average, minimum, and maximum values of evaluation metrics for \chkG{five latent codes \chkE{sampled} from the codebook}. 
%\chkE{We then multiplied the inferred flow field by $\frac{1}{\beta}$} \chkK{so that the motion \chkA{speed} \chkE{becomes} the same as \chkE{that of} the training data.}
The previous \chkE{method~\cite{Prediction-ECCV-2018}} based on VAE can also synthesize different future sequences, 
and thus we sampled different noises five times \chkP{from the normally distributed latent space} \chkK{for generating} five \chkE{sequences}\chkA{, which are used to calculate the metrics in the same manner as ours.}
%and thus we sampled different noises \chkP{from the latent-space normal distribution} five times \chkK{for generating} five \chkE{sequences}\chkA{, which are used to calculate the metrics in the same manner as ours.}
%The previous \chkE{method~\cite{Prediction-ECCV-2018}} based on VAE can also synthesize different future sequences, and thus we sampled different noises in \chkE{the} latent space five times \chkK{for generating} five \chkE{sequences}\chkA{, which are used to calculate the metrics in the same manner as ours.}
Figure~\ref{fig:graph} shows \chkP{the} frame-by-frame RMSEs and perceptual dissimilarities for each method. The solid lines and error bars denote average, minimum, and maximum values of the metrics computed from the different future sequences.
\chkK{\chkE{The increasing trends} in both graphs imply} that \chkP{long-term prediction} is challenging. 
%\chkK{\chkE{The increasing trends} in both graphs imply} that generating more distant future frames is challenging. 
Nevertheless, our method outperforms \chkP{the} state-of-the-art \chkP{method} \chkK{in that \chkP{ours} can} generate higher-resolution and longer sequences. 
%Nevertheless, our method outperforms \chkD{other} state-of-the-art methods \chkK{in that it can} generate higher-resolution and longer sequences. 
In particular, \chkK{our results} are perceptually more similar to the ground-truth sequences\chkD{,} even \chkD{when} generated \chkD{with} different parameters. 

\subsection{Comparisons on Appearance Manipulation}
%\subsection{Comparisons with Appearance Manipulation Methods}
\label{sec:AppearanceComparisons}

\begin{figure}
   \includegraphics[width=1.\linewidth, clip]{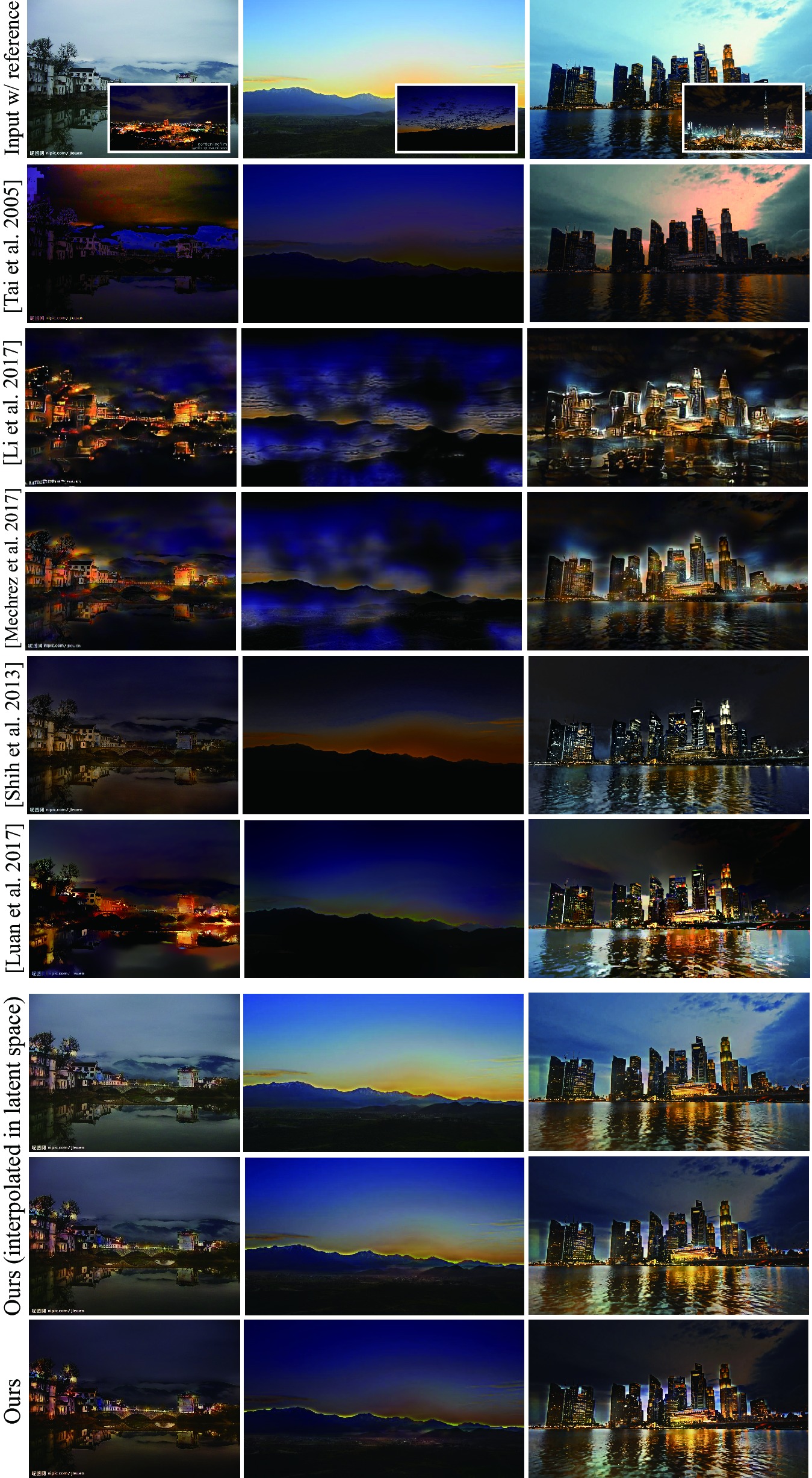}
   \caption{
   Comparison\chkP{s} with previous methods \chkE{for appearance manipulation}. From left to right, the output image sizes are $700\times525$, $700\times394$, and $700\times394$\chkE{, respectively}. 
   \chkV{Input photos: Shih et al.~\shortcite{DBLP:journals/tog/ShihPDF13}.}
   }\label{fig:res4}
\end{figure}
\chkE{We further compared our appearance-only results with those of previous \chkP{color/style} transfer methods.}
%\chkE{We further compared our appearance-only results with those of previous color and style transfer methods that can handle relatively longer predictions than \cite{xiong2018learning,Prediction-ECCV-2018}.}
Figure~\ref{fig:res4} shows \chkA{the} results of appearance transfer obtained using the source and target images \chkA{(inset)} in the top row. \chkK{For the local} color transfer~\cite{DBLP:conf/cvpr/TaiJT05} in the second row, the appearance variations are monotonic and inconsistent with the scene structures. Style transfer based on WCT~\cite{DBLP:conf/nips/LiFYWLY17} in \chkA{the third} row can handle \chkA{more diverse} appearance variations but some artifacts can be observed. Although these artifacts are alleviated by \chkP{solving} \chkA{the} screened Poisson \chkD{equation}~\cite{DBLP:conf/bmvc/MechrezSZ17} \chkA{as shown in the fourth row}, the results are still unnatural. On the other hand, the example-based hallucination~\cite{DBLP:journals/tog/ShihPDF13} \chkA{(fifth row)} and deep photo style transfer ~\cite{DBLP:conf/cvpr/LuanPSB17} \chkA{(sixth row)} successfully transfer the target appearances. 
\chkK{These methods, however, require a target video and an additional semantic segmentation map\chkE{, respectively}}. 
\chkP{Even worse, when applied to videos, frame-by-frame optimization will cause flickering artifacts, and key-frame interpolation cannot be used with dynamic objects.}
%\chkD{In addition,} \chkK{their optimization of an output image often cause \chkA{flickering} artifacts}.
\chkK{\chkE{Our results} in the bottom row are generated} without any additional inputs\chkE{, except for latent code\chkP{s}} encoded from target image\chkP{s}.
%\chkK{\chkE{Our results} in the bottom row are generated} without any additional inputs\chkE{, except for a latent code} encoded from a target image in tens of milliseconds.
\chkE{\chkP{Thanks to latent codes,} natural and smooth interpolation is possible in the latent space, as demonstrated \chkP{from} the seventh \chkP{to ninth} rows where the appearance changes from the source to the target.}
%\chkE{The use of latent codes is advantageous in that natural and smooth interpolation is possible in the latent space, as demonstrated in the seventh and eighth rows where the appearance changes from the source to the target.}
\chkG{Moreover, \chkE{we} can dispense even with target image\chkP{s} if latent codes are specified from the codebook or are predicted from source image\chkP{s} via LSTM prediction (see Appendix~\ref{sec:LSTM}). }

\begin{figure*}
   \includegraphics[width=1.\linewidth, clip]{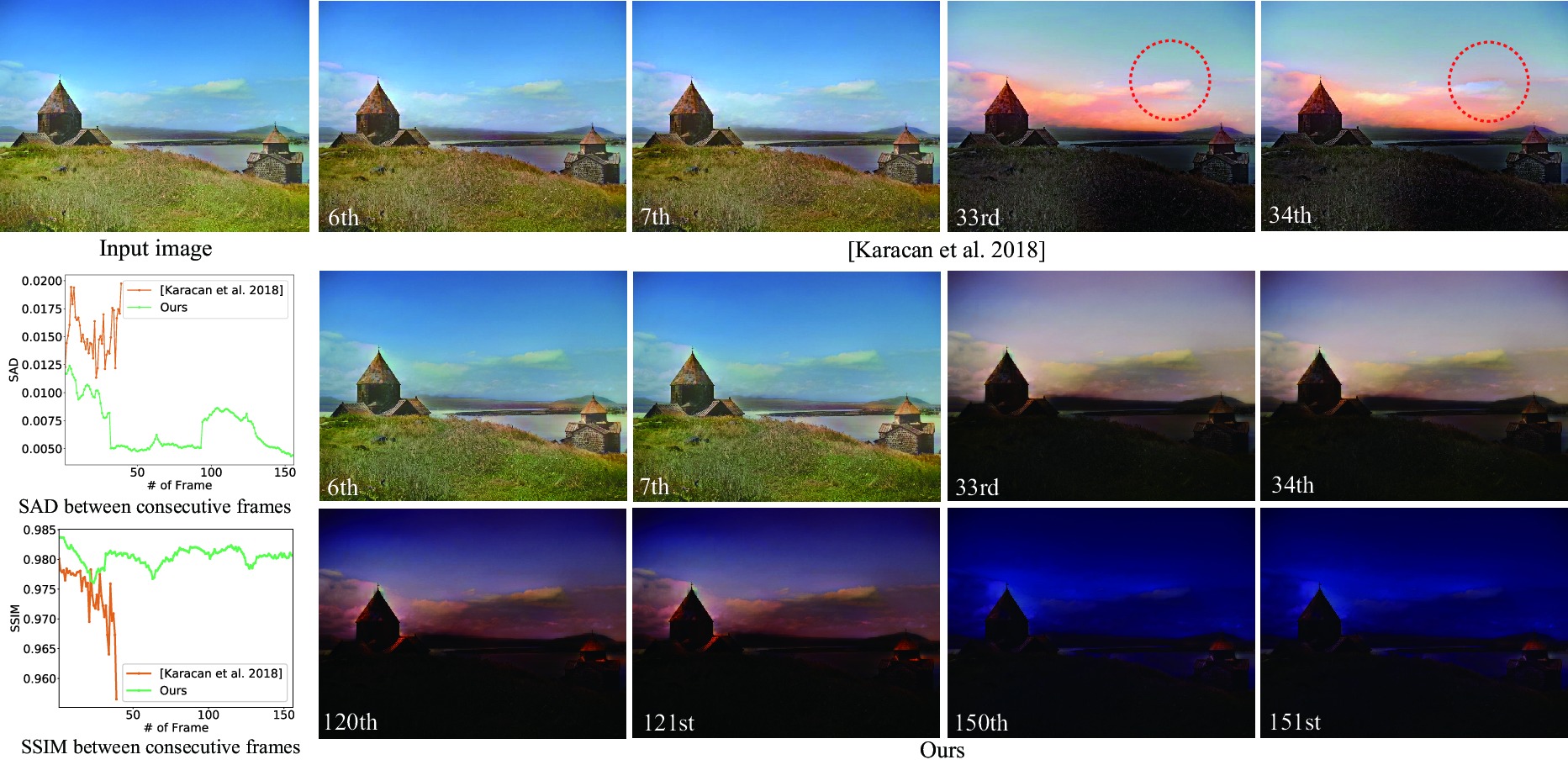}
   \caption{
   \chkE{Comparison}\chkP{s} with \chkE{the} state-of-the-art attribute manipulation \chkE{by Karacan et al.~\shortcite{DBLP:journals/corr/abs-1808-07413}}. The red circles \chkE{indicate} flicker artifacts in \chkE{their results}. In contrast, our method can reproduce smoothly-varying appearances. We also tried to quantitatively visualize this difference based on \chkD{the sum of absolute distance} and \chkD{structural similarity} between consecutive frames. The output \chkE{resolution} is $640\times480$. 
   \chkV{Input photo: Heretiq/Wikipedia.com.}
   }\label{fig:res5}
\end{figure*}

\begin{figure*}
   \includegraphics[width=1.\linewidth, clip]{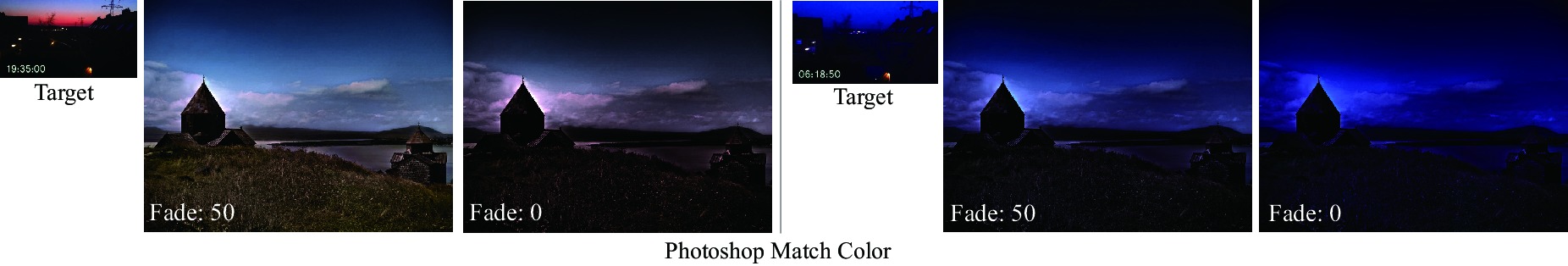}
   \caption{\chkQ{
   Comparisons with commercial appearance editing software (Photoshop Match Color). \chkR{The} input image is the same as \chkR{that in} Figure~\ref{fig:res5}, 
%   and the target images are the same as ones used for latent code extraction of our method.
   and the target images are \chkR{also used to extract the latent codes for our method}.
      \chkR{The fade} parameter (\chkR{from 0 to 100}) can control the degree of color transfer. }
}\label{fig:matchcolor}
\end{figure*}

\begin{figure*}
   \includegraphics[width=1.\linewidth, clip]{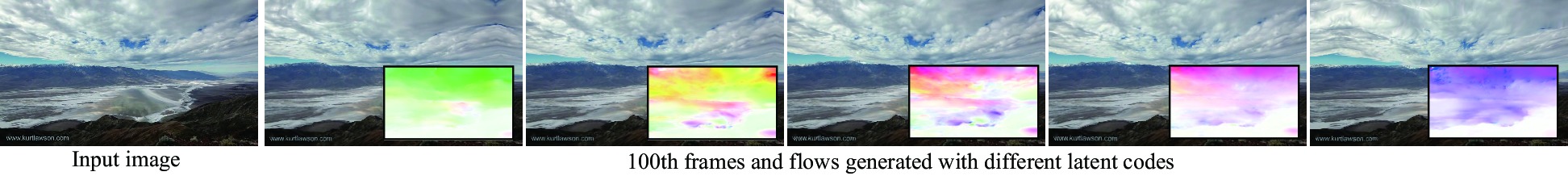}
   \caption{
   Effects of \chkG{latent codes for motion.} \chkE{The input latent codes are extracted via \chkR{self-supervised} learning, and sorted along the first principle axis in the codebook. We can see that the output frames are also aligned according to the input latent codes.} The output \chkE{resolution} is $640\times360$. 
   \chkV{Input photo: echoesLA (zleuiAR2syI)/Youtube.com.}
   }\label{fig:res2}
\end{figure*}

To the best of our knowledge, there \chkE{are} no \chkE{methods using} generative models \chkE{for} appearance variation, 
\chkK{except for \chkE{the} recent \chkP{one} \chkE{for} manipulating} image attributes~\cite{DBLP:journals/corr/abs-1808-07413}.
This method can be applied to generation of videos containing appearance transitions \chkP{by gradually changing attributes}. 
%This method can be applied to generation of videos containing appearance transitions \chkA{via gradual attribute changes}. 
Therefore, in Figure~\ref{fig:res5}, we compared our appearance predictor with their method using the input image and results \chkA{on their project page}. In our results, we selected the latent codes that yield appearances \chkD{similar to those of the} compared results. As demonstrated in the first row, \chkA{for each output frame}, the compared method can generate semantically plausible appearances \chkA{that match the image content}.
\chkK{Their sequence, however, contains \chkA{flickering} artifacts \chkE{due to the two-stage synthesis where temporal consistency is difficult to impose as is}}.
In the second and third rows, we can see that our \chkE{result is free from noticeable} artifacts. 
\chkS{We can generate temporally-coherent animations thanks to keyframe interpolation in the latent space, unlike the compared method. }
Also, we tried to quantitatively \chkA{visualize such artifacts} by computing \chkD{the sum of absolute differences} and \chkD{structural similarity} between the consecutive frames\chkD{,} as shown \chkD{at} the lower left. 
\chkA{The resultant values imply \chkD{that} our method \chkE{allows smoother transition than} the compared method.}
\chkQ{In addition, we compared commercial appearance editing software (Photoshop Match Color). Figure~\ref{fig:matchcolor} demonstrates that local appearance transitions cannot be reproduced \chkR{by the software}, unlike our method.}
\chkR{The halo artifacts behind the cottage roof in Figure~\ref{fig:matchcolor} is stronger than ours in Figure~\ref{fig:res5}. 
Note that, in the supplemental video, the resultant animation using the commercial software was generated by repeatedly applying color transfer to intermediate frames using multiple target images.
In this case, the halo artifact is reduced unexpectedly, but the global color variation becomes monotonic due to error accumulation.
Our method can avoid such error accumulation thanks to the latent-space interpolation.}

\subsection{Controlling Future Variations}
\label{sec:OutputControllability}

\subsubsection{Effects of \chkG{latent codes}.}

\begin{figure}
   \includegraphics[width=1.\linewidth, clip]{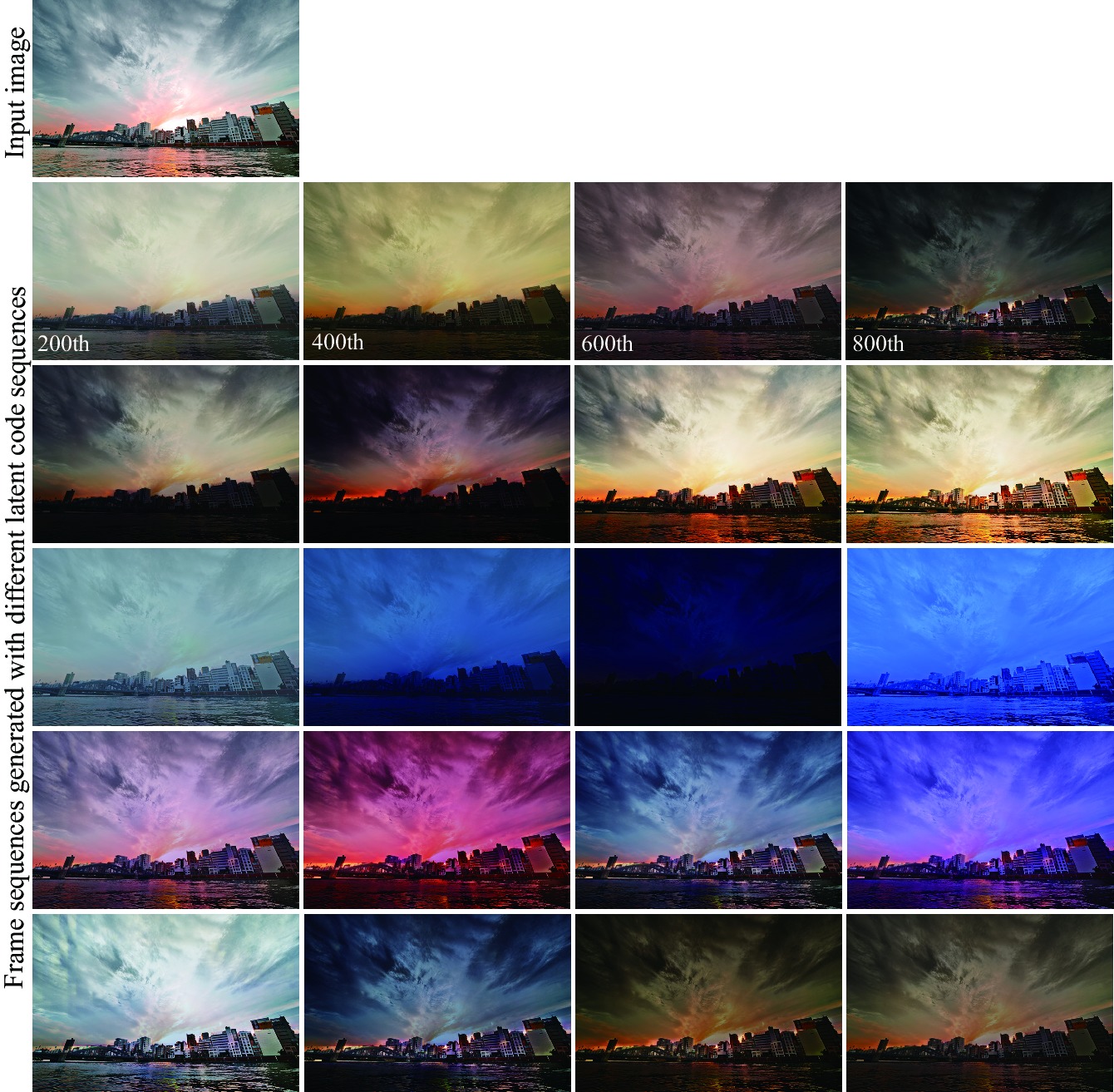}
   \caption{
   Effects of \chkG{latent codes for appearance. \chkE{In a frame sequence in each row, diverse appearances are synthesized from} a different latent code sequence in \chkE{the} codebook.} The output \chkE{resolution} is $700\times466$. 
   \chkV{Input photo: Jezael Melgoza/Unsplash.com. }
   }\label{fig:res3}
\end{figure}

To verify that our method can \chkP{handle} future uncertainty and \chkG{can} learn  meaningful latent \chkG{space in an unsupervised manner}, \chkE{i.e., without any ground-truth labels such as wind directions or time labels (e.g., ``daytime'' or ``night''),} we investigated how \chkG{latent codes} affect outputs. 
\chkE{Figure~\ref{fig:res2} shows the examples of motion.
Here we sorted latent codes in the codebook according to the first principle component and applied them to the same input image.
As we can see from the optical flows, our method generates similar motions from similar latent codes, while retaining a wide variety of motions.}
\chkE{For appearance, a sequence $\{\mathbf{z}^\mathcal{A}\}$ contains time-varying latent codes, and thus we can see that similar consecutive latent codes yield smooth transition in a time series (please check our supplemental video). 
Figure~\ref{fig:res3} also demonstrates that diverse appearances} (e.g., \chkE{sunset, twilight, and} night) \chkG{can be reproduced \chkE{from} the same \chkE{input image with different latent code sequences.}}

\chkE{Whereas direct use of latent codes from the codebook yields natural transition (e.g., from daytime to sunset in appearance) because they are extracted from real time-lapse videos, we also provide means to indirectly specify latent codes, namely, using arrow annotations for motion and image patches for appearance, as explained below.}

\begin{figure}
   \includegraphics[width=1.\linewidth, clip]{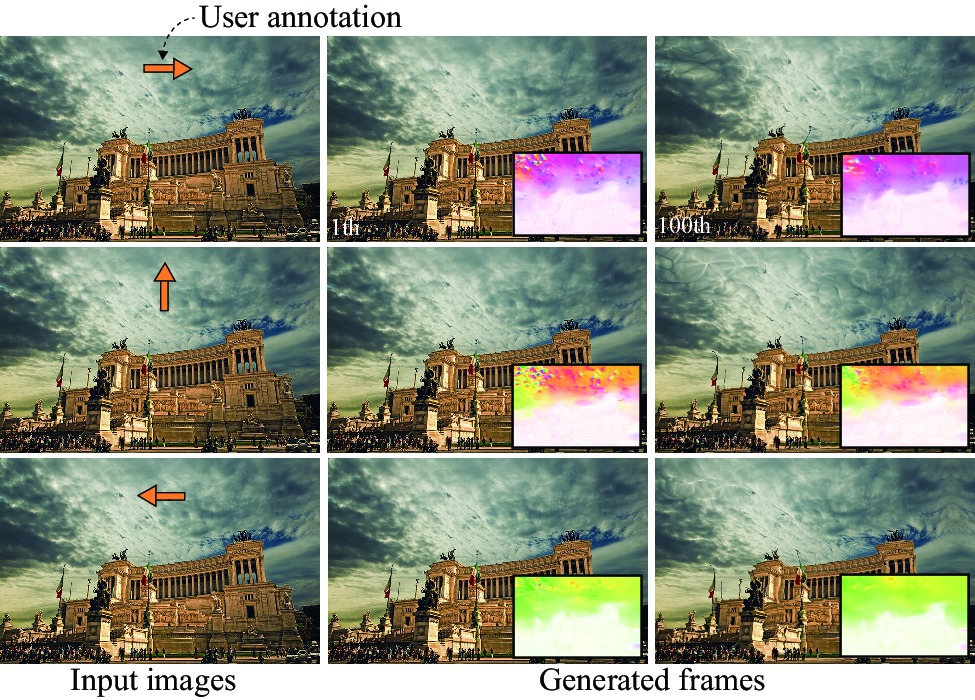}
   \caption{
    \chkE{Motion specification using arrow annotations.} The output \chkE{resolution} is $640\times411$. 
    \chkV{Input photo: Pixabay/Pexels.com}
   }\label{fig:assign_direction}
\end{figure}

\subsubsection{\chkE{Motion control using arrow annotations}}
\chkE{We offer arrow annotations for specifying flow directions of motion}, as shown in the left column in Figure~\ref{fig:assign_direction}.
We represent these sparse annotations as 2D maps \chkE{$\mathbf{U}^\mathcal{M} \in [-1,1]^{w \times h \times 2}$}, where pixels corresponding to the arrows have \chkD{the} specified flow vectors.
Given $\textbf{U}^\mathcal{M}$ and an input image $\textbf{I}^\mathcal{M}$, 
\chkP{an optimum latent code} $\hat{\textbf{z}}^\mathcal{M}$ is \chkP{obtained via optimization w.r.t. $\mathbf{z}^\mathcal{M}$ while fixing the network parameters of the motion predictor $P^\mathcal{M}$, as done in \cite{DBLP:conf/cvpr/GatysEB16}:}
%\chkL{the optimum latent vector $\hat{\textbf{z}}^\mathcal{M}$ is iteratively sought by computing the gradient of a minimization function, which is formulated as:} 
\begin{equation}
\hat{\textbf{z}}^\mathcal{M} = \argmin_{\textbf{z}^\mathcal{M}} \|\max(\textbf{0}, \textbf{M} \circ (\textbf{1} - D(\textbf{U}^\mathcal{M}, P^\mathcal{M}(\textbf{I}^\mathcal{M}, \textbf{z}^\mathcal{M}))-\textbf{m}))\|_2^2, \label{eq:OptMotion}
\end{equation}
where the function $D$ gives a map containing \chkD{the} cosine of \chkD{an} angle between two flows for each pixel.
%where the function $D$ gives a map containing \chkD{the} cosine of \chkD{an} angle between two flows for each pixel, \chkL{and the gradient $\partial P^\mathcal{M} / \partial \textbf{z}^\mathcal{M}$ is obtained via backpropagation} \chkH{with fixed network parameters \chkL{of the trained predictor}}, as done in \cite{DBLP:conf/cvpr/GatysEB16}.
The mask \chkE{$\mathbf{M} \in \{0, 1\}^{w \times h}$} is used to compute error on \chkA{arrows} only, and $\textbf{m} \in \mathbb{R}^{w \times h}$ is a constant margin map \chkH{(\chkP{having} 0.5 for each pixel)} that allows a certain level of difference between estimated flows and specified flows. 
\chkP{We used the Adam optimizer~\cite{DBLP:journals/corr/KingmaB14} for this optimization.}
Using $\hat{\textbf{z}}^\mathcal{M}$ and $\textbf{I}^\mathcal{M}$, 
we recurrently \chkP{predict} image sequences containing motions \chkD{similar to the directions of} the annotations.
%we recurrently compute forward propagation to obtain image sequences containing motions \chkD{similar to the directions of} the annotations.
The middle and right columns in Figure~\ref{fig:assign_direction} demonstrate that entire flow fields are plausibly generated using the sparse annotations. \chkQ{Optimization for each user edit took about seven seconds.}

\begin{figure}
   \includegraphics[width=1.\linewidth, clip]{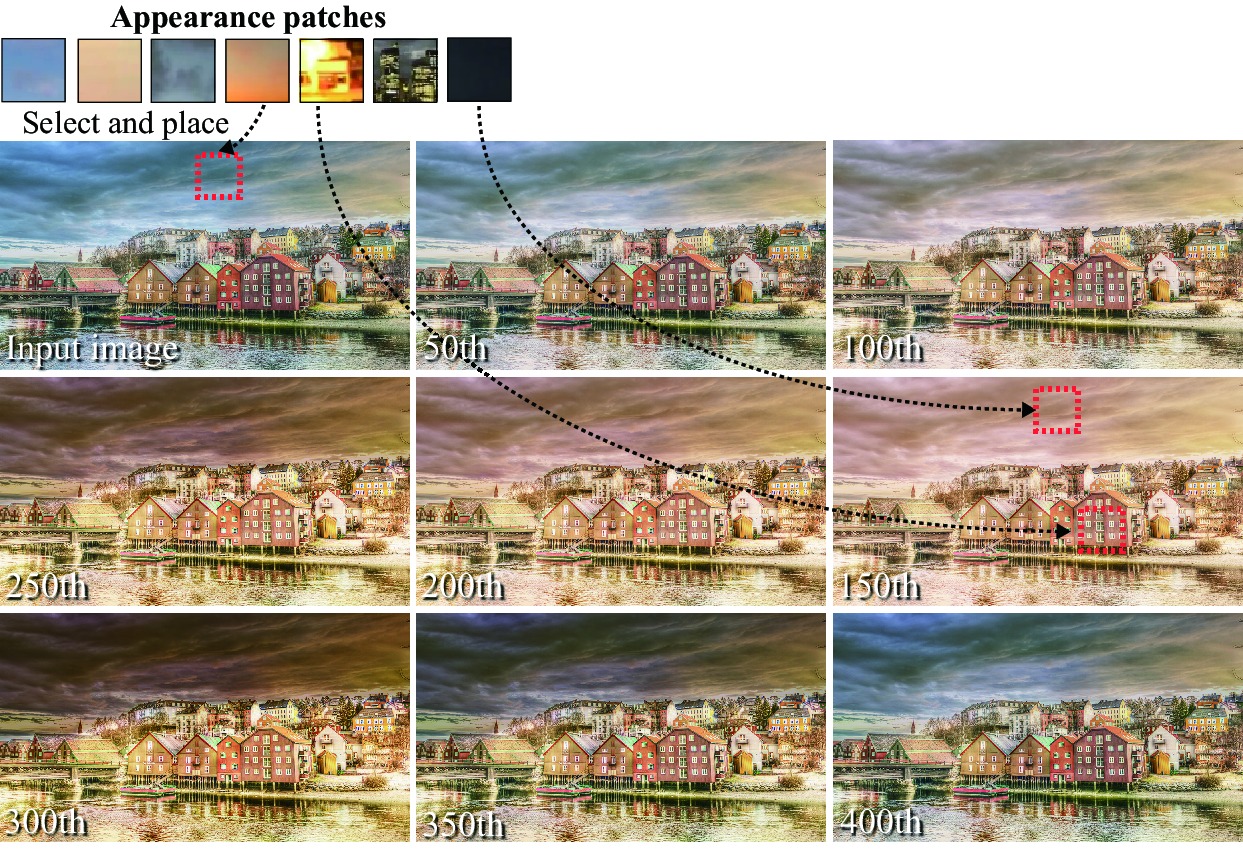}
   \caption{
   \chkE{Appearance specification using image patches.} The output \chkE{resolution} is $1\chkD{,}024\times571$. 
   \chkV{Input photo: Pixabay/Pexels.com}
   }\label{fig:assign_color}
\end{figure}

\subsubsection{\chkE{Appearance control using image patches}}
\chkE{We offer a means to specify the appearance at specific positions and frames using image patches.}
Using a map \chkE{$\mathbf{U}^\mathcal{A} \in [-1,1]^{w \times h \times 3}$} containing information of placed patches, 
\chkP{an optimum} latent code $\hat{\textbf{z}}^\mathcal{A}$ is \chkP{obtained}\chkH{, 
%\chkD{the} latent code $\hat{\textbf{z}}^\mathcal{A}$ is \chkH{optimized, 
similarly to Equation~(\ref{eq:OptMotion})}:
\begin{equation}
\hat{\textbf{z}}^\mathcal{A} \!\! = \argmin_{\textbf{z}^\mathcal{A}} \|\textbf{M}\circ (\textbf{U}^\mathcal{A} \!\! - \textsc{ColorTransfer}(P^\mathcal{A}(\textbf{I}^\mathcal{A}\!\!,  \textbf{z}^\mathcal{A}), \textbf{I}^\mathcal{A})\|_2^2. 
\end{equation}
Figure~\ref{fig:assign_color} shows the results obtained via latent space interpolation between the
input image and $\hat{\textbf{z}}^\mathcal{A}$. 
\chkE{We} can further change its appearance by placing multiple appearance patches as shown in the \chkA{right image in the middle row}. 
\chkD{Cyclic} animations can also be created via interpolation between the final and input images \chkA{as demonstrated in the 300th to 400th \chkD{frames}. }

\subsection{Ablation Study}
\label{sec:ablation}
\begin{figure}
   \includegraphics[width=1.\linewidth, clip]{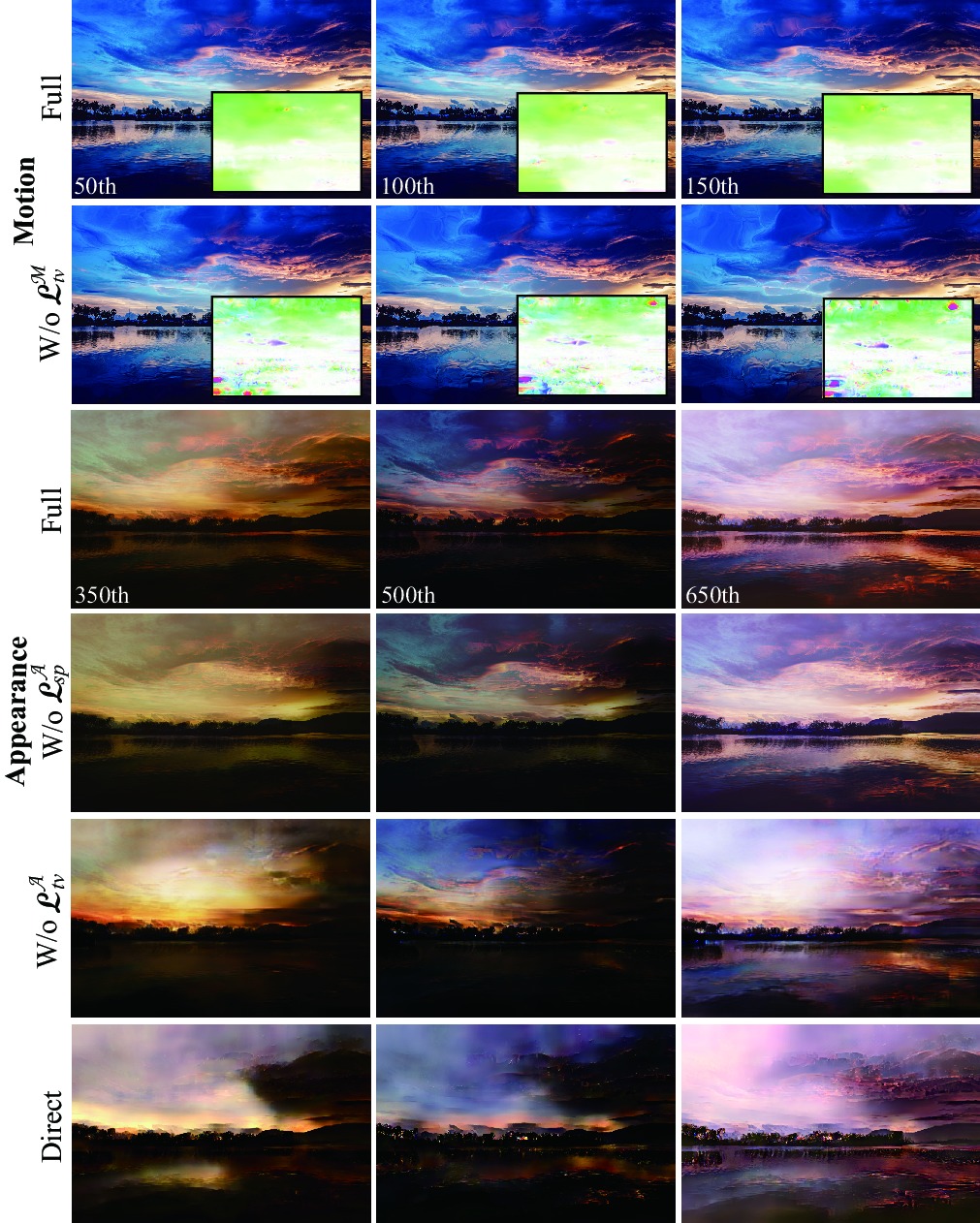}
   \caption{\chkE{Ablation study of loss functions. \chkQ{The input image is the same as that in Figure~\ref{fig:res1_2}.} ``Direct'' means direct inference of output frames without color transfer.} The output \chkE{resolution} is $1\chkD{,}024\times683$. 
   %\chkV{Input photo: Fancycrave.com/Pexels.com.}
   }\label{fig:ablation}
\end{figure}

We conducted an ablation study to investigate the effectiveness of our loss functions. 
Figure~\ref{fig:ablation} shows comparisons between the generated frames with and without each of the loss functions. 
Without the \chkA{motion} regularization term $\mathcal{L}^\mathcal{M}_{tv}$, the resultant flow fields in the third row are often inconsistent with the scene structure due to overfitting. 
For the same reason, the lack of the \chkA{appearance} regularization term $\mathcal{L}^\mathcal{A}_{tv}$ also causes noticeable artifacts \chkE{(see our supplemental video)} in the generated frames in the sixth row. 
Moreover, without the loss function $\mathcal{L}^\mathcal{A}_{sp}$ for learning spatial color distributions, the appearances \chkD{vary uniformly,} as shown in the fifth row,
and the \chkP{partially-reddish sky due to sunset} \chkD{is} not sufficiently reproduced. 
%and the \chkA{sunset that makes the sky region around the ridge line reddish} \chkD{is} not sufficiently reproduced. 
In contrast, we can see that the resultant frames generated with the full losses in the second and fourth rows are more stable than the others. 

\chkC{\textit{Direct} in Figure~\ref{fig:ablation} means that the output images were inferred directly without color transfer functions. 
For this, we used a CNN with the same architecture as the appearance predictor except \chkE{for the three-channel output}. 
Although this CNN was trained with the full losses (the TV loss was applied to network outputs), the results are less natural than the others. }

\subsection{Discussion}
\label{sec:discussion}

\paragraph{\chkQ{Do latent codes need regularization?}}

\begin{figure}
   \includegraphics[width=1.\linewidth, clip]{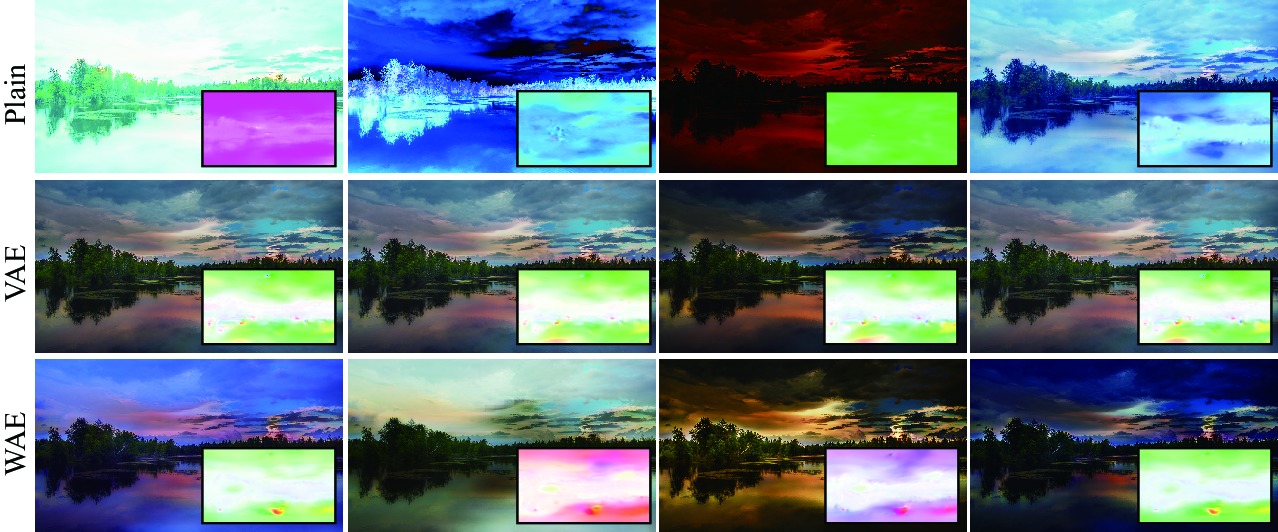}
   \caption{
    \chkE{Comparison of with and without regularizations} used \chkE{in} VAE and WAE. \chkE{The appearance} and motion (insets) \chkE{in each column are} inferred using \chkE{the same} latent code randomly sampled from a Gaussian distribution. The input image is the same as \chkD{that in} Figure~\ref{fig:teaser}.
    %\chkV{Input photo: Pixabay/Pexels.com.}
   }\label{fig:regularize_z}
\end{figure}

To \chkG{make search of latent codes in a latent space more stable}, there is an additional training option for adopting regularizers used by \chkA{Variational Auto-Encoder (VAE)}~\cite{DBLP:journals/corr/KingmaW13} and \chkA{Wasserstein Auto-Encoder (WAE)}~\cite{DBLP:journals/corr/abs-1711-01558}. 
\chkE{Whereas we regard the direct use of stored latent codes as the default choice because they yield plausible results, VAE and WAE allow us to select latent codes from regularized latent space, without referring to the codebook.}
In Figure~\ref{fig:regularize_z}, the predictors trained with these regularizers generated the results using latent codes sampled from a Gaussian distribution. In particular, the \chkD{WAE regularizer} is effective for generating more various outputs than \chkD{the VAE regularizer} because latent codes of different examples \chkD{can} stay far away from each other. In contrast, we can see that the models trained without these regularizers failed to generate plausible results from Gaussian latent codes. Although the regularization for training latent codes is not essential \chkE{in our case} \chkA{because we use the codebook} \chkE{and how to select appropriate sequences of latent codes for appearance from regularized latent space is not clear without the codebook}, it might be useful for \chkE{future} applications. 

\begin{figure}
   \includegraphics[width=1.\linewidth, clip]{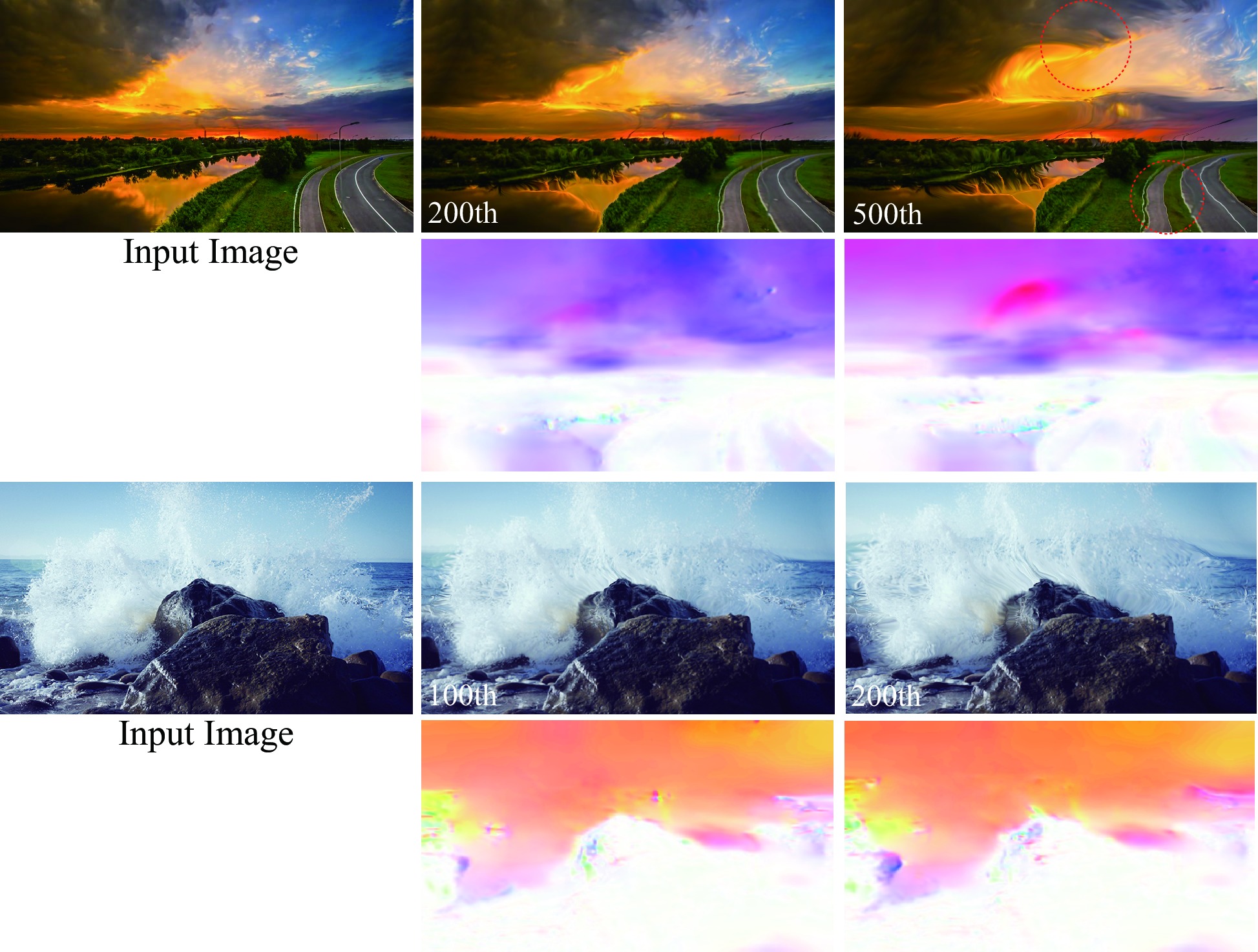}
   \caption{
    \chkE{Failure cases. Very long-term prediction causes distortions (red circles in the top right), and non-uniform motions are difficult to reproduce (bottom rows).} The \chkE{output resolutions} are $640\times360$ \chkE{(top)} and $600\times338$ \chkE{(bottom)}, respectively. 
    \chkV{Input photos: Race The World (jZOLRAIUW2s)/Youtube.com and Justin Leibow/Unsplash.com. }
   }\label{fig:limitations}
\end{figure}

\paragraph{Limitations}
Although our motion predictor can generate \chkD{an} unlimited number of frames, very long-term prediction \chkE{causes} unnatural \chkE{distortions} because predicted frames are reconstructed only from an input image. 
The first row in \chkA{Figure}~\ref{fig:limitations} shows an example, where the clouds in the 500th predicted frame are unnaturally stretched \chkE{and the border of the road is deformed,} compared \chkE{to those in} the 200th \chkD{frame}. 
Also, our method \chkE{erroneously generates \chkR{uni-directional} motion even for objects that \chkP{should} exhibit \chkR{scattered}} motions such as splashes, 
as shown in the third and fourth rows in \chkA{Figure}~\ref{fig:limitations}.
\chkR{There is still room for improvement in handling specific targets; cloud motions sometimes look unnatural, and mirror images of the sky on the water surface do not move synchronously in our results.}
\chkU{When the user controls cloud motions (Section 6.4.2), the reflected motions on water surfaces are also changed but do not necessarily move consistently with the clouds. }
\chkE{These artifacts might be alleviated by introducing \chkP{specialized} loss function\chkP{(s)} \chkR{(e.g., physically- or geometrically-inspired loss functions for clouds and mirror images)} \chkP{and} training data \chkP{for each target}.}
%\chkE{These artifacts might be alleviated by introducing other loss functions specialized for each target with specialized training data.}

\chkA{There is also} a trade-off between the diversity of output videos and \chkD{the} generalization ability of the models.
To handle more \chkO{various} motions and appearances, a straightforward solution is to reduce the regularization weights \chkK{while \chkE{restricting} unnatural deformations and artifacts \chkE{to} a tolerable level}.
%\chkQ{Additionally, reproducing occurrence animations including building being built and flower growing is outside the scope of our method because our method depends on image conversion using intermediate maps and does not generate output frames directly. } Nevertheless, our method can handle various \chkK{outdoor} scenery, which can be taken \chkD{anywhere} and are ubiquitous on the Web. 

%Although we mainly focus on landscape animations \chkT{especially for skies and waters} and put other types of animations where something appears (e.g., flower florescence or building construction) outside the scope, we believe our scope is sufficiently large to handle various landscapes, which is a popular subject as introduced in Section~1. 
We mainly focus on landscape animations\chkT{, especially of skies and waters,} and put other types of animations where something appears (e.g., flower florescence or building construction) outside the scope. 
\chkT{Nonetheless, we believe that our scope covers a wide variety of landscape videos and our motion predictor can also handle other types of motions (e.g., crowd motions seen from a distance) that can be well described with flow fields.}
%\chkT{We believe that our scope is sufficiently large to handle various landscapes and that our motion predictor can also handle other motions (e.g., crowd motions seen from a distance) that can be well described with flow fields. However, as all other machine learning techniques, our method cannot learn such motions without appropriate training datasets.}

\section{Conclusion}
\label{sec:Conclusion}
%This paper \chkE{has presented} \chkA{a} method that can create diverse and realistic \chkK{time-lapse videos} from a single image. 
This paper \chkE{has presented} \chkA{a} method that can create \chkR{a plausible animation} \chkR{at high resolution} from a single \chkR{scenery} image. 
We demonstrated the effectiveness of our method by qualitatively and quantitatively comparing it with not only \chkP{the} state-of-the-art video prediction models but also other appearance manipulation methods. 
To the best of our knowledge, it is unprecedented to \chkE{synthesize} \chkQ{high-resolution videos} \chkE{with separated control over} motion and appearance variations. 
This was accomplished by self-supervised, \chkT{decoupled} learning and latent\chkE{-}space interpolation.
\chkP{Our} method \chkE{can} generate \chkD{images with} higher-resolution and \chkE{longer-term} sequences than previous methods.
This \chkE{advantage comes from the indirect} image \chkE{synthesis} \chkP{using} intermediate maps \chkE{predicted via training with} regularization\chkE{,} rather than directly generating output frames \chkD{as done in} previous methods. 
The output sequences \chkE{can} also be controlled using latent codes \chkE{extracted during training}, which \chkE{can be specified} not only \chkE{directly from the codebook} but also \chkE{indirectly via simple} annotations.

\chkA{One} future direction \chkP{is} to improve the proposed model to create higher-quality animations. 
For example, \chkA{additional information \chkD{for} semantic segmentation might be helpful \chkD{for improving} the performance\chkD{,} as \chkD{done in} existing style transfer \chkD{methods}~\cite{DBLP:conf/cvpr/LuanPSB17}. Our current method does not adopt \chkD{this} approach to avoid \chkD{the} influence of segmentation error. }
\chkD{Occlusion} information~\chkK{\cite{Wang_2018_CVPR} could} be incorporated \chkE{explicitly} into training of the motion predictor. 
%\chkP{Acceleration of inference would be also beneficial to faster feedback.}
%\chkC{More intuitive interfaces also should be developed to \chkE{control} output videos with minimum \chkE{interactions}.}
\chkC{We \chkE{believe} that our work \chkE{has taken} a \chkE{significant} step in \chkE{single-image synthesis of videos and will inspire} successive work \chkE{for} diverse animations.}

\begin{acks}
\chkT{
The authors would like to thank the anonymous reviewers for giving insightful comments. Many thanks also go to Dr. Kyosuke Nishida for discussion and encouragement. This work was supported by JSPS KAKENHI Grant Number 17K12689. 
%For our accompanying video, input images courtesy of Per Erik Sviland, echoesLA, Caleb Minear, Race The World, Pixabay, Numerius Negidius, Jezael Melgoza, Justin Leibow, Kevin Jarrett, Fancycrave.com, Caleb Minear, Ran Berkovich, David Holzmann, James Wheeler, Frans Van Heerden, Dawid Zawila, mohammad alizade, Agnieszka Mordaunt, Tim Mossholder, Tim Peterson, Riley Pope, SoloTravelGoals, Sebastian B, and Samuel Scrimshaw.
}
\end{acks}

\bibliographystyle{ACM-Reference-Format}
\bibliography{sample-acmtog}


\begin{thebibliography}{58}

%%% the Web address.

\ifx \showCODEN    \undefined \def \showCODEN     #1{\unskip}     \fi
\ifx \showDOI      \undefined \def \showDOI       #1{#1}\fi
\ifx \showISBNx    \undefined \def \showISBNx     #1{\unskip}     \fi
\ifx \showISBNxiii \undefined \def \showISBNxiii  #1{\unskip}     \fi
\ifx \showISSN     \undefined \def \showISSN      #1{\unskip}     \fi
\ifx \showLCCN     \undefined \def \showLCCN      #1{\unskip}     \fi
\ifx \shownote     \undefined \def \shownote      #1{#1}          \fi
\ifx \showarticletitle \undefined \def \showarticletitle #1{#1}   \fi
\ifx \showURL      \undefined \def \showURL       {\relax}        \fi
% invisible to TeX
\providecommand\bibfield[2]{#2}
\providecommand\bibinfo[2]{#2}
\providecommand\natexlab[1]{#1}
\providecommand\showeprint[2][]{arXiv:#2}

\bibitem[\protect\citeauthoryear{Babaeizadeh, Finn, Erhan, Campbell, and
  Levine}{Babaeizadeh et~al\mbox{.}}{2018}]%
        {ICLR08Babaeizadeh}
\bibfield{author}{\bibinfo{person}{Mohammad Babaeizadeh},
  \bibinfo{person}{Chelsea Finn}, \bibinfo{person}{Dumitru Erhan},
  \bibinfo{person}{Roy~H. Campbell}, {and} \bibinfo{person}{Sergey Levine}.}
  \bibinfo{year}{2018}\natexlab{}.
\newblock \showarticletitle{Stochastic Variational Video Prediction}.
\newblock  (\bibinfo{date}{4} \bibinfo{year}{2018}).
\newblock


\bibitem[\protect\citeauthoryear{Bai, Agarwala, Agrawala, and Ramamoorthi}{Bai
  et~al\mbox{.}}{2012}]%
        {DBLP:journals/tog/BaiAAR12}
\bibfield{author}{\bibinfo{person}{Jiamin Bai}, \bibinfo{person}{Aseem
  Agarwala}, \bibinfo{person}{Maneesh Agrawala}, {and} \bibinfo{person}{Ravi
  Ramamoorthi}.} \bibinfo{year}{2012}\natexlab{}.
\newblock \showarticletitle{Selectively de-animating video}.
\newblock \bibinfo{journal}{{\em {ACM} Trans. Graph.\/}} \bibinfo{volume}{31},
  \bibinfo{number}{4} (\bibinfo{year}{2012}), \bibinfo{pages}{66:1--66:10}.
\newblock


\bibitem[\protect\citeauthoryear{Byeon, Wang, Srivastava, and
  Koumoutsakos}{Byeon et~al\mbox{.}}{2018}]%
        {DBLP:conf/eccv/ByeonWSK18}
\bibfield{author}{\bibinfo{person}{Wonmin Byeon}, \bibinfo{person}{Qin Wang},
  \bibinfo{person}{Rupesh~Kumar Srivastava}, {and} \bibinfo{person}{Petros
  Koumoutsakos}.} \bibinfo{year}{2018}\natexlab{}.
\newblock \showarticletitle{ContextVP: Fully Context-Aware Video Prediction}.
  In \bibinfo{booktitle}{{\em Computer Vision - {ECCV} 2018 - 15th European
  Conference, Munich, Germany, September 8-14, 2018, Proceedings, Part {XVI}}}.
  \bibinfo{pages}{781--797}.
\newblock


\bibitem[\protect\citeauthoryear{Chuang, Goldman, Zheng, Curless, Salesin, and
  Szeliski}{Chuang et~al\mbox{.}}{2005}]%
        {DBLP:journals/tog/ChuangGZCSS05}
\bibfield{author}{\bibinfo{person}{Yung{-}Yu Chuang}, \bibinfo{person}{Dan~B.
  Goldman}, \bibinfo{person}{Ke~Colin Zheng}, \bibinfo{person}{Brian Curless},
  \bibinfo{person}{David Salesin}, {and} \bibinfo{person}{Richard Szeliski}.}
  \bibinfo{year}{2005}\natexlab{}.
\newblock \showarticletitle{Animating pictures with stochastic motion
  textures}.
\newblock \bibinfo{journal}{{\em {ACM} Trans. Graph.\/}} \bibinfo{volume}{24},
  \bibinfo{number}{3} (\bibinfo{year}{2005}), \bibinfo{pages}{853--860}.
\newblock


\bibitem[\protect\citeauthoryear{Denton and Birodkar}{Denton and
  Birodkar}{2017}]%
        {DBLP:conf/nips/DentonB17}
\bibfield{author}{\bibinfo{person}{Emily~L. Denton} {and}
  \bibinfo{person}{Vighnesh Birodkar}.} \bibinfo{year}{2017}\natexlab{}.
\newblock \showarticletitle{Unsupervised Learning of Disentangled
  Representations from Video}. In \bibinfo{booktitle}{{\em Advances in Neural
  Information Processing Systems 30: Annual Conference on Neural Information
  Processing Systems 2017, 4-9 December 2017, Long Beach, CA, {USA}}}.
  \bibinfo{pages}{4417--4426}.
\newblock


\bibitem[\protect\citeauthoryear{Dosovitskiy, Fischer, Ilg, H{\"{a}}usser,
  Hazirbas, Golkov, van~der Smagt, Cremers, and Brox}{Dosovitskiy
  et~al\mbox{.}}{2015}]%
        {DBLP:conf/iccv/DosovitskiyFIHH15}
\bibfield{author}{\bibinfo{person}{Alexey Dosovitskiy},
  \bibinfo{person}{Philipp Fischer}, \bibinfo{person}{Eddy Ilg},
  \bibinfo{person}{Philip H{\"{a}}usser}, \bibinfo{person}{Caner Hazirbas},
  \bibinfo{person}{Vladimir Golkov}, \bibinfo{person}{Patrick van~der Smagt},
  \bibinfo{person}{Daniel Cremers}, {and} \bibinfo{person}{Thomas Brox}.}
  \bibinfo{year}{2015}\natexlab{}.
\newblock \showarticletitle{FlowNet: Learning Optical Flow with Convolutional
  Networks}. In \bibinfo{booktitle}{{\em 2015 {IEEE} International Conference
  on Computer Vision, {ICCV} 2015, Santiago, Chile, December 7-13, 2015}}.
  \bibinfo{pages}{2758--2766}.
\newblock


\bibitem[\protect\citeauthoryear{Gao, Xiong, and Grauman}{Gao
  et~al\mbox{.}}{2017}]%
        {DBLP:journals/corr/abs-1712-04109}
\bibfield{author}{\bibinfo{person}{Ruohan Gao}, \bibinfo{person}{Bo Xiong},
  {and} \bibinfo{person}{Kristen Grauman}.} \bibinfo{year}{2017}\natexlab{}.
\newblock \showarticletitle{Im2Flow: Motion Hallucination from Static Images
  for Action Recognition}.
\newblock \bibinfo{journal}{{\em CoRR\/}}  \bibinfo{volume}{abs/1712.04109}
  (\bibinfo{year}{2017}).
\newblock
\showeprint[arxiv]{1712.04109}


\bibitem[\protect\citeauthoryear{Gatys, Ecker, and Bethge}{Gatys
  et~al\mbox{.}}{2016}]%
        {DBLP:conf/cvpr/GatysEB16}
\bibfield{author}{\bibinfo{person}{Leon~A. Gatys},
  \bibinfo{person}{Alexander~S. Ecker}, {and} \bibinfo{person}{Matthias
  Bethge}.} \bibinfo{year}{2016}\natexlab{}.
\newblock \showarticletitle{Image Style Transfer Using Convolutional Neural
  Networks}. In \bibinfo{booktitle}{{\em 2016 {IEEE} Conference on Computer
  Vision and Pattern Recognition, {CVPR} 2016, Las Vegas, NV, USA, June 27-30,
  2016}}. \bibinfo{pages}{2414--2423}.
\newblock


\bibitem[\protect\citeauthoryear{Hao, Huang, and Belongie}{Hao
  et~al\mbox{.}}{2018}]%
        {DBLP:conf/cvpr/HaoHB18}
\bibfield{author}{\bibinfo{person}{Zekun Hao}, \bibinfo{person}{Xun Huang},
  {and} \bibinfo{person}{Serge~J. Belongie}.} \bibinfo{year}{2018}\natexlab{}.
\newblock \showarticletitle{Controllable Video Generation With Sparse
  Trajectories}. In \bibinfo{booktitle}{{\em 2018 {IEEE} Conference on Computer
  Vision and Pattern Recognition, {CVPR} 2018, Salt Lake City, UT, USA, June
  18-22, 2018}}. \bibinfo{pages}{7854--7863}.
\newblock


\bibitem[\protect\citeauthoryear{Hays and Efros}{Hays and Efros}{2007}]%
        {DBLP:journals/tog/HaysE07}
\bibfield{author}{\bibinfo{person}{James Hays} {and} \bibinfo{person}{Alexei~A.
  Efros}.} \bibinfo{year}{2007}\natexlab{}.
\newblock \showarticletitle{Scene completion using millions of photographs}.
\newblock \bibinfo{journal}{{\em {ACM} Trans. Graph.\/}} \bibinfo{volume}{26},
  \bibinfo{number}{3} (\bibinfo{year}{2007}), \bibinfo{pages}{4}.
\newblock


\bibitem[\protect\citeauthoryear{He, Zhang, Ren, and Sun}{He
  et~al\mbox{.}}{2015}]%
        {DBLP:journals/pami/HeZR015}
\bibfield{author}{\bibinfo{person}{Kaiming He}, \bibinfo{person}{Xiangyu
  Zhang}, \bibinfo{person}{Shaoqing Ren}, {and} \bibinfo{person}{Jian Sun}.}
  \bibinfo{year}{2015}\natexlab{}.
\newblock \showarticletitle{Spatial Pyramid Pooling in Deep Convolutional
  Networks for Visual Recognition}.
\newblock \bibinfo{journal}{{\em {IEEE} Trans. Pattern Anal. Mach. Intell.\/}}
  \bibinfo{volume}{37}, \bibinfo{number}{9} (\bibinfo{year}{2015}),
  \bibinfo{pages}{1904--1916}.
\newblock


\bibitem[\protect\citeauthoryear{Johnson, Alahi, and Fei{-}Fei}{Johnson
  et~al\mbox{.}}{2016}]%
        {DBLP:conf/eccv/JohnsonAF16}
\bibfield{author}{\bibinfo{person}{Justin Johnson}, \bibinfo{person}{Alexandre
  Alahi}, {and} \bibinfo{person}{Li Fei{-}Fei}.}
  \bibinfo{year}{2016}\natexlab{}.
\newblock \showarticletitle{Perceptual Losses for Real-Time Style Transfer and
  Super-Resolution}. In \bibinfo{booktitle}{{\em Computer Vision - {ECCV} 2016
  - 14th European Conference, Amsterdam, The Netherlands, October 11-14, 2016,
  Proceedings, Part {II}}}. \bibinfo{pages}{694--711}.
\newblock


\bibitem[\protect\citeauthoryear{Karacan, Akata, Erdem, and Erdem}{Karacan
  et~al\mbox{.}}{2018}]%
        {DBLP:journals/corr/abs-1808-07413}
\bibfield{author}{\bibinfo{person}{Levent Karacan}, \bibinfo{person}{Zeynep
  Akata}, \bibinfo{person}{Aykut Erdem}, {and} \bibinfo{person}{Erkut Erdem}.}
  \bibinfo{year}{2018}\natexlab{}.
\newblock \showarticletitle{Manipulating Attributes of Natural Scenes via
  Hallucination}.
\newblock \bibinfo{journal}{{\em CoRR\/}}  \bibinfo{volume}{abs/1808.07413}
  (\bibinfo{year}{2018}).
\newblock
\showeprint[arxiv]{1808.07413}


\bibitem[\protect\citeauthoryear{Karras, Aila, Laine, and Lehtinen}{Karras
  et~al\mbox{.}}{2018}]%
        {PGGAN}
\bibfield{author}{\bibinfo{person}{Tero Karras}, \bibinfo{person}{Timo Aila},
  \bibinfo{person}{Samuli Laine}, {and} \bibinfo{person}{Jaakko Lehtinen}.}
  \bibinfo{year}{2018}\natexlab{}.
\newblock \showarticletitle{Progressive Growing of GANs for Improved Quality,
  Stability, and Variation}. In \bibinfo{booktitle}{{\em ICLR 2018}}.
\newblock


\bibitem[\protect\citeauthoryear{Kingma and Ba}{Kingma and Ba}{2014}]%
        {DBLP:journals/corr/KingmaB14}
\bibfield{author}{\bibinfo{person}{Diederik~P. Kingma} {and}
  \bibinfo{person}{Jimmy Ba}.} \bibinfo{year}{2014}\natexlab{}.
\newblock \showarticletitle{Adam: {A} Method for Stochastic Optimization}.
\newblock \bibinfo{journal}{{\em CoRR\/}}  \bibinfo{volume}{abs/1412.6980}
  (\bibinfo{year}{2014}).
\newblock
\showeprint[arxiv]{1412.6980}


\bibitem[\protect\citeauthoryear{Kingma and Welling}{Kingma and
  Welling}{2013}]%
        {DBLP:journals/corr/KingmaW13}
\bibfield{author}{\bibinfo{person}{Diederik~P. Kingma} {and}
  \bibinfo{person}{Max Welling}.} \bibinfo{year}{2013}\natexlab{}.
\newblock \showarticletitle{Auto-Encoding Variational Bayes}.
\newblock \bibinfo{journal}{{\em CoRR\/}}  \bibinfo{volume}{abs/1312.6114}
  (\bibinfo{year}{2013}).
\newblock
\showeprint[arxiv]{1312.6114}
\showURL{%
\url{http://arxiv.org/abs/1312.6114}}


\bibitem[\protect\citeauthoryear{Krizhevsky, Sutskever, and Hinton}{Krizhevsky
  et~al\mbox{.}}{2012}]%
        {DBLP:conf/nips/KrizhevskySH12}
\bibfield{author}{\bibinfo{person}{Alex Krizhevsky}, \bibinfo{person}{Ilya
  Sutskever}, {and} \bibinfo{person}{Geoffrey~E. Hinton}.}
  \bibinfo{year}{2012}\natexlab{}.
\newblock \showarticletitle{ImageNet Classification with Deep Convolutional
  Neural Networks}. In \bibinfo{booktitle}{{\em Advances in Neural Information
  Processing Systems 25: 26th Annual Conference on Neural Information
  Processing Systems 2012. Proceedings of a meeting held December 3-6, 2012,
  Lake Tahoe, Nevada, United States.}} \bibinfo{pages}{1106--1114}.
\newblock


\bibitem[\protect\citeauthoryear{Laffont, Ren, Tao, Qian, and Hays}{Laffont
  et~al\mbox{.}}{2014}]%
        {DBLP:journals/tog/LaffontRTQH14}
\bibfield{author}{\bibinfo{person}{Pierre{-}Yves Laffont},
  \bibinfo{person}{Zhile Ren}, \bibinfo{person}{Xiaofeng Tao},
  \bibinfo{person}{Chao Qian}, {and} \bibinfo{person}{James Hays}.}
  \bibinfo{year}{2014}\natexlab{}.
\newblock \showarticletitle{Transient attributes for high-level understanding
  and editing of outdoor scenes}.
\newblock \bibinfo{journal}{{\em {ACM} Trans. Graph.\/}} \bibinfo{volume}{33},
  \bibinfo{number}{4} (\bibinfo{year}{2014}), \bibinfo{pages}{149:1--149:11}.
\newblock


\bibitem[\protect\citeauthoryear{Li, Fang, Yang, Wang, Lu, and Yang}{Li
  et~al\mbox{.}}{2017}]%
        {DBLP:conf/nips/LiFYWLY17}
\bibfield{author}{\bibinfo{person}{Yijun Li}, \bibinfo{person}{Chen Fang},
  \bibinfo{person}{Jimei Yang}, \bibinfo{person}{Zhaowen Wang},
  \bibinfo{person}{Xin Lu}, {and} \bibinfo{person}{Ming{-}Hsuan Yang}.}
  \bibinfo{year}{2017}\natexlab{}.
\newblock \showarticletitle{Universal Style Transfer via Feature Transforms}.
  In \bibinfo{booktitle}{{\em Advances in Neural Information Processing Systems
  30: Annual Conference on Neural Information Processing Systems 2017, 4-9
  December 2017, Long Beach, CA, {USA}}}. \bibinfo{pages}{385--395}.
\newblock


\bibitem[\protect\citeauthoryear{Li, Fang, Yang, Wang, Lu, and Yang}{Li
  et~al\mbox{.}}{2018}]%
        {Prediction-ECCV-2018}
\bibfield{author}{\bibinfo{person}{Yijun Li}, \bibinfo{person}{Chen Fang},
  \bibinfo{person}{Jimei Yang}, \bibinfo{person}{Zhaowen Wang},
  \bibinfo{person}{Xin Lu}, {and} \bibinfo{person}{Ming-Hsuan Yang}.}
  \bibinfo{year}{2018}\natexlab{}.
\newblock \showarticletitle{Flow-Grounded Spatial-Temporal Video Prediction
  from Still Images}. In \bibinfo{booktitle}{{\em European Conference on
  Computer Vision}}.
\newblock


\bibitem[\protect\citeauthoryear{Liao, Finch, and Hoppe}{Liao
  et~al\mbox{.}}{2015}]%
        {DBLP:journals/tog/LiaoFH15}
\bibfield{author}{\bibinfo{person}{Jing Liao}, \bibinfo{person}{Mark Finch},
  {and} \bibinfo{person}{Hugues Hoppe}.} \bibinfo{year}{2015}\natexlab{}.
\newblock \showarticletitle{Fast computation of seamless video loops}.
\newblock \bibinfo{journal}{{\em {ACM} Trans. Graph.\/}} \bibinfo{volume}{34},
  \bibinfo{number}{6} (\bibinfo{year}{2015}), \bibinfo{pages}{197:1--197:10}.
\newblock


\bibitem[\protect\citeauthoryear{Liao, Joshi, and Hoppe}{Liao
  et~al\mbox{.}}{2013}]%
        {DBLP:journals/tog/LiaoJH13}
\bibfield{author}{\bibinfo{person}{Zicheng Liao}, \bibinfo{person}{Neel Joshi},
  {and} \bibinfo{person}{Hugues Hoppe}.} \bibinfo{year}{2013}\natexlab{}.
\newblock \showarticletitle{Automated video looping with progressive dynamism}.
\newblock \bibinfo{journal}{{\em {ACM} Trans. Graph.\/}} \bibinfo{volume}{32},
  \bibinfo{number}{4} (\bibinfo{year}{2013}), \bibinfo{pages}{77:1--77:10}.
\newblock


\bibitem[\protect\citeauthoryear{Lotter, Kreiman, and Cox}{Lotter
  et~al\mbox{.}}{2017}]%
        {ICLR07Lotter}
\bibfield{author}{\bibinfo{person}{William Lotter}, \bibinfo{person}{Gabriel
  Kreiman}, {and} \bibinfo{person}{David~D. Cox}.}
  \bibinfo{year}{2017}\natexlab{}.
\newblock \showarticletitle{Deep Predictive Coding Networks for Video
  Prediction and Unsupervised Learning}.
\newblock  (\bibinfo{date}{4} \bibinfo{year}{2017}).
\newblock


\bibitem[\protect\citeauthoryear{Luan, Paris, Shechtman, and Bala}{Luan
  et~al\mbox{.}}{2017}]%
        {DBLP:conf/cvpr/LuanPSB17}
\bibfield{author}{\bibinfo{person}{Fujun Luan}, \bibinfo{person}{Sylvain
  Paris}, \bibinfo{person}{Eli Shechtman}, {and} \bibinfo{person}{Kavita
  Bala}.} \bibinfo{year}{2017}\natexlab{}.
\newblock \showarticletitle{Deep Photo Style Transfer}. In
  \bibinfo{booktitle}{{\em 2017 {IEEE} Conference on Computer Vision and
  Pattern Recognition, {CVPR} 2017, Honolulu, HI, USA, July 21-26, 2017}}.
  \bibinfo{pages}{6997--7005}.
\newblock


\bibitem[\protect\citeauthoryear{Martin{-}Brualla, Gallup, and
  Seitz}{Martin{-}Brualla et~al\mbox{.}}{2015}]%
        {DBLP:journals/tog/Seitz15}
\bibfield{author}{\bibinfo{person}{Ricardo Martin{-}Brualla},
  \bibinfo{person}{David Gallup}, {and} \bibinfo{person}{Steven~M. Seitz}.}
  \bibinfo{year}{2015}\natexlab{}.
\newblock \showarticletitle{Time-lapse mining from internet photos}.
\newblock \bibinfo{journal}{{\em {ACM} Trans. Graph.\/}} \bibinfo{volume}{34},
  \bibinfo{number}{4} (\bibinfo{year}{2015}), \bibinfo{pages}{62:1--62:8}.
\newblock


\bibitem[\protect\citeauthoryear{Mathieu, Couprie, and Lecun}{Mathieu
  et~al\mbox{.}}{2016}]%
        {ICLR06Mathieu}
\bibfield{author}{\bibinfo{person}{Michael Mathieu}, \bibinfo{person}{Camille
  Couprie}, {and} \bibinfo{person}{Yann Lecun}.}
  \bibinfo{year}{2016}\natexlab{}.
\newblock \showarticletitle{Deep multi-scale video prediction beyond mean
  square error}. In \bibinfo{booktitle}{{\em ICLR'06}}.
\newblock


\bibitem[\protect\citeauthoryear{Mechrez, Shechtman, and
  Zelnik{-}Manor}{Mechrez et~al\mbox{.}}{2017}]%
        {DBLP:conf/bmvc/MechrezSZ17}
\bibfield{author}{\bibinfo{person}{Roey Mechrez}, \bibinfo{person}{Eli
  Shechtman}, {and} \bibinfo{person}{Lihi Zelnik{-}Manor}.}
  \bibinfo{year}{2017}\natexlab{}.
\newblock \showarticletitle{Photorealistic Style Transfer with Screened Poisson
  Equation}. In \bibinfo{booktitle}{{\em British Machine Vision Conference
  2017, {BMVC} 2017, London, UK, September 4-7, 2017}}.
\newblock


\bibitem[\protect\citeauthoryear{Oh, Joo, Joshi, Wang, Kweon, and Kang}{Oh
  et~al\mbox{.}}{2017}]%
        {DBLP:conf/iccv/OhJJWKK17}
\bibfield{author}{\bibinfo{person}{Tae{-}Hyun Oh}, \bibinfo{person}{Kyungdon
  Joo}, \bibinfo{person}{Neel Joshi}, \bibinfo{person}{Baoyuan Wang},
  \bibinfo{person}{In~So Kweon}, {and} \bibinfo{person}{Sing~Bing Kang}.}
  \bibinfo{year}{2017}\natexlab{}.
\newblock \showarticletitle{Personalized Cinemagraphs Using Semantic
  Understanding and Collaborative Learning}. In \bibinfo{booktitle}{{\em {IEEE}
  International Conference on Computer Vision, {ICCV} 2017, Venice, Italy,
  October 22-29, 2017}}. \bibinfo{pages}{5170--5179}.
\newblock


\bibitem[\protect\citeauthoryear{Okabe, Anjyo, Igarashi, and Seidel}{Okabe
  et~al\mbox{.}}{2009}]%
        {DBLP:journals/cgf/OkabeAIS09}
\bibfield{author}{\bibinfo{person}{Makoto Okabe}, \bibinfo{person}{Ken{-}ichi
  Anjyo}, \bibinfo{person}{Takeo Igarashi}, {and} \bibinfo{person}{Hans{-}Peter
  Seidel}.} \bibinfo{year}{2009}\natexlab{}.
\newblock \showarticletitle{Animating Pictures of Fluid using Video Examples}.
\newblock \bibinfo{journal}{{\em Comput. Graph. Forum\/}} \bibinfo{volume}{28},
  \bibinfo{number}{2} (\bibinfo{year}{2009}), \bibinfo{pages}{677--686}.
\newblock


\bibitem[\protect\citeauthoryear{Okabe, Anjyo, and Onai}{Okabe
  et~al\mbox{.}}{2011}]%
        {DBLP:journals/cgf/OkabeAO11}
\bibfield{author}{\bibinfo{person}{Makoto Okabe}, \bibinfo{person}{Ken Anjyo},
  {and} \bibinfo{person}{Rikio Onai}.} \bibinfo{year}{2011}\natexlab{}.
\newblock \showarticletitle{Creating Fluid Animation from a Single Image using
  Video Database}.
\newblock \bibinfo{journal}{{\em Comput. Graph. Forum\/}} \bibinfo{volume}{30},
  \bibinfo{number}{7} (\bibinfo{year}{2011}), \bibinfo{pages}{1973--1982}.
\newblock


\bibitem[\protect\citeauthoryear{Okabe, Dobashi, and Anjyo}{Okabe
  et~al\mbox{.}}{2018}]%
        {DBLP:journals/vc/OkabeDA18}
\bibfield{author}{\bibinfo{person}{Makoto Okabe}, \bibinfo{person}{Yoshinori
  Dobashi}, {and} \bibinfo{person}{Ken Anjyo}.}
  \bibinfo{year}{2018}\natexlab{}.
\newblock \showarticletitle{Animating pictures of water scenes using video
  retrieval}.
\newblock \bibinfo{journal}{{\em The Visual Computer\/}} \bibinfo{volume}{34},
  \bibinfo{number}{3} (\bibinfo{year}{2018}), \bibinfo{pages}{347--358}.
\newblock


\bibitem[\protect\citeauthoryear{Prashnani, Noorkami, Vaquero, and
  Sen}{Prashnani et~al\mbox{.}}{2017}]%
        {DBLP:journals/cgf/PrashnaniNVS17}
\bibfield{author}{\bibinfo{person}{Ekta Prashnani}, \bibinfo{person}{Maneli
  Noorkami}, \bibinfo{person}{Daniel Vaquero}, {and} \bibinfo{person}{Pradeep
  Sen}.} \bibinfo{year}{2017}\natexlab{}.
\newblock \showarticletitle{A Phase-Based Approach for Animating Images Using
  Video Examples}.
\newblock \bibinfo{journal}{{\em Comput. Graph. Forum\/}} \bibinfo{volume}{36},
  \bibinfo{number}{6} (\bibinfo{year}{2017}), \bibinfo{pages}{303--311}.
\newblock


\bibitem[\protect\citeauthoryear{Ranjan and Black}{Ranjan and Black}{2017}]%
        {DBLP:conf/cvpr/RanjanB17}
\bibfield{author}{\bibinfo{person}{Anurag Ranjan} {and}
  \bibinfo{person}{Michael~J. Black}.} \bibinfo{year}{2017}\natexlab{}.
\newblock \showarticletitle{Optical Flow Estimation Using a Spatial Pyramid
  Network}. In \bibinfo{booktitle}{{\em 2017 {IEEE} Conference on Computer
  Vision and Pattern Recognition, {CVPR} 2017, Honolulu, HI, USA, July 21-26,
  2017}}. \bibinfo{pages}{2720--2729}.
\newblock


\bibitem[\protect\citeauthoryear{Ranzato, Szlam, Bruna, Mathieu, Collobert, and
  Chopra}{Ranzato et~al\mbox{.}}{2014}]%
        {DBLP:journals/corr/RanzatoSBMCC14}
\bibfield{author}{\bibinfo{person}{Marc'Aurelio Ranzato},
  \bibinfo{person}{Arthur Szlam}, \bibinfo{person}{Joan Bruna},
  \bibinfo{person}{Micha{\"{e}}l Mathieu}, \bibinfo{person}{Ronan Collobert},
  {and} \bibinfo{person}{Sumit Chopra}.} \bibinfo{year}{2014}\natexlab{}.
\newblock \showarticletitle{Video (language) modeling: a baseline for
  generative models of natural videos}.
\newblock \bibinfo{journal}{{\em CoRR\/}}  \bibinfo{volume}{abs/1412.6604}
  (\bibinfo{year}{2014}).
\newblock
\showeprint[arxiv]{1412.6604}


\bibitem[\protect\citeauthoryear{Reinhard, Ashikhmin, Gooch, and
  Shirley}{Reinhard et~al\mbox{.}}{2001}]%
        {DBLP:journals/cga/ReinhardAGS01}
\bibfield{author}{\bibinfo{person}{Erik Reinhard}, \bibinfo{person}{Michael
  Ashikhmin}, \bibinfo{person}{Bruce Gooch}, {and} \bibinfo{person}{Peter
  Shirley}.} \bibinfo{year}{2001}\natexlab{}.
\newblock \showarticletitle{Color Transfer between Images}.
\newblock \bibinfo{journal}{{\em {IEEE} Computer Graphics and Applications\/}}
  \bibinfo{volume}{21}, \bibinfo{number}{5} (\bibinfo{year}{2001}),
  \bibinfo{pages}{34--41}.
\newblock


\bibitem[\protect\citeauthoryear{Ren, Yan, Ni, Liu, Yang, and Zha}{Ren
  et~al\mbox{.}}{2017}]%
        {DBLP:conf/aaai/RenYNLYZ17}
\bibfield{author}{\bibinfo{person}{Zhe Ren}, \bibinfo{person}{Junchi Yan},
  \bibinfo{person}{Bingbing Ni}, \bibinfo{person}{Bin Liu},
  \bibinfo{person}{Xiaokang Yang}, {and} \bibinfo{person}{Hongyuan Zha}.}
  \bibinfo{year}{2017}\natexlab{}.
\newblock \showarticletitle{Unsupervised Deep Learning for Optical Flow
  Estimation}. In \bibinfo{booktitle}{{\em Proceedings of the Thirty-First
  {AAAI} Conference on Artificial Intelligence, February 4-9, 2017, San
  Francisco, California, {USA.}}} \bibinfo{pages}{1495--1501}.
\newblock


\bibitem[\protect\citeauthoryear{Revaud, Weinzaepfel, Harchaoui, and
  Schmid}{Revaud et~al\mbox{.}}{2016}]%
        {DBLP:journals/ijcv/RevaudWHS16}
\bibfield{author}{\bibinfo{person}{J{\'{e}}r{\^{o}}me Revaud},
  \bibinfo{person}{Philippe Weinzaepfel}, \bibinfo{person}{Za{\"{\i}}d
  Harchaoui}, {and} \bibinfo{person}{Cordelia Schmid}.}
  \bibinfo{year}{2016}\natexlab{}.
\newblock \showarticletitle{DeepMatching: Hierarchical Deformable Dense
  Matching}.
\newblock \bibinfo{journal}{{\em International Journal of Computer Vision\/}}
  \bibinfo{volume}{120}, \bibinfo{number}{3} (\bibinfo{year}{2016}),
  \bibinfo{pages}{300--323}.
\newblock


\bibitem[\protect\citeauthoryear{Ronneberger, P.Fischer, and Brox}{Ronneberger
  et~al\mbox{.}}{2015}]%
        {RFB15a}
\bibfield{author}{\bibinfo{person}{O. Ronneberger},
  \bibinfo{person}{P.Fischer}, {and} \bibinfo{person}{T. Brox}.}
  \bibinfo{year}{2015}\natexlab{}.
\newblock \showarticletitle{U-Net: Convolutional Networks for Biomedical Image
  Segmentation}. In \bibinfo{booktitle}{{\em Medical Image Computing and
  Computer-Assisted Intervention (MICCAI)}} {\em (\bibinfo{series}{LNCS})},
  Vol.~\bibinfo{volume}{9351}. \bibinfo{publisher}{Springer},
  \bibinfo{pages}{234--241}.
\newblock
\showURL{%
\url{http://lmb.informatik.uni-freiburg.de/Publications/2015/RFB15a}}
\newblock
\shownote{(available on arXiv:1505.04597 [cs.CV]).}


\bibitem[\protect\citeauthoryear{Sch{\"{o}}dl, Szeliski, Salesin, and
  Essa}{Sch{\"{o}}dl et~al\mbox{.}}{2000}]%
        {DBLP:conf/siggraph/SchodlSSE00}
\bibfield{author}{\bibinfo{person}{Arno Sch{\"{o}}dl}, \bibinfo{person}{Richard
  Szeliski}, \bibinfo{person}{David Salesin}, {and} \bibinfo{person}{Irfan~A.
  Essa}.} \bibinfo{year}{2000}\natexlab{}.
\newblock \showarticletitle{Video textures}. In \bibinfo{booktitle}{{\em
  Proceedings of the 27th Annual Conference on Computer Graphics and
  Interactive Techniques, {SIGGRAPH} 2000, New Orleans, LA, USA, July 23-28,
  2000}}. \bibinfo{pages}{489--498}.
\newblock


\bibitem[\protect\citeauthoryear{Sch{\"{u}}ldt, Laptev, and
  Caputo}{Sch{\"{u}}ldt et~al\mbox{.}}{2004}]%
        {DBLP:conf/icpr/SchuldtLC04}
\bibfield{author}{\bibinfo{person}{Christian Sch{\"{u}}ldt},
  \bibinfo{person}{Ivan Laptev}, {and} \bibinfo{person}{Barbara Caputo}.}
  \bibinfo{year}{2004}\natexlab{}.
\newblock \showarticletitle{Recognizing Human Actions: {A} Local {SVM}
  Approach}. In \bibinfo{booktitle}{{\em 17th International Conference on
  Pattern Recognition, {ICPR} 2004, Cambridge, UK, August 23-26, 2004.}}
  \bibinfo{pages}{32--36}.
\newblock


\bibitem[\protect\citeauthoryear{Shi, Chen, Wang, Yeung, Wong, and Woo}{Shi
  et~al\mbox{.}}{2015}]%
        {DBLP:conf/nips/ShiCWYWW15}
\bibfield{author}{\bibinfo{person}{Xingjian Shi}, \bibinfo{person}{Zhourong
  Chen}, \bibinfo{person}{Hao Wang}, \bibinfo{person}{Dit{-}Yan Yeung},
  \bibinfo{person}{Wai{-}Kin Wong}, {and} \bibinfo{person}{Wang{-}chun Woo}.}
  \bibinfo{year}{2015}\natexlab{}.
\newblock \showarticletitle{Convolutional {LSTM} Network: {A} Machine Learning
  Approach for Precipitation Nowcasting}. In \bibinfo{booktitle}{{\em Advances
  in Neural Information Processing Systems 28: Annual Conference on Neural
  Information Processing Systems 2015, December 7-12, 2015, Montreal, Quebec,
  Canada}}. \bibinfo{pages}{802--810}.
\newblock


\bibitem[\protect\citeauthoryear{Shih, Paris, Durand, and Freeman}{Shih
  et~al\mbox{.}}{2013}]%
        {DBLP:journals/tog/ShihPDF13}
\bibfield{author}{\bibinfo{person}{Yi{-}Chang Shih}, \bibinfo{person}{Sylvain
  Paris}, \bibinfo{person}{Fr{\'{e}}do Durand}, {and}
  \bibinfo{person}{William~T. Freeman}.} \bibinfo{year}{2013}\natexlab{}.
\newblock \showarticletitle{Data-driven hallucination of different times of day
  from a single outdoor photo}.
\newblock \bibinfo{journal}{{\em {ACM} Trans. Graph.\/}} \bibinfo{volume}{32},
  \bibinfo{number}{6} (\bibinfo{year}{2013}), \bibinfo{pages}{200:1--200:11}.
\newblock


\bibitem[\protect\citeauthoryear{Simonyan and Zisserman}{Simonyan and
  Zisserman}{2014}]%
        {Simonyan14c}
\bibfield{author}{\bibinfo{person}{K. Simonyan} {and} \bibinfo{person}{A.
  Zisserman}.} \bibinfo{year}{2014}\natexlab{}.
\newblock \showarticletitle{Very Deep Convolutional Networks for Large-Scale
  Image Recognition}.
\newblock \bibinfo{journal}{{\em CoRR\/}}  \bibinfo{volume}{abs/1409.1556}
  (\bibinfo{year}{2014}).
\newblock


\bibitem[\protect\citeauthoryear{Srivastava, Mansimov, and
  Salakhutdinov}{Srivastava et~al\mbox{.}}{2015}]%
        {DBLP:conf/icml/SrivastavaMS15}
\bibfield{author}{\bibinfo{person}{Nitish Srivastava}, \bibinfo{person}{Elman
  Mansimov}, {and} \bibinfo{person}{Ruslan Salakhutdinov}.}
  \bibinfo{year}{2015}\natexlab{}.
\newblock \showarticletitle{Unsupervised Learning of Video Representations
  using LSTMs}. In \bibinfo{booktitle}{{\em Proceedings of the 32nd
  International Conference on Machine Learning, {ICML} 2015, Lille, France,
  6-11 July 2015}}. \bibinfo{pages}{843--852}.
\newblock


\bibitem[\protect\citeauthoryear{Tai, Jia, and Tang}{Tai et~al\mbox{.}}{2005}]%
        {DBLP:conf/cvpr/TaiJT05}
\bibfield{author}{\bibinfo{person}{Yu{-}Wing Tai}, \bibinfo{person}{Jiaya Jia},
  {and} \bibinfo{person}{Chi{-}Keung Tang}.} \bibinfo{year}{2005}\natexlab{}.
\newblock \showarticletitle{Local Color Transfer via Probabilistic Segmentation
  by Expectation-Maximization}. In \bibinfo{booktitle}{{\em 2005 {IEEE}
  Computer Society Conference on Computer Vision and Pattern Recognition
  {(CVPR} 2005), 20-26 June 2005, San Diego, CA, {USA}}}.
  \bibinfo{pages}{747--754}.
\newblock


\bibitem[\protect\citeauthoryear{Tolstikhin, Bousquet, Gelly, and
  Sch{\"{o}}lkopf}{Tolstikhin et~al\mbox{.}}{2017}]%
        {DBLP:journals/corr/abs-1711-01558}
\bibfield{author}{\bibinfo{person}{Ilya~O. Tolstikhin},
  \bibinfo{person}{Olivier Bousquet}, \bibinfo{person}{Sylvain Gelly}, {and}
  \bibinfo{person}{Bernhard Sch{\"{o}}lkopf}.} \bibinfo{year}{2017}\natexlab{}.
\newblock \showarticletitle{Wasserstein Auto-Encoders}.
\newblock \bibinfo{journal}{{\em CoRR\/}}  \bibinfo{volume}{abs/1711.01558}
  (\bibinfo{year}{2017}).
\newblock
\showeprint[arxiv]{1711.01558}
\showURL{%
\url{http://arxiv.org/abs/1711.01558}}


\bibitem[\protect\citeauthoryear{Tsai, Shen, Lin, Sunkavalli, and Yang}{Tsai
  et~al\mbox{.}}{2016}]%
        {DBLP:journals/tog/TsaiSLSY16}
\bibfield{author}{\bibinfo{person}{Yi{-}Hsuan Tsai}, \bibinfo{person}{Xiaohui
  Shen}, \bibinfo{person}{Zhe Lin}, \bibinfo{person}{Kalyan Sunkavalli}, {and}
  \bibinfo{person}{Ming{-}Hsuan Yang}.} \bibinfo{year}{2016}\natexlab{}.
\newblock \showarticletitle{Sky is not the limit: semantic-aware sky
  replacement}.
\newblock \bibinfo{journal}{{\em {ACM} Trans. Graph.\/}} \bibinfo{volume}{35},
  \bibinfo{number}{4} (\bibinfo{year}{2016}), \bibinfo{pages}{149:1--149:11}.
\newblock


\bibitem[\protect\citeauthoryear{Vondrick, Pirsiavash, and Torralba}{Vondrick
  et~al\mbox{.}}{2016}]%
        {DBLP:conf/nips/VondrickPT16}
\bibfield{author}{\bibinfo{person}{Carl Vondrick}, \bibinfo{person}{Hamed
  Pirsiavash}, {and} \bibinfo{person}{Antonio Torralba}.}
  \bibinfo{year}{2016}\natexlab{}.
\newblock \showarticletitle{Generating Videos with Scene Dynamics}. In
  \bibinfo{booktitle}{{\em Advances in Neural Information Processing Systems
  29: Annual Conference on Neural Information Processing Systems 2016, December
  5-10, 2016, Barcelona, Spain}}. \bibinfo{pages}{613--621}.
\newblock


\bibitem[\protect\citeauthoryear{Walker, Gupta, and Hebert}{Walker
  et~al\mbox{.}}{2015}]%
        {DBLP:conf/iccv/WalkerGH15}
\bibfield{author}{\bibinfo{person}{Jacob Walker}, \bibinfo{person}{Abhinav
  Gupta}, {and} \bibinfo{person}{Martial Hebert}.}
  \bibinfo{year}{2015}\natexlab{}.
\newblock \showarticletitle{Dense Optical Flow Prediction from a Static Image}.
  In \bibinfo{booktitle}{{\em 2015 {IEEE} International Conference on Computer
  Vision, {ICCV} 2015, Santiago, Chile, December 7-13, 2015}}.
  \bibinfo{pages}{2443--2451}.
\newblock


\bibitem[\protect\citeauthoryear{Wang, Liu, Zhu, Tao, Kautz, and
  Catanzaro}{Wang et~al\mbox{.}}{2018a}]%
        {wang2018pix2pixHD}
\bibfield{author}{\bibinfo{person}{Ting-Chun Wang}, \bibinfo{person}{Ming-Yu
  Liu}, \bibinfo{person}{Jun-Yan Zhu}, \bibinfo{person}{Andrew Tao},
  \bibinfo{person}{Jan Kautz}, {and} \bibinfo{person}{Bryan Catanzaro}.}
  \bibinfo{year}{2018}\natexlab{a}.
\newblock \showarticletitle{High-Resolution Image Synthesis and Semantic
  Manipulation with Conditional GANs}. In \bibinfo{booktitle}{{\em Proceedings
  of the IEEE Conference on Computer Vision and Pattern Recognition}}.
\newblock


\bibitem[\protect\citeauthoryear{Wang, Yang, Yang, Zhao, Wang, and Xu}{Wang
  et~al\mbox{.}}{2018b}]%
        {Wang_2018_CVPR}
\bibfield{author}{\bibinfo{person}{Yang Wang}, \bibinfo{person}{Yi Yang},
  \bibinfo{person}{Zhenheng Yang}, \bibinfo{person}{Liang Zhao},
  \bibinfo{person}{Peng Wang}, {and} \bibinfo{person}{Wei Xu}.}
  \bibinfo{year}{2018}\natexlab{b}.
\newblock \showarticletitle{Occlusion Aware Unsupervised Learning of Optical
  Flow}. In \bibinfo{booktitle}{{\em The IEEE Conference on Computer Vision and
  Pattern Recognition (CVPR)}}.
\newblock


\bibitem[\protect\citeauthoryear{Weinzaepfel, Revaud, Harchaoui, and
  Schmid}{Weinzaepfel et~al\mbox{.}}{2013}]%
        {DBLP:conf/iccv/WeinzaepfelRHS13}
\bibfield{author}{\bibinfo{person}{Philippe Weinzaepfel},
  \bibinfo{person}{J{\'{e}}r{\^{o}}me Revaud}, \bibinfo{person}{Za{\"{\i}}d
  Harchaoui}, {and} \bibinfo{person}{Cordelia Schmid}.}
  \bibinfo{year}{2013}\natexlab{}.
\newblock \showarticletitle{DeepFlow: Large Displacement Optical Flow with Deep
  Matching}. In \bibinfo{booktitle}{{\em {IEEE} International Conference on
  Computer Vision, {ICCV} 2013, Sydney, Australia, December 1-8, 2013}}.
  \bibinfo{pages}{1385--1392}.
\newblock


\bibitem[\protect\citeauthoryear{Wu, Dong, Kong, Mei, Paul, and Zhang}{Wu
  et~al\mbox{.}}{2013}]%
        {DBLP:journals/cgf/WuDKMPZ13}
\bibfield{author}{\bibinfo{person}{Fuzhang Wu}, \bibinfo{person}{Weiming Dong},
  \bibinfo{person}{Yan Kong}, \bibinfo{person}{Xing Mei},
  \bibinfo{person}{Jean{-}Claude Paul}, {and} \bibinfo{person}{Xiaopeng
  Zhang}.} \bibinfo{year}{2013}\natexlab{}.
\newblock \showarticletitle{Content-Based Colour Transfer}.
\newblock \bibinfo{journal}{{\em Comput. Graph. Forum\/}} \bibinfo{volume}{32},
  \bibinfo{number}{1} (\bibinfo{year}{2013}), \bibinfo{pages}{190--203}.
\newblock


\bibitem[\protect\citeauthoryear{Xiong, Luo, Ma, Liu, and Luo}{Xiong
  et~al\mbox{.}}{2018}]%
        {xiong2018learning}
\bibfield{author}{\bibinfo{person}{Wei Xiong}, \bibinfo{person}{Wenhan Luo},
  \bibinfo{person}{Lin Ma}, \bibinfo{person}{Wei Liu}, {and}
  \bibinfo{person}{Jiebo Luo}.} \bibinfo{year}{2018}\natexlab{}.
\newblock \showarticletitle{Learning to Generate Time-Lapse Videos Using
  Multi-Stage Dynamic Generative Adversarial Networks}. In
  \bibinfo{booktitle}{{\em Computer Vision and Pattern Recognition (CVPR), 2018
  IEEE Conference on}}.
\newblock


\bibitem[\protect\citeauthoryear{Xue, Wu, Bouman, and Freeman}{Xue
  et~al\mbox{.}}{2016}]%
        {DBLP:conf/nips/Xue0BF16}
\bibfield{author}{\bibinfo{person}{Tianfan Xue}, \bibinfo{person}{Jiajun Wu},
  \bibinfo{person}{Katherine~L. Bouman}, {and} \bibinfo{person}{Bill Freeman}.}
  \bibinfo{year}{2016}\natexlab{}.
\newblock \showarticletitle{Visual Dynamics: Probabilistic Future Frame
  Synthesis via Cross Convolutional Networks}. In \bibinfo{booktitle}{{\em
  Advances in Neural Information Processing Systems 29: Annual Conference on
  Neural Information Processing Systems 2016, December 5-10, 2016, Barcelona,
  Spain}}. \bibinfo{pages}{91--99}.
\newblock


\bibitem[\protect\citeauthoryear{Zhang, Isola, Efros, Shechtman, and
  Wang}{Zhang et~al\mbox{.}}{2018}]%
        {Zhang_2018_CVPR}
\bibfield{author}{\bibinfo{person}{Richard Zhang}, \bibinfo{person}{Phillip
  Isola}, \bibinfo{person}{Alexei~A. Efros}, \bibinfo{person}{Eli Shechtman},
  {and} \bibinfo{person}{Oliver Wang}.} \bibinfo{year}{2018}\natexlab{}.
\newblock \showarticletitle{The Unreasonable Effectiveness of Deep Features as
  a Perceptual Metric}. In \bibinfo{booktitle}{{\em The IEEE Conference on
  Computer Vision and Pattern Recognition (CVPR)}}.
\newblock


\bibitem[\protect\citeauthoryear{Zhou and Berg}{Zhou and Berg}{2016}]%
        {DBLP:conf/eccv/ZhouB16}
\bibfield{author}{\bibinfo{person}{Yipin Zhou} {and} \bibinfo{person}{Tamara~L.
  Berg}.} \bibinfo{year}{2016}\natexlab{}.
\newblock \showarticletitle{Learning Temporal Transformations from Time-Lapse
  Videos}. In \bibinfo{booktitle}{{\em Computer Vision - {ECCV} 2016 - 14th
  European Conference, Amsterdam, The Netherlands, October 11-14, 2016,
  Proceedings, Part {VIII}}}. \bibinfo{pages}{262--277}.
\newblock


\bibitem[\protect\citeauthoryear{Zhu, Zhang, Pathak, Darrell, Efros, Wang, and
  Shechtman}{Zhu et~al\mbox{.}}{2017}]%
        {DBLP:conf/nips/ZhuZPDEWS17}
\bibfield{author}{\bibinfo{person}{Jun{-}Yan Zhu}, \bibinfo{person}{Richard
  Zhang}, \bibinfo{person}{Deepak Pathak}, \bibinfo{person}{Trevor Darrell},
  \bibinfo{person}{Alexei~A. Efros}, \bibinfo{person}{Oliver Wang}, {and}
  \bibinfo{person}{Eli Shechtman}.} \bibinfo{year}{2017}\natexlab{}.
\newblock \showarticletitle{Toward Multimodal Image-to-Image Translation}. In
  \bibinfo{booktitle}{{\em Advances in Neural Information Processing Systems
  30: Annual Conference on Neural Information Processing Systems 2017, 4-9
  December 2017, Long Beach, CA, {USA}}}. \bibinfo{pages}{465--476}.
\newblock


\end{thebibliography}

\appendix
\section{Implementation Details of Our DNNs}
\label{sec:ImplementationDetails}

\subsection{Network Architectures}
\label{sec:NetworkArchitecture}
\begin{table*}[t]
\begin{center}
\caption{
Network architecture of the motion predictor and appearance predictor. Concat($\textbf{z}$) means concatenation between each pixel of input feature maps and $\textbf{z}$. For 2D convolutional layers and residual blocks, C is \chkE{the} number of channels, K is \chkE{the} kernel \chkG{width and height}, S is \chkE{the} stride, and P is \chkE{the} padding. Upsample(2) means magnifying input feature maps \chkE{twice} using nearest\chkE{-}neighbor interpolation. 
}
\label{table:generator}
\begin{tabular}{l|l|l}
\hline
Components                & Layers  &   Specifications  \\ \hline \hline
                         & conv1   & Concat($\textbf{z}$), Conv2D(C(128), K(5), S(2), P(2)), LeakyReLU(0.1)                                                \\
Downsampling & conv2   & Concat($\textbf{z}$), Conv2D(C(256), K(3), S(2), P(1)), InstanceNorm(256), LeakyReLU(0.1)                             \\
                         & conv3   & Concat($\textbf{z}$), Conv2D(C(512), K(3), S(2), P(1)), InstanceNorm(512), LeakyReLU(0.1)                             \\ \hline
Residual  & res1, $\ldots$, res5 & ResBlock2D(C(512), K(3), S(1), P(1))                                                         \\ \hline
                         & upconv1 & Concat(conv3), Upsample(2), Conv2D(C(256), K(3), S(1), P(1)), InstanceNorm(256), LeakyReLU(0.1) \\
Upsampling  & upconv2 & Concat(conv2), Upsample(2), Conv2D(C(128), K(3), S(1), P(1)), InstanceNorm(128), LeakyReLU(0.1) \\
                         & upconv3 & Concat(conv1), Upsample(2), Conv2D(C(3) or C(6), K(5), S(1), P(2))                             \\ \hline
\end{tabular}
\end{center}
\end{table*}

\begin{table}[t]
\begin{center}
\caption{Network architecture of the motion encoder and the appearance encoder. \chkE{Notations are the same as those in Table~\ref{table:generator}.}}
\label{table:encoder}
\begin{tabular}{l|l|l}
\hline
Component & Layers  &  Specifications                  \\ \hline \hline
          & conv1   & Conv2D(C(64), K(4), S(2), P(1))    \\
          & res1    & ResBlock2D(C(128), K(3), S(1), P(1)) \\
Encoder   & res2    & ResBlock2D(C(192), K(3), S(1), P(1)) \\
          & res3    & ResBlock2D(C(256), K(3), S(1), P(1)) \\
          & pooling & LeakyReLU(0.2), AvgPool2D(8,8)                             \\
          & fc      & Linear(8)                          \\ \hline       
\end{tabular}
\end{center}
\end{table}

\chkE{Table~\ref{table:generator} summarizes the architecture of our motion and appearance predictors $P^{\mathcal{M}}$ and $P^{\mathcal{A}}$, whereas Table~\ref{table:encoder} shows the network architecture of our motion and appearance encoders $E^{\mathcal{M}}$ and $E^{\mathcal{A}}$.}

\subsection{Training Algorithms}
\label{sec:TrainingDetails}
The training procedures are summarized in Algorithms~1~and~2. The motion and appearance predictors $P^{\mathcal{M}}$ and $P^{\mathcal{A}}$ are trained using \chkD{the} time-lapse video datasets $\mathcal{D^M}$ and $\mathcal{D^A}$. \chkD{These} datasets contain $N$ video clips $\mathcal{S}_i~(i=1, ..., N)$, each of which consists of $T_i$ images $\mathcal{S}_i = \{\textbf{I}^{i}_1, \textbf{I}^{i}_2, ..., \textbf{I}^i_{T_i}\}$ in a time series. 
%\chkS{All video clips in datasets are trained using separated minibatches $\{\mathcal{B}\}$, each of which contains multiple video clips, and thus it is guaranteed that latent codes are learned per training video in a single epoch. }
\chkR{Note that, in the training of our motion predictor (Algorithm~1), each minibatch $\{\mathcal{B}\}$ uses frames only from a specific set of video clips $\{S_j\}$ randomly selected in each epoch, and a latent code is learned and saved for each training video.}

\begin{figure}[t]
\begin{center}
\begin{tabular}{l}\hline
\textbf{Algorithm 2.} Training \chkE{of} Motion Predictor \\\hline
\textbf{Input}: $\mathcal{D^M} = \{\mathcal{S}_1, \mathcal{S}_2, ..., \mathcal{S}_N\}$, $\mathcal{S}_i = \{\textbf{I}^{i}_1, \textbf{I}^{i}_2, ..., \textbf{I}^{i}_{T_i}\}$\\ 
~1:  {\bf for} $i = 1$ {\bf to} $N$ \textbf{do} \\
~2: \hspace{1em} Common motion field $\textbf{B}^\mathcal{M}_{\mathcal{S}_i}$ for $\mathcal{S}_i$ $\leftarrow \textbf{0}$\\
~3: \textbf{endfor} \\
~4:  \textbf{for each} epoch \textbf{do} \\
~5: \hspace{1em} \textbf{for each} minibatch $\mathcal{B} = \{\mathcal{S}_j\} \subset \mathcal{D^M}$  \textbf{do} \\
~6: \hspace{2em} $\mathcal{L^M} \leftarrow 0$\\
~7: \hspace{2em} \textbf{for each} $\mathcal{S}_j$ in $\mathcal{B}$  \textbf{do} \\
~8: \hspace{3em} $t$ $\leftarrow$ {\sc RandomSample}$([1, T_j])$\\
~9: \hspace{3em} $\textbf{I}^\mathcal{M}_t$, $\textbf{I}^\mathcal{M}_{t+1}$ $\leftarrow \textbf{I}_t^j, \textbf{I}_{t+1}^j \in \mathcal{S}_j$\\
10: \hspace{3em} $\textbf{z}^\mathcal{M}$ $\leftarrow$ $E^{\mathcal{M}}(\textbf{B}^\mathcal{M}_{\mathcal{S}_j})$\\
11: \hspace{3em} $\hat{\textbf{B}}^\mathcal{M}_{t+1}$ $\leftarrow$ $\frac{1}{\beta}$ $P^{\mathcal{M}}(\textbf{I}^\mathcal{M}_{t}, \textbf{z}^\mathcal{M})$\\
12: \hspace{3em} $\hat{\textbf{O}}^\mathcal{M}_{t+1}$ $\leftarrow$ {\sc Warp($\hat{\textbf{B}}^\mathcal{M}_{t+1}$, $\textbf{I}^\mathcal{M}_{t}$)}\\
13: \hspace{3em} $\mathcal{L}^{\mathcal{M}}$ $\leftarrow$ $\mathcal{L}^{\mathcal{M}}$ + $\lambda^\mathcal{M}_{p}\!\mathcal{L}^\mathcal{M}_p$ + $\lambda^\mathcal{M}_{tv}\!\mathcal{L}^\mathcal{M}_{tv}$ \\
14: \hspace{3em} $\textbf{B}^\mathcal{M}_{\mathcal{S}_j}$ $\leftarrow$ $\hat{\textbf{B}}^\mathcal{M}_{t+1}$ \\
15: \hspace{2em} \textbf{endfor}\\
16: \hspace{2em} $E^{\mathcal{M}}$, $P^{\mathcal{M}}$ $\leftarrow$ {\sc Optimize}$(\frac{\partial \mathcal{L}^{\mathcal{M}}}{\partial E^{\mathcal{M}}})$,  {\sc Optimize}$(\frac{\partial \mathcal{L}^{\mathcal{M}}}{\partial P^{\mathcal{M}}})$ \\
17: \hspace{1em} \textbf{endfor} \\
18: \textbf{endfor} \\
\hline
\end{tabular}
\end{center}
\end{figure}

\begin{figure}[t]
\begin{center}
\begin{tabular}{l}\hline
\textbf{Algorithm 3.} Training \chkE{of} Appearance Predictor \\\hline
\textbf{Input}: $\mathcal{D^A} = \{\mathcal{S}_1, \mathcal{S}_2, ..., \mathcal{S}_N\}$, $\mathcal{S}_i = \{\textbf{I}^{i}_1, \textbf{I}^{i}_2, ..., \textbf{I}^{i}_{T_i}\}$\\ 
~1:  \textbf{for each} epoch \textbf{do} \\
~2: \hspace{1em} \textbf{for each} minibatch $\mathcal{B} = \{\mathcal{S}_j\} \subset \mathcal{D^A}$  \textbf{do} \\
~3: \hspace{2em} $\mathcal{L^A} \leftarrow 0$\\
~4: \hspace{2em} \textbf{for each} $\mathcal{S}_j \in \mathcal{B}$  \textbf{do} \\
~5: \hspace{3em} $t, \tau$ $\leftarrow$ {\sc RandomSample}$([1, T_j])$\\
~6: \hspace{3em}  $\textbf{I}^\mathcal{A}$, $\textbf{I}^\mathcal{A}_{\tau}$ $\leftarrow \textbf{I}_t^j, \textbf{I}_{\tau}^j \in \mathcal{S}_j$\\
~7: \hspace{3em} $\textbf{z}^\mathcal{A}_{\tau}$ $\leftarrow$ $E^{\mathcal{A}}(\textbf{I}^\mathcal{A}_{\tau})$\\
~8: \hspace{3em} $\hat{\textbf{C}}^\mathcal{A}_{\tau}$ $\leftarrow$ $P^{\mathcal{A}}(\textbf{I}^\mathcal{A}, \textbf{z}^\mathcal{A}_{\tau})$\\
~9: \hspace{3em} $\hat{\textbf{O}}^\mathcal{A}_{\tau}$ $\leftarrow$ {\sc ColorTransfer}($\hat{\textbf{C}}^\mathcal{A}_{\tau}$, $\textbf{I}^\mathcal{A})$\\
10: \hspace{3em} $\mathcal{L}^{\mathcal{A}}$ $\leftarrow$ $\mathcal{L}^{\mathcal{A}}$ +  $\lambda^\mathcal{A}_{s}\!\mathcal{L}^\mathcal{A}_{s}$ +  $\lambda^\mathcal{A}_{sp}\!\mathcal{L}^\mathcal{A}_{sp}$ + $\lambda^\mathcal{A}_{c}\!\mathcal{L}^\mathcal{A}_c$ + $\lambda^\mathcal{A}_{tv}\!\mathcal{L}^\mathcal{A}_{tv}$ \\
11: \hspace{2em} \textbf{endfor}\\
12: \hspace{2em} $E^{\mathcal{A}}$, $P^{\mathcal{A}}$ $\leftarrow$ {\sc Optimize}$(\frac{\partial \mathcal{L}^{\mathcal{A}}}{\partial E^{\mathcal{A}}})$, {\sc Optimize}$(\frac{\partial \mathcal{L}^{\mathcal{A}}}{\partial P^{\mathcal{A}}})$ \\
13: \hspace{1em} \textbf{endfor} \\
14: \textbf{endfor} \\
\hline
\end{tabular}
\end{center}
\end{figure}

\section{Latent code prediction using LSTM}
\label{sec:LSTM}
\begin{figure*}[t]
   \includegraphics[width=1.\linewidth, clip]{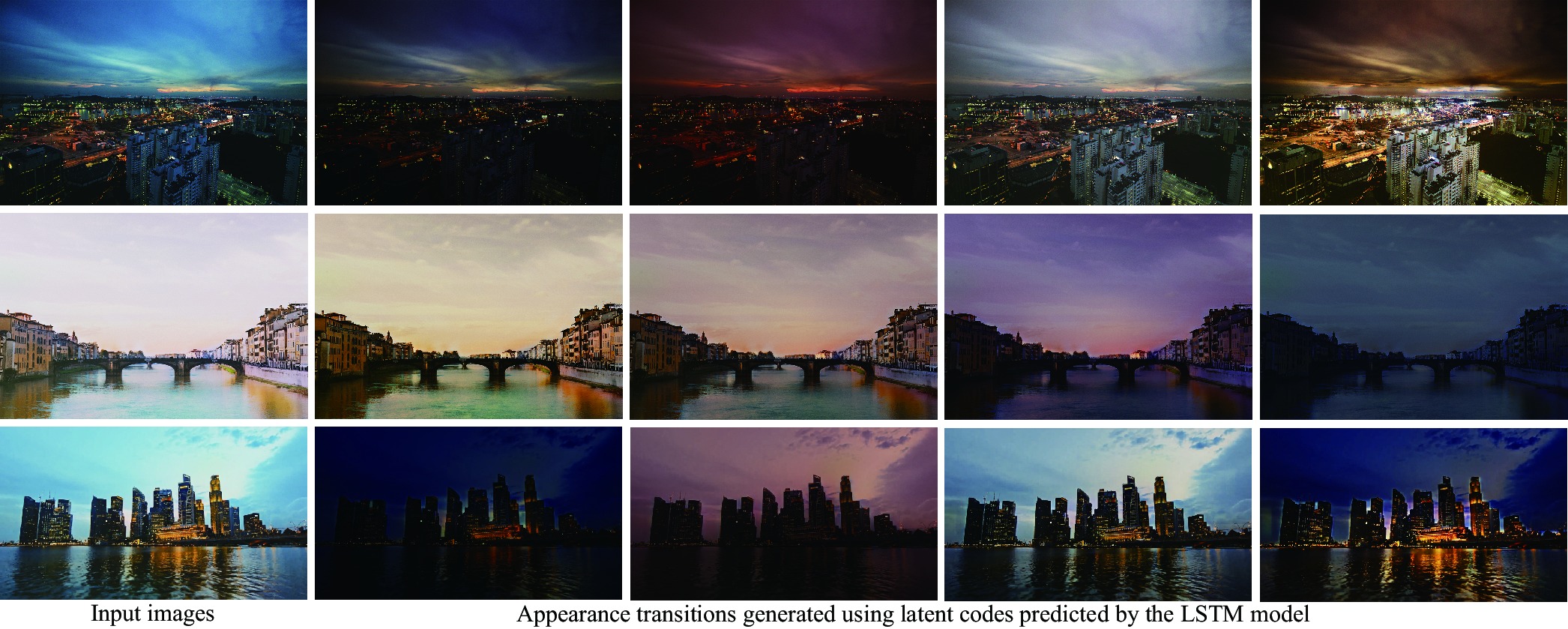}
   \caption{
    \chkE{Appearance predictions only from input images. In each row, the latent code for the first frame is encoded using the appearance encoder, and successive latent codes are predicted recurrently by the LSTM model.} Various appearance transitions \chkE{are} reproduced \chkE{from different} latent code sequence\chkE{s}, \chkE{each of which varies smoothly} via latent\chkE{-}space interpolation. From top to bottom, the output \chkE{resolutions} are $1\chkD{,}024\times683$, $1\chkD{,}080\times720$, and $700\times394$. 
    \chkV{Input photos: :DC Snapshots/Unsplash.com, Domenico Loia/Unsplash.com, and Shih et al.~\shortcite{DBLP:journals/tog/ShihPDF13}.}
}\label{fig:lstm_prediction}
\end{figure*}

To generate \chkE{latent code sequences for appearance} without using the codebook \chkE{in the inference phase}, we can use \chkE{a} simple LSTM neural \chkE{network that predicts} future latent codes \chkE{recurrently}. 
\chkE{The LSTM model is trained in advance using} latent code sequences in the codebook.
\chkE{In the inference phase, the first latent code is encoded by the appearance encoder, and successive latent codes are predicted by the LSTM model recurrently.}
The network \chkE{architecture} is shown in Table~\ref{table:lstm}. 
Figure~\ref{fig:lstm_prediction} shows the \chkE{resultant video sequences with latent code} sequences predicted \chkE{only} from input images \chkE{in the left}.

\section{Generalizability}
\chkQ{
%提案手法は基本的に雲の動きや一日のアピアランスの変化を対象としているが、さらなる一般化可能性についても調査した。
%Although our \chkR{main} focus is on generating cloud-like motion and one-day appearance transition, we further investigated the generalizability of our method. 
\chkR{Whereas most of our results contain} cloud-like motion and one-day appearance transition \chkR{simply because time-lapse videos in available datasets typically capture such scenes}, we further investigated the generalizability of our method. 
}

\begin{table}[t]
\begin{center}
\caption{Network architecture of the LSTM model for \chkE{predicting} latent codes for appearance. }
\label{table:lstm}
\begin{tabular}{l|l|l}
\hline
Component & Layers  &  Specifications  \\ \hline \hline
          & fc   & Linear(128) \\
LSTM model    & lstm    & LSTM(128)\\
	     & fc   &  Linear(8)  \\ \hline       
\end{tabular}
\end{center}
\end{table}

\begin{figure*}[t]
   \includegraphics[width=1.\linewidth, clip]{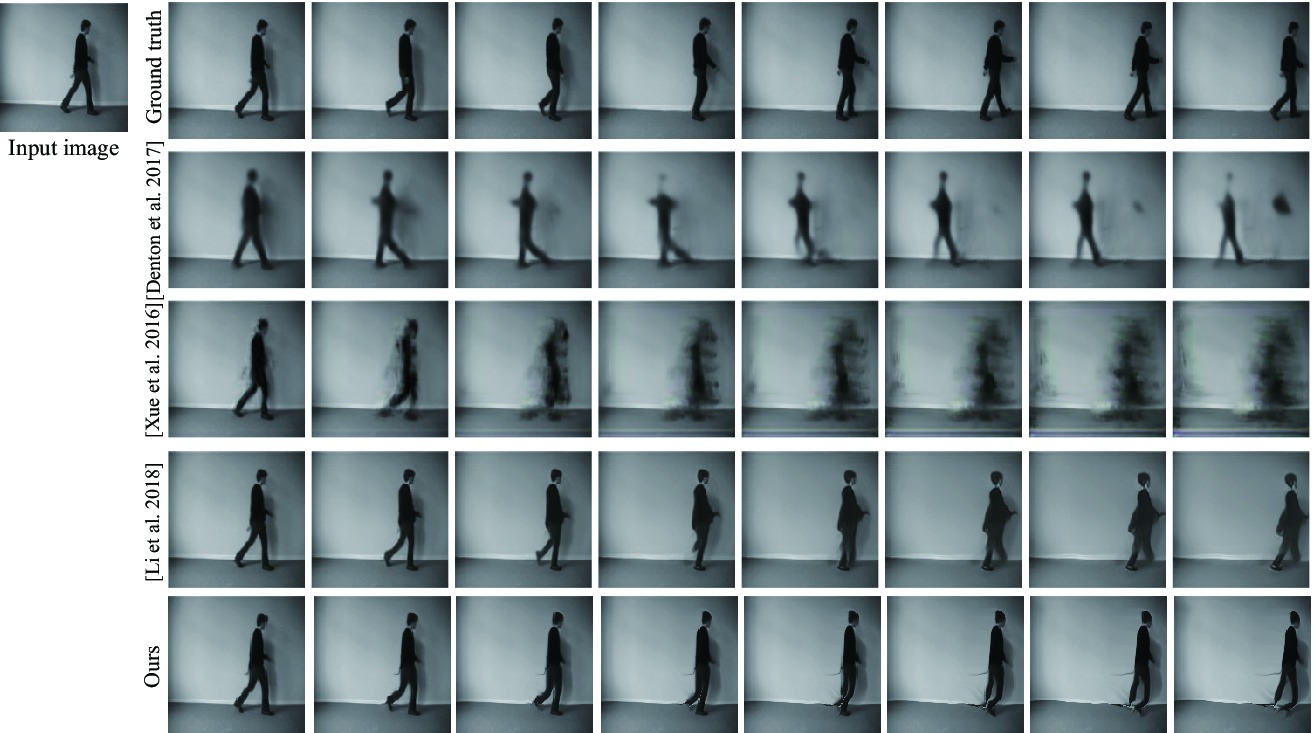}
   \caption{\chkQ{
%    Walking motion generation comparisons using a KTH dataset~\cite{DBLP:conf/icpr/SchuldtLC04}. }
    \chkR{Comparison of gait motions generated from the} KTH dataset~\cite{DBLP:conf/icpr/SchuldtLC04}. 
    %\vspace{3mm}
    }
}\label{fig:walking}
\end{figure*}

%図~\ref{fig:walking}は人物の歩行動画のKTH データセット~\cite{DBLP:conf/icpr/SchuldtLC04}を、モーション生成器に学習させて、既存手法と比較した結果である。提案手法は、歩行の進行方向や生成される画像のそれらしさの点において、既存手法のいくつか\cite{DBLP:conf/nips/Xue0BF16,DBLP:conf/nips/DentonB17}よりも高品質な結果を生成できているのがわかる。また、フレームの前半は3DCNNに基づく最新手法~\cite{Prediction-ECCV-2018}にも匹敵する外観である。しかし、提案手法は長期的状態依存性を考慮していないため、フレームの後半の足が交差した後にどちらの足が前に出るか推定するのが難しい。提案手法は潜在変数で結果を制御しながら、従来手法に比べてより高解像度で多くのフレームを生成できるが、このような長期的な状態依存性を考慮した予測は今後の課題である。
\chkQ{
%Figure~\ref{fig:walking} shows walking motion comparisons generated using a KTH dataset~\cite{DBLP:conf/icpr/SchuldtLC04}. 
Figure~\ref{fig:walking} \chkR{compares gait motions} generated \chkR{from the} KTH dataset~\cite{DBLP:conf/icpr/SchuldtLC04}. 
%Our method can produce higher-quality results than some of the existing methods~\cite{DBLP:conf/nips/Xue0BF16,DBLP:conf/nips/DentonB17} in terms of the direction of walking and the plausibility of the generated frames. 
Our method \chkR{yields more plausible} results than \cite{DBLP:conf/nips/Xue0BF16,DBLP:conf/nips/DentonB17}. 
%In addition, the quality of the first half frames of ours is comparable to that of state-of-the-art~\cite{Prediction-ECCV-2018}. 
\chkR{The} quality of \chkR{our first-half} frames is comparable to that of \chkR{the} state-of-the-art~\cite{Prediction-ECCV-2018}. 
%Meanwhile, our latter\chkR{-}half frames indicate that our method does not consider long-term dependency because it cannot properly predict which leg comes forward from the crossed legs in the single frame. 
Meanwhile, our latter\chkR{-}half frames indicate that \chkR{there is room for improvement in the prediction of which leg precedes next after the leg-crossed pose. Modeling} long-term dependency \chkR{to handle such a situation is left for future work.} 
}

\begin{figure*}[t]
   \includegraphics[width=1.\linewidth, clip]{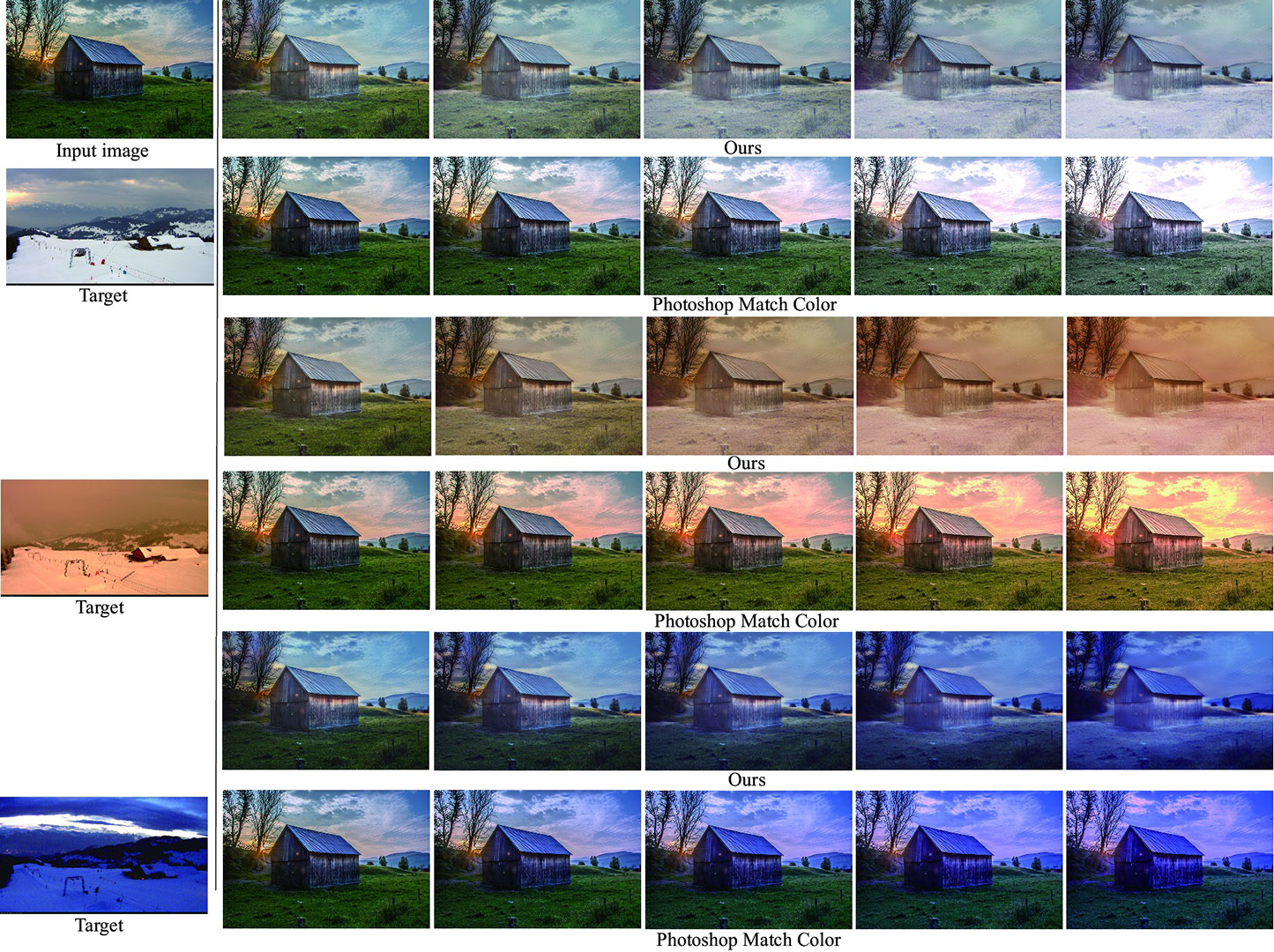}
   \caption{\chkQ{
%    Winter appearance generation comparisons using a transient attribute dataset~\cite{DBLP:journals/tog/LaffontRTQH14}.} 
    \chkR{Comparison of season transitions into winter, generated from the} transient attribute dataset~\cite{DBLP:journals/tog/LaffontRTQH14}.} 
    \chkV{Input photo: Kevin Jarrett/Unsplash.com.}
    }\label{fig:winter}
\end{figure*}

\chkQ{
%図~\ref{fig:winter}はアピアランス生成器にLaffont の季節画像データセット~\cite{DBLP:journals/tog/LaffontRTQH14}を学習させて、冬への変化の様子をMatch color と比較している。提案手法は滑らかなcolor transfer map を推定するため、窓枠の灯りのような高周波の変化は扱えないものの、一日の変化の他にもMatch color と比べてローカルな色変化を再現できる。
%Figure~\ref{fig:winter} shows winter appearance comparisons generated using a transient attribute dataset~\cite{DBLP:journals/tog/LaffontRTQH14}. 
Figure~\ref{fig:winter} \chkR{compares season transitions into winter, generated from the} transient attribute dataset~\cite{DBLP:journals/tog/LaffontRTQH14}. 
%While our method cannot handle high-frequency appearance variations such as lighted window because of the smoothing constraint of the color transfer maps, it can handle more locally-varying appearance transition than general image editing software. 
\chkR{Our transition sequences contain more wider variations in spatially-local appearance and are more faithful to the target images than those of Photoshop Match Color, even in different times of the day.}
%\chkQ{Our method can also reproduce appearance of different times of the day while preserving the structure of the same scene. }
%同じシーンの時間変化を、ちゃんとシーンの構造を保って再現できている
%While our method cannot handle high-frequency appearance variations such as lighted window because of the smoothing constraint of the color transfer maps, it can handle more locally-varying appearance transition than general image editing software. 
% 窓の明かりは画像が小さすぎて確認しづらいので、言及しなくてよいのでは。(金森)
}

\section{User study}
\label{sec:UserStudy}
\chkQ{
%提案手法が自然な動きとアピランスの変化を再現できるか定量的に評価するために、ユーザスタディを実施した。
%比較対象は既存手法~\cite{xiong2018learning,Prediction-ECCV-2018}と商用ソフト（Plotagraph、After Effects、またはPhotoshop）である。既存手法とはフェアに比較するために、既存手法の結果も提案手法と同じようにループさせ、我々の結果の解像度も既存手法と同じ$128\times128$ に縮小した。商用ソフトによる動画生成結果は、Youtube やVimeo に公開されたアーティストによる手作業の編集結果を利用し、提案手法に対しても同様の入力画像を用いた。これらの結果にはアピアランスは含めれていないため、アピアランスについては別途Photoshnop のMatch color を用いて追加で評価した。なお、アピアランスに関してはどちらも写実性が高いと思われるので、写実性についての評価ではなく、参照画像に対する忠実さを評価した。我々はx人のユーザに、それぞれの手法のクリップを並べた動画20シーン（既存手法との比較10シーン、商用ソフトとの比較10シーン)を順番に提示した。それぞれの手法の結果に対して、1-to-4 の範囲で「動き」と「色の変化」として自然さ(or 忠実性)についてスコアリングを依頼した。
%We conducted user studies to quantitatively validate our method can reproduce plausible motion and appearance variations. 
We conducted \chkR{a user study for subjective validation of the plausibility of our results}. 
%We compared our method with commercial software (Plotagraph, After Effects, and Photoshop) as well as the previous methods~\cite{xiong2018learning,Prediction-ECCV-2018}. 
We compared our method with commercial software (Plotagraph, After Effects, and Photoshop) \chkT{that requires manual annotations (e.g., static and movable regions plus fine flow directions)} as well as the previous methods~\cite{xiong2018learning,Prediction-ECCV-2018}. 
We used 20 different scenes (\chkR{ten} for \chkR{comparisons with} the previous methods and \chkR{the other ten} for commercial software). 
For fair comparisons with the previous methods~\cite{xiong2018learning,Prediction-ECCV-2018}, we made \chkR{their} results looped in \chkR{the same way as} \chkR{ours} and minified our results to the same size ($128 \times 128$) as their results. 
%For comparisons with commercial software, we collected results manually edited by artists, from Youtube and Vimeo, and our method used the same input image. 
For comparisons with commercial software, we collected \chkR{manually-created animations} from Youtube and Vimeo, and \chkR{generated our results from} the same input \chkR{images}. 
Because the \chkR{collected animations} do not contain appearance transition, we created two \chkR{more} results containing only appearance transition using Photoshop Match Color. 
\chkR{The evaluation criteria are i) plausibility w.r.t. motion and appearance transition for motion-added animations and ii) faithfulness against target images for appearance-only animations (i.e., comparisons with Photoshop Match Color); appearance-only results are highly plausible in any methods, and thus we omitted plausibility for them.}
%For appearance comparisons with commercial software that uses reference frames (which are also used to extract latent codes for our method), we evaluated the faithfulness of resultant videos to reference frames because the plausibility of our and the compared results seem to be high. 
%We showed 11 users videos containing three (or two) clips generated by our method and the compared methods only once, and then asked them to score clips on a 1-to-4 scale ranging from ``implausible (or unfaithful)'' to ``plausible (or faithful)'' in terms of motion and color variations. 
\chkR{We requested 11 subjects to score video} clips on a 1-to-4 scale ranging from ``implausible (or unfaithful)'' to ``plausible (or faithful)'' \chkR{after they watch each clip only once. The movie used in this user study is submitted as a supplemental material.}
}

\begin{figure*}[t]
   \includegraphics[width=1.\linewidth, clip]{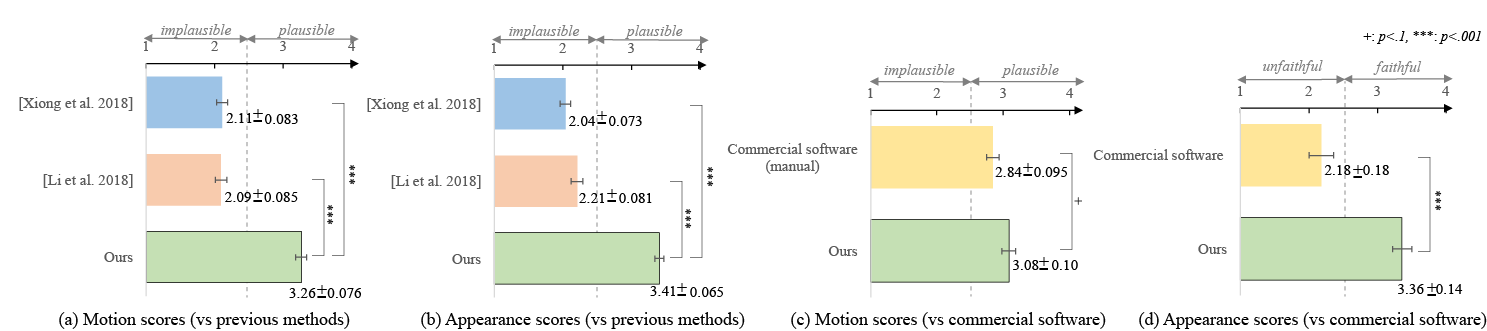}
   \caption{
%    \chkQ{User study results confirming that our method produces natural motion and appearance transition by comparing the previous methods and commercial software. The error bars represent standard deviation. } 
    \chkR{Statistics of our user study. The graphs indicate that our method yields more plausible results than} \chkQ{the previous methods and commercial software. The error bars represent standard errors. The results marked with * show statistically\chkR{-}significant differences (paired t-test). }}\label{fig:user_study}
\end{figure*}

\chkQ{
%図~\ref{fig:user_study} に各手法に対するスコアの平均値と標準偏差を示す。既存手法との比較から、動きとアピアランスの両方の点において、提案手法は格段に高品質なアニメーションを生成できている。さらに、手動の動画生成ソフトと比べても遜色ない結果であり、自動のアピアランス生成(Match color) と比べても自然な結果を再現できることを確認した。
%Figure~\ref{fig:user_study}~(a) and (b) show that our method significantly outperforms the previous methods in terms of plausibility of motion and appearance. 
\chkR{Figure~\ref{fig:user_study} summarizes the statistics of the user study. The graphs in Figure~\ref{fig:user_study}~(a) and (b) indicate that} our method significantly outperforms the previous methods in terms of plausibility of motion and appearance.
%In Figure~\ref{fig:user_study}~(c), we can confirm that the motion scores of ours are slightly better than those of manual commercial software. 
In Figure~\ref{fig:user_study}~(c), we can confirm that \chkR{our} motion scores are slightly better than those \chkR{manually created using} commercial software. 
%Figure~\ref{fig:user_study}~(d) shows that the our method can more faithfully reproduce reference-based results and can handle more locally-varying appearance than commercial software, as demonstrated in Figures~\ref{fig:matchcolor} and~\ref{fig:winter}. 
Figure~\ref{fig:user_study}~(d) shows that our method can reproduce \chkR{target styles more faithfully} and can handle \chkR{wider variations in} appearance than commercial software, as demonstrated in Figures~\ref{fig:res5}, \ref{fig:matchcolor}, and~\ref{fig:winter}. 
%To evaluate this quantitatively, we conducted a second study to estimate the faithfulness of appearance transfer. We asked additional $x$ users to score the clips in terms of faithfulness for given source images. The faithfulness score of ours was $xx\pm x$, and that of commercial software was $xx\pm xx$. 
%We provide the resultant videos used in our user studies in the supplemental material. 
}

\end{document}